\documentclass[usenatbib]{mnras}
\usepackage[utf8]{inputenc}
\usepackage{aas_macros}
\usepackage{amsfonts}
\usepackage{amsmath}
\usepackage{amssymb}
\usepackage{graphicx}
\usepackage{booktabs}
\usepackage[dvipsnames]{xcolor}
\usepackage{hyperref}
\hypersetup{colorlinks=true,allcolors=teal}

\defcitealias{2021MNRAS.507..632P}{PS21}
\defcitealias{2015JCAP...12..049S}{ST15}

\newcommand{\Mh}{\ensuremath{h^{-1}M_{\odot}}}

\newcommand{\Mpch}{\ensuremath{h^{-1}{\rm Mpc}}}

\newcommand{\eqn}[1]{equation~\eqref{#1}}

\newcommand{\figref}[1]{Fig.~\ref{#1}}
\newcommand{\secref}[1]{section~\ref{#1}}

\newcommand{\be}{\begin{equation}}
\newcommand{\ee}{\end{equation}}

\title[Quasi-adiabatic relaxation]{The quasi-adiabatic relaxation of haloes in the IllustrisTNG and EAGLE cosmological simulations} 
\author[Velmani \& Paranjape]{
 Premvijay Velmani$^{1}$\thanks{E-mail: premv@iucaa.in} \& 
 Aseem Paranjape$^{1}$\thanks{E-mail: aseem@iucaa.in},
\\  
 $^1$ Inter-University Centre for Astronomy \& Astrophysics, Ganeshkhind, Post Bag 4, Pune 411007, India}

\begin{document}
\label{firstpage}
\pagerange{\pageref{firstpage}--\pageref{lastpage}}
\maketitle

\begin{abstract}
The dark matter content of a gravitationally bound halo is known to be affected by the galaxy and gas it hosts. %
We characterise this response for haloes spanning over four orders of magnitude in mass in the hydrodynamical simulation suites IllustrisTNG and EAGLE. %
We present simple fitting functions in the spherically averaged quasi-adiabatic relaxation framework that accurately capture the dark matter response over the full range of halo mass and halo-centric distance we explore. %
We show that commonly employed schemes, which consider the relative change in radius $r_f/r_i-1$ of a spherical dark matter shell to be a function of only the relative change in its mass $M_i/M_f-1$, do not accurately describe the measured response of most haloes in IllustrisTNG and EAGLE. %
Rather, $r_f/r_i$ additionally explicitly depends upon halo-centric distance $r_f/R_{\rm vir}$ for haloes with virial radius $R_{\rm vir}$, being very similar between IllustrisTNG and EAGLE and across halo mass. %
We also account for a previously unmodelled effect, likely driven by feedback-related outflows, in which shells having $r_f/r_i\simeq1$ (i.e., no relaxation) have $M_i/M_f$ significantly different from unity. %
Our results are immediately applicable to a number of semi-analytical tools for modelling galactic and large-scale structure. %
We also study the dependence of this response on several halo and galaxy properties beyond total mass, finding that it is primarily related to halo concentration and star formation rate. %
We discuss possible extensions of these results to build a deeper physical understanding of the small-scale connection between dark matter and baryons. %
\end{abstract}

\begin{keywords}
galaxies: formation - cosmology: theory, dark matter - methods: numerical
\end{keywords}

\section{Introduction}
\label{sec:intro}
\noindent
A detailed understanding of the formation and evolution of galaxies and their interaction with their environment is a pressing open problem. 
In the Lambda-cold dark matter ($\Lambda$CDM) picture of the Universe, the primary environment of any galaxy is provided by a `halo' of dark matter surrounding it \citep[e.g.,][]{wr78}. 
Therefore, understanding the dynamical behaviour of these dark haloes during the evolution of the galaxies they host is a key ingredient needed for building a complete picture of galaxy evolution.

Dark haloes form through the gravitational collapse of overdensities that developed from tiny fluctuations in the initial distribution of matter (e.g., \citealp[][]{1974ApJ...187..425P}; for a review see \citealp{2002PhR...372....1C}). 
The properties of dark haloes identified in gravity-only cosmological $N$-body simulations have been extensively studied in the literature. For example, while these haloes are known to be triaxial \citep[][]{1988ApJ...327..507F}, their sphericalised mass profiles are found to have a universal form (\citealp{1996ApJ...462..563N,1997ApJ...490..493N}, hereafter, NFW; see also \citealp{2010MNRAS.402...21N}). 
However, as galaxies and clusters of galaxies form within these haloes from various non-gravitational baryonic interactions, their gravitational coupling to the dark matter can affect the spatial distribution and evolution of the latter. Understanding this response of a halo's dark matter content to the baryons it hosts is then critical for understanding the coupled evolution of haloes and galaxies. 

Early work treated this response using adiabatic invariants
\citep[][]{osti6457593,1984MNRAS.211..753B,1986ApJ...301...27B,1987ApJ...318...15R}. 
Using simplifying assumptions such as spherical symmetry, no shell crossing, angular momentum conservation with circular orbits for dark matter particles, \citet[][]{1986ApJ...301...27B} derived a simple formula that quantifies the adiabatic relaxation of the dark matter mass profile in terms of the final baryonic distribution (we discuss this in detail later). Besides the change in their mass profiles, dark haloes can also become more spherical as a result of galaxy formation \citep[][]{1994ApJ...431..617D}.
Currently, the most robust technique to understand the consequences of gas assembly and galaxy formation on dark matter structure is the use of high-resolution cosmological hydrodynamical (zoom) simulations, using `sub-grid' recipes for modelling very small-scale astrophysics such as feedback from stellar/supernovae activity or the effects of active galactic nuclei (AGN) (see, e.g., OWLS, \citealp{2010MNRAS.402.1536S}, Illustris; \citealp{2014MNRAS.445..175G}; FIRE, \citealp{2014MNRAS.445..581H}; EAGLE, \citealp{2015MNRAS.446..521S}; Horizon-AGN,
\citealp[][]{2017MNRAS.467.4739K}; SIMBA,
\citealp[][]{2019MNRAS.486.2827D}; IllustrisTNG, \citealp{2019ComAC...6....2N}). In such simulations, the response of the dark matter to the presence of baryons in a halo can be ascertained by comparing a halo in the full hydrodynamical simulation to a matched `partner' halo in a collisionless, gravity-only simulation performed using the same initial random fluctuations. 
Using this technique, it was found that a simple adiabatic contraction model like that of \citet[][]{1986ApJ...301...27B} is an inaccurate description of the response in a variety of simulations \citep[see, e.g.,][]{2004ApJ...616...16G,2006PhRvD..74l3522G,2010MNRAS.402..776P,2010MNRAS.406..922T,2010MNRAS.405.2161D,2010MNRAS.407..435A,2011MNRAS.414..195T,2016MNRAS.461.2658D,2019A&A...622A.197A,2022MNRAS.511.3910F}. Even in a simpler setting, where star formation and feedback effects are ignored, \citet{2010MNRAS.407..435A} found that the halo responds to the condensation of baryons to the center by becoming more spherical and compact, while the change in its mass profile is found to be significantly less than the prediction of the idealized adiabatic relaxation model.

Several authors have investigated whether discarding some of the assumptions of the idealized model of \citet[][]{1986ApJ...301...27B} can help reconcile with the simulation results. \citet{2004ApJ...616...16G} considered non-spherical orbits for the dark matter particles and suggested a simple modification to the original model. This modified empirical formula shows wide variation in its parameters across haloes from different simulations and at different redshifts \citep[][]{2006PhRvD..74l3522G,2010MNRAS.405.2161D}. On the other hand, \citet[][]{2005ApJ...634...70S} accounted for these random motions within the halo using invariant action integrals, following the method described in \citet{1980ApJ...242.1232Y}. This model gives a reasonably good approximation of the response even in modern hydrodynamical simulations \citep{2020MNRAS.495...12C}; 
however, making predictions using this model requires access to orbital phase space information of the halo, which may not always be feasible.

Physically, one expects that the overall response of the halo is mediated by a combination of different astrophysical processes that occur in the galaxy. Feedback processes are known to reduce the contraction of the halo significantly; e.g.,
supernova-driven winds can completely transform the inner density profile of the dark matter halo
\citep[][]{1996MNRAS.283L..72N}. This may be the key in reconciling the observation of dark matter cores at the center of various galaxies with the cuspy haloes found in gravity-only $\Lambda$CDM simulations \citep[see][for a review]{2014Natur.506..171P}.
However, such feedback effects do not always produce dark matter cores from cusps, rather, this can depend on the amount of gas ejected, the mass loss time scale and the frequency of starburst events 
(see, e.g., \citealp{2011ApJ...736L...2O,2014ApJ...793...46O,2012MNRAS.421.3464P}, and also \citealp{bfln18}).
In massive haloes hosting galaxy groups or clusters, while the formation of powerful AGN in the central galaxy can strongly suppress star formation,
it can still significantly reduce the adiabatic contraction of the halo \citep[][]{2011MNRAS.414..195T}.
Moreover, the fluctuation in gravitational potential due to such feedback can expel the dark matter from the inner halo producing inner cores \citep[][]{2012MNRAS.422.3081M}.

Different aspects of the halo response, such as the change in its mass profile, shape, phase space distribution and substructure population, have been explored to date in a variety of hydrodynamical simulations \citep[see, e.g.,][]{2004ApJ...611L..73K,2008ApJ...681.1076D,2014MNRAS.441.2986D,2015MNRAS.451.1247S,2017MNRAS.466.3876Z,2017MNRAS.472.4343C,2019MNRAS.484..476C,2021arXiv210900012C,2021MNRAS.501.5679C,2020MNRAS.494.4291C,freundlich+20,riggs+22}.
Understanding the nature of this response (or `baryonic backreaction') is critical in building accurate and robust models of halo shapes and sizes, for use in interpreting the results of upcoming large-volume surveys (\citealp{2015JCAP...12..049S,2018MNRAS.480.3962C,2021MNRAS.503.3596A}; see also \citealp{velliscig+14,hwvh15,mead+15}) as well as the detailed prediction of rotation curves and related statistics \citep{2021MNRAS.507..632P,2021arXiv211200026P}.
The goal of the present work is to perform a systematic, statistical study of this dark matter response in high-resolution hydrodynamical simulations incorporating realistic feedback and quantify it using simple analytical forms, including the sensitivity of this response to halo-centric distance and halo and galaxy properties. To this end, we use the publicly available suites of simulations from the IllustrisTNG and EAGLE projects.

The rest of this article is organised as follows.
In \secref{sec:sim-and-teq}, we briefly discuss the simulations used and the techniques employed in this work. In \secref{sec:results} we present our results. These are expected to be relevant for a variety of problems; for example, the change in the dark matter density profile of the halo caused by the galaxy formation affects the rotation curve of the galaxy.
We discuss such applications in \secref{sec:applic} and conclude in \secref{sec:conclusion}. 
Throughout, we will use $\log$ and $\ln$ to denote the base-10 and natural logarithm, respectively.

\section{Simulations and Techniques}
\label{sec:sim-and-teq}
In this section, we describe the numerical simulations used in this work and the generation of matched halo catalog as well as the methods employed in our study of halo response to galaxy formation.

\subsection{Simulations}
We use two suites of cosmological hydrodynamical simulations, namely the IllustrisTNG \citep[][]{2019ComAC...6....2N} and the EAGLE \citep[][]{2017arXiv170609899T} suites that were run with different (yet state-of-the-art) prescriptions for baryonic processes involved in galaxy formation. We isolate the effects of galaxy formation process on dark matter halo by comparing these hydrodynamical simulations with their corresponding gravity-only runs that evolve the same initial cosmological volumes but treating baryons as collisionless.

\subsubsection{IllustrisTNG simulations}
This suite of simulations by the TNG collaboration project were run with the \textsc{arepo} code \citep[][]{2020ApJS..248...32W}, simulating the hydrodynamics using moving mesh defined by Voronoi tessellation \citep[][]{2010MNRAS.401..791S}. This follows an updated model of galaxy formation based on the results from the original Illustris simulation; including all major baryonic processes like cooling, star formation, stellar feedback and AGN feedback 
\citep[see][for details]{2017MNRAS.465.3291W,2018MNRAS.473.4077P}. In addition to these, this updated model of TNG includes a cosmic magnetic field.
The TNG suite has three cosmological boxes TNG50, TNG100 and TNG300 with a periodic box size of $35 \Mpch$, $75 \Mpch$ and $200 \Mpch$, respectively, assuming a cosmology consistent with results from \cite{2016A&A...594A..13P}. Initial conditions of these simulations were generated with the Zel'dovich approximation \citep[][]{1970A&A.....5...84Z} at redshift $z=127$ using the \textsc{N-GenIC} code \citep[][]{2015ascl.soft02003S}, for the realization selected by $\chi^2$-minimization for cumulative halo mass function. 
Results from this simulation have been compared against observations and found to have reasonably realistic galaxies \citep[see][]{2018MNRAS.475..624N,2018MNRAS.475..648P,2018MNRAS.475..676S,2018MNRAS.477.1206N,2018MNRAS.480.5113M,2019MNRAS.490.3196P,2019MNRAS.490.3234N}.
In order to statistically study the response of a wide range of haloes to galaxy formation, we use the highest resolution runs of all three cosmological boxes; while the smallest box TNG50 provides sufficient resolution to resolve low-mass haloes, we need the large box TNG300 to get sufficient number of cluster-scale haloes. 
Throughout this work, we use the redshift $z=0.01$ data 
from IllustrisTNG for both hydrodynamical as well as the corresponding gravity-only runs.

\subsubsection{EAGLE simulations}
This suite of cosmological simulations was performed with a modified version of the \textsc{gadget-3} code that evolves the gas using smoothed particle hydrodynamics \citep[][]{2005MNRAS.364.1105S}. Initial conditions for these simulations were generated using the \textsc{ic\_2lpt\_gen} code following \cite{2010MNRAS.403.1859J}. 
From this suite, we use the $z=0$ data of L0025N0752 and L0100N1504 boxes that were run with EAGLE's reference model of galaxy formation, along with their corresponding gravity-only runs. This reference model includes sub-grid prescriptions for various baryonic processes like cooling, star formation and feedbacks  \citep[see][for more details]{2015MNRAS.446..521S,2015MNRAS.450.1937C}. While this model differs from that of IllustrisTNG in many aspects (e.g. it doesn't include a cosmic magnetic field), this suite of simulations has also produced realistic galaxies \citep[see e.g.][]{2015MNRAS.448.2941S,2015MNRAS.450.4486F,2015MNRAS.452.2879T}

\subsubsection{Halo and galaxy properties}
\label{sec:halo-props}
Along with the particle data, 
both these simulation projects provide a catalogue of Friends-Of-Friends (FOF) group haloes, found with a linking length of 0.2 times the interparticle spacing \citep[see][for specifics]{2016A&C....15...72M,2019ComAC...6....2N}.
For each halo, we take the co-moving position of the minimum of the gravitational potential within that halo as its centre, 
and define its `virial' radius $R_{\rm vir}$ as the halo-centric radius enclosing a total density of 200 times the critical density of the Universe:  $R_{\rm vir}\equiv R_{\rm 200c}$. Throughout, we consider the total mass $M$ of each halo to be the mass enclosed inside its virial radius, i.e. $M\equiv M_{200c}$. 
In addition, we have a catalogue of gravitationally bound substructures identified by the \textsc{subfind} code \citep{2001MNRAS.328..726S} within those FOF haloes. 
A single FOF group can have more than one subhalo, the one containing the central particle is considered as the central subhalo associated with that FOF halo.

For the simulations from TNG suite, we also focus on the following halo properties in addition to their mass in our study of halo response to galaxy formation. 
For the haloes in the gravity-only runs, we define the concentration as $c=2.1626 \times R_{\rm vir}/R_{V_{\rm{max}}}$, where $R_{V_{\rm{max}}}$ is the radius at which the rotation curve attains its peak. If the sphericalised mass profile of these haloes had a perfectly NFW form, this concentration would be exactly equal to the standard NFW concentration defined in terms of NFW scale radius \citep[see equation 5 of ][]{1996ApJ...462..563N}. 
This definition of concentration is convenient since it does not require any statistical fit to the measured halo profile.
For the haloes in hydrodynamic simulations, we consider 
three different properties, namely, \textbf{gas fraction} $(f_g)$ of the whole FOF group, and the \textbf{stellar mass fraction} $(f_{\ast})$ and specific star formation rate \textbf{(SSFR)} of the central subhalo associated to each of them. 
\begin{align}
    f_{g} = \frac{M_{g}^{\rm{F}}}{M^{\rm{F}}}\,; \quad
    f_{\ast} = \frac{M_{\ast}^{\rm{S}}}{M^{\rm{S}}}\,; \quad {\rm SSFR} = \frac{\rm{SFR}}{M_{\ast}^{\rm{S}}}\,.
\label{eq:galpropdefs}
\end{align}
Here $M^{\rm{F}}$ ($M_{g}^{\rm{F}}$) denotes the total mass of all (gas) particles in the FOF group, whereas $M^{\rm{S}}$ ($M_{\ast}^{\rm{S}}$) denotes the total mass of all (stellar) particles in the associated central subhalo. 
Finally, SFR denotes the sum of star formation rate of all gas cells in the central subhalo. The above definitions allow us to track the response of dark matter to the gas content of the full halo and the stellar content and activity of its central galaxy. We have also checked that using the stellar content and activity of the full halo leads to qualitatively similar results as when using the central galaxy alone.

\subsection{Halo matching}
\label{sec:methods-match}
To study how a dark halo responded to the galaxy forming in it, we need to first reliably match the catalogue of haloes found in the hydrodynamic simulation (which includes galaxy formation physics) to those found in its gravity-only run. For various numerical reasons a given halo in the hydro run may not have a true match in the halo catalogue of gravity-only run. So we first try to obtain an exhaustive catalogue of matched haloes that will be used to build a statistical description of this halo response.

\subsubsection{Matching procedure}
We match the haloes using the particle data associated with the haloes;
while the mass of each dark matter particle differs between the hydrodynamical and gravity-only runs, the \emph{number} of dark matter particles within the same initial periodic box is the same for each of the five pairs of simulations that we consider in IllustrisTNG and EAGLE. So a given particle in a hydro simulation has originated from the same region as the particle of the same ID in its corresponding gravity-only run.
For any given pair of haloes, with one in the hydrodynamical simulation and the other in the corresponding gravity-only run, we define the matching fraction of each of those two haloes (with respect to the other) 
as the fraction of its dark matter particles that are also present in the other halo. Below, we describe how we use 
these matching fractions to decide if the given pair can be considered a valid matched pair.

\begin{figure}
    \centering
    \includegraphics[width=\linewidth]{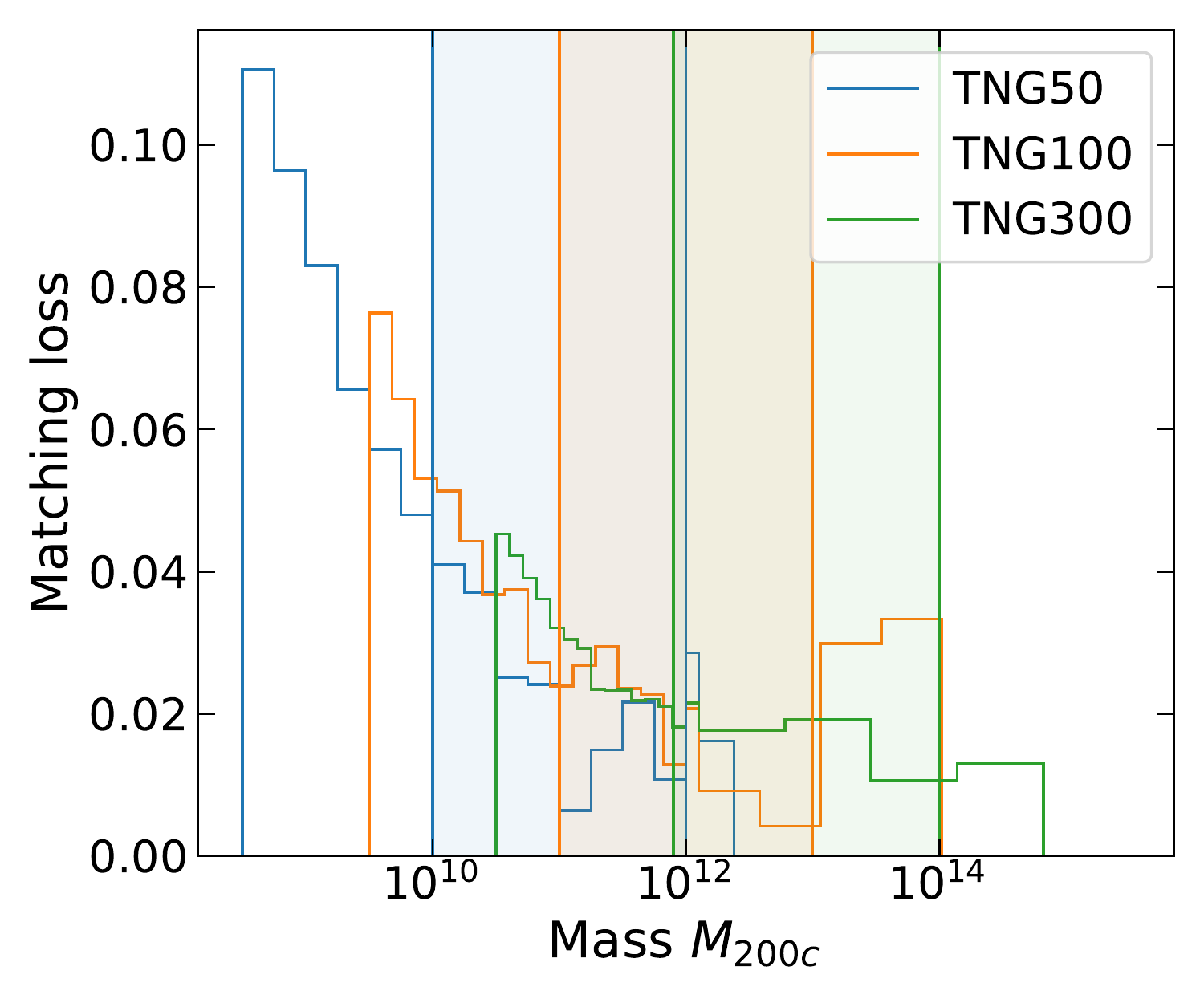}
    \caption{Fraction of haloes in the TNG hydrodynamical simulations that have not found a match is shown as a function of mass $M_{200c}$. coloured vertical bands represent the mass range relevant for this work in each of the three TNG simulation boxes.}
    \label{fig:matching-loss-all}
\end{figure}

\begin{figure*}
\centering
\includegraphics[clip,trim={0.5cm 0cm 2cm 0.5cm}, width=\linewidth]{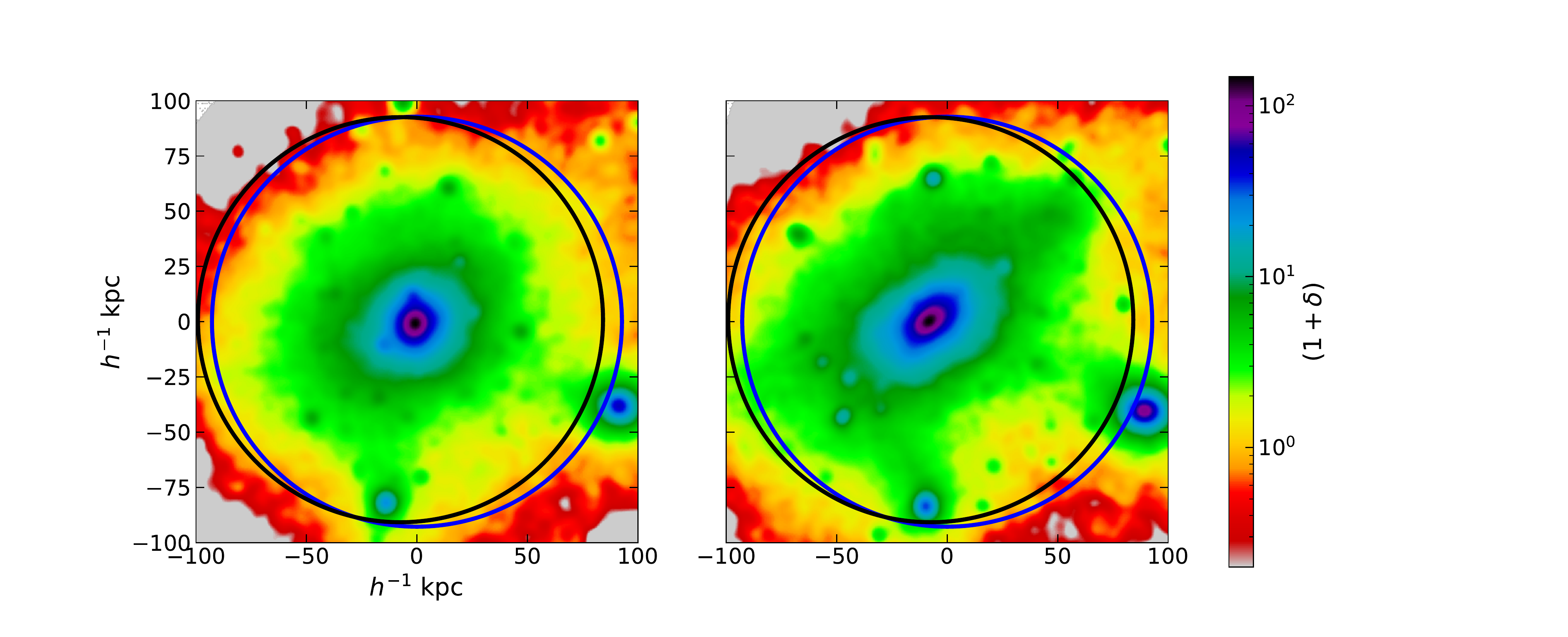}
\includegraphics[clip,trim={0.5cm 0cm 2cm 0.5cm}, width=\linewidth]{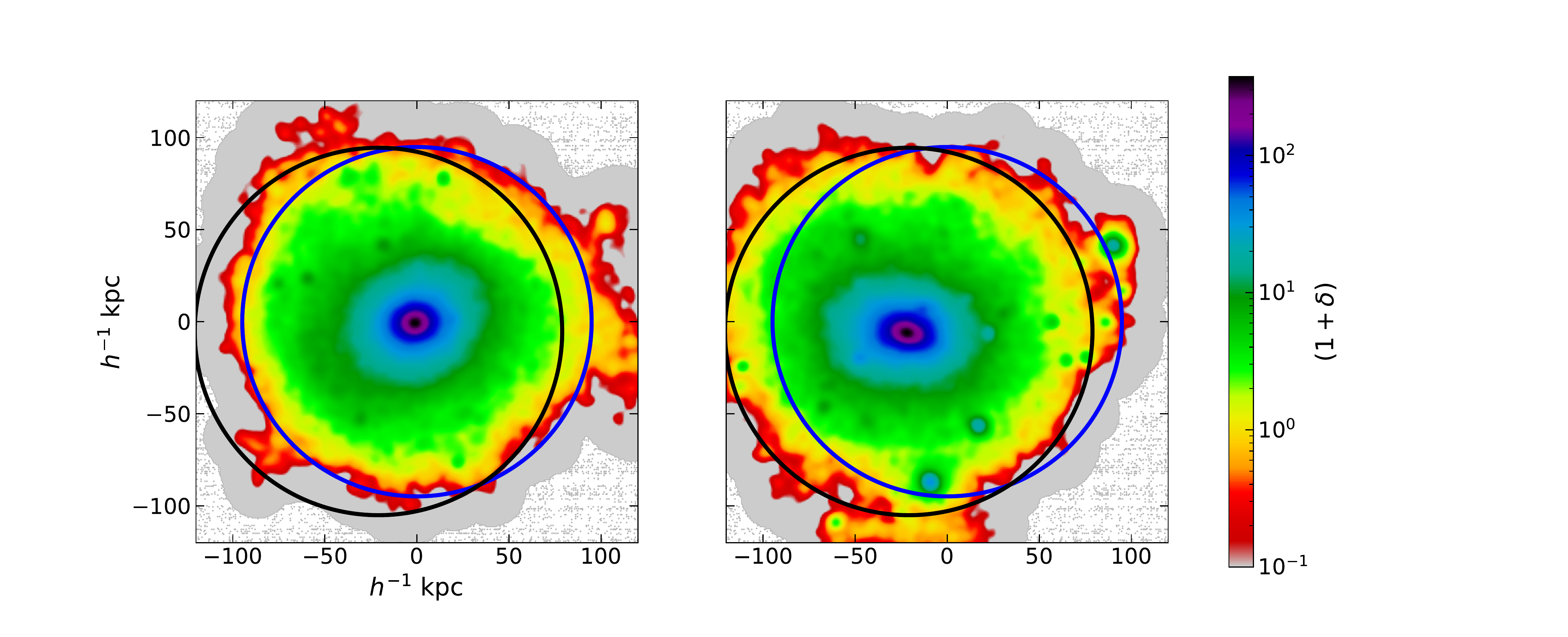}
\caption{Visually inspecting two matched FOF halo pairs, one each from IllustrisTNG \emph{(top row)} and EAGLE \emph{(bottom row)} using 2D-projected dark matter density field around the center of the hydrodynamical halo in a thick slice. The \emph{left panel} shows the halo in the hydrodynamical simulation and the \emph{right panel} shows the corresponding matched pair in the gravity-only simulation. The black circle shows the virial boundary of the gravity-only halo and blue circle shows that of the hydrodynamical halo. In both cases, the hydrodynamical halo is noticeably more spherical and compact than its gravity-only counterpart, with a spatially offset center. See text for a discussion.} 
\label{fig:single-halo-pair}
\end{figure*}

Computing the matching fractions 
for every pair of haloes is computationally expensive with $O(n^2)$ for millions of haloes in the catalogue. We decrease the complexity to $O(n)$ by supplying, for each halo in the hydro-simulation, an ordered list of most probable match candidates in the gravity-only run. These candidate lists of gravity-only haloes are generated and ranked based on the spatial positions of the haloes and their masses using a KD-tree based neighbour finding algorithm, implemented using \texttt{scipy.spatial.KDTree}.
For each halo in the hydro run, we test if the matching fraction of this halo with respect to any of the haloes in its match candidate list exceeds the value of 0.5. This ensures that at most one halo is selected as a match for each of the hydrodynamical haloes, so that our matched catalogue of halo pairs will be a subset of the source catalogue without repetitions.
While we match as many haloes as possible, it is also important to ensure that false matches don't plague our study. Based on the results in Appendix \ref{sec:apndx-matching}, we therefore additionally require that in a valid matched pair, the gravity-only FOF halo must also have a matching fraction of more than 0.5 with respect to the hydrodynamical halo.
Our FOF based matching can be compared with
\citet[][]{2018MNRAS.481.1950L} where a similar matching algorithm has been followed for central subhaloes.

\subsubsection{Halo pair catalogue}
We generate a matched catalogue of haloes for each of the five simulations studied in this work, including FOF haloes resolved with more than 1000 particles\footnote{Mass resolution in the gravity-only runs are $7 \times 10^7 \Mh$, $8.9 \times 10^6 \Mh$ and $5.4 \times 10^5$ for TNG300, TNG100 and TNG50 respectively; whereas $9.7 \times 10^6 \Mh$ for the L100 and $1.21 \times 10^6 \Mh$ for the L25 simulation of EAGLE.}. The fraction of hydrodynamical haloes that fails to be part of the matched catalogue is shown in \figref{fig:matching-loss-all} in bins of halo mass for the IllustrisTNG simulations. %
In the mass range in which the halo samples are selected for this work (see \secref{sec:results-mass}), more than 96\% of the haloes in hydrodynamical simulation have been assigned a match in the gravity-only run. The small fraction of unmatched haloes primarily reside in dense environments 
where our algorithm presumably fails due to
the inherent issues with 3D FOF algorithm in dealing with mergers. Similar results hold for the EAGLE simulations as well. 
For illustrative purpose, a visual representation of two randomly chosen halo pairs, one each from IllustrisTNG and EAGLE, is shown in \figref{fig:single-halo-pair}.

\subsection{Methods to study the halo response}
\label{sec:method}
The response of dark matter halo to galaxy formation has primarily two aspects, contraction or expansion of the halo towards the centre and a change in its triaxial shape. In this work, we study the former aspect of the halo response, by focusing on spherically averaged mass profiles. 
The illustrative haloes shown in \figref{fig:single-halo-pair} become more compact and spherical in the hydrodynamical simulation that includes galaxy formation.
Also notice that there is an offset between the center-of-potential locations of matched pairs of haloes. These offsets are likely correlated with the halo tidal environment and will be interesting to follow-up in future work.

\subsubsection{Mass profiles}
\label{subsec:massprofiles}

The overall expansion and/or contraction of dark matter in response to galaxy formation can be studied through the differences in spherically averaged mass profiles between matched haloes. For the dark matter, these radial profiles are obtained by adding up the mass of all dark matter particles contained within concentric spherical shells. In addition to these, we also need baryon mass profiles in modelling the dark matter response. While stellar mass profiles are computed in a similar fashion as dark matter,
for the gas mass profiles we use a Gaussian kernel to assign mass enclosed to each of the spherical shells\footnote{Throughout this work, we consider concentric shells defined by their radii and the mass enclosed by such a shell is the mass in the sphere bounded by that shell.
}. 
The width of this Gaussian kernel was taken to match the SPH smoothing length for the EAGLE simulation, whereas for IllustrisTNG we use the cube root of the Voronoi cell volume to define the kernel smoothing scale. We have tested that our results are robust to differences in the choice of this kernel.

\subsubsection{Quasi-adiabatic relaxation model}
\label{sec:methods-adiab}
The impact of galaxy formation on the dark halo is expected to be primarily an adiabatic relaxation of dark matter particle orbits in response to baryon condensation \citep[][]{1986ApJ...301...27B}. We start by discussing this simplified model and study more complex effects such as the impact of baryonic feedback processes below.
Assuming that the dark matter halo is spherical and doesn't undergo shell crossing while baryons condense towards the centre, the adiabatic relaxation of any given dark matter shell is determined by the change in baryonic mass within that shell. 
Consider a shell enclosing a \emph{dark matter} mass $M_i^d(r_i)$ in radius $r_i$ in the unrelaxed halo. After relaxation, the radius of the shell changes to $r_f$. By definition, the dark matter mass $M_f^d(r_f)$ enclosed in $r_f$ in the relaxed halo is simply
\be 
M_f^d(r_f) = M_i^d(r_i)\,.
\label{eq:DMmass}
\ee
The \emph{total} mass $M_i(r_i)$ enclosed in $r_i$ in the unrelaxed halo, on the other hand, does not necessarily equal the total mass $M_f(r_f)$ enclosed in $r_f$ in the relaxed halo.
If angular momentum were to be conserved and the dark matter particle orbits stay circular, then the amount of relaxation of the shell is completely determined by the change in this total mass within the shell
\citep[][]{1986ApJ...301...27B},
\begin{align}
    r_i \,M_i(r_i) = r_f \,M_f(r_f) %
    \implies 
\frac{r_f}{r_i} = \frac{M_i(r_i)}{M_f(r_f)}\,. 
\label{eq:AR}
\end{align}
Extending this idealised scenario, quasi-adiabatic relaxation models consider the relaxation ratio $r_f/r_i$ as a function of the mass ratio $M_i/M_f$.
\begin{align}
\frac{r_f}{r_i} &= 1 + \chi \left( \frac{M_i(r_i)}{M_f(r_f)} \right) 
\label{eq:qAR}
\end{align}
For example, the baryonification procedures in \cite{2015JCAP...12..049S,2021MNRAS.503.4147P} include dark matter response as a quasi-adiabatic relaxation with
\be
\chi(y) = q\,(y-1)\,.
\label{eq:chi-linear}
\ee

\begin{figure*}
    \centering
    \includegraphics[width=\linewidth,trim={0 0 2cm 0},clip]{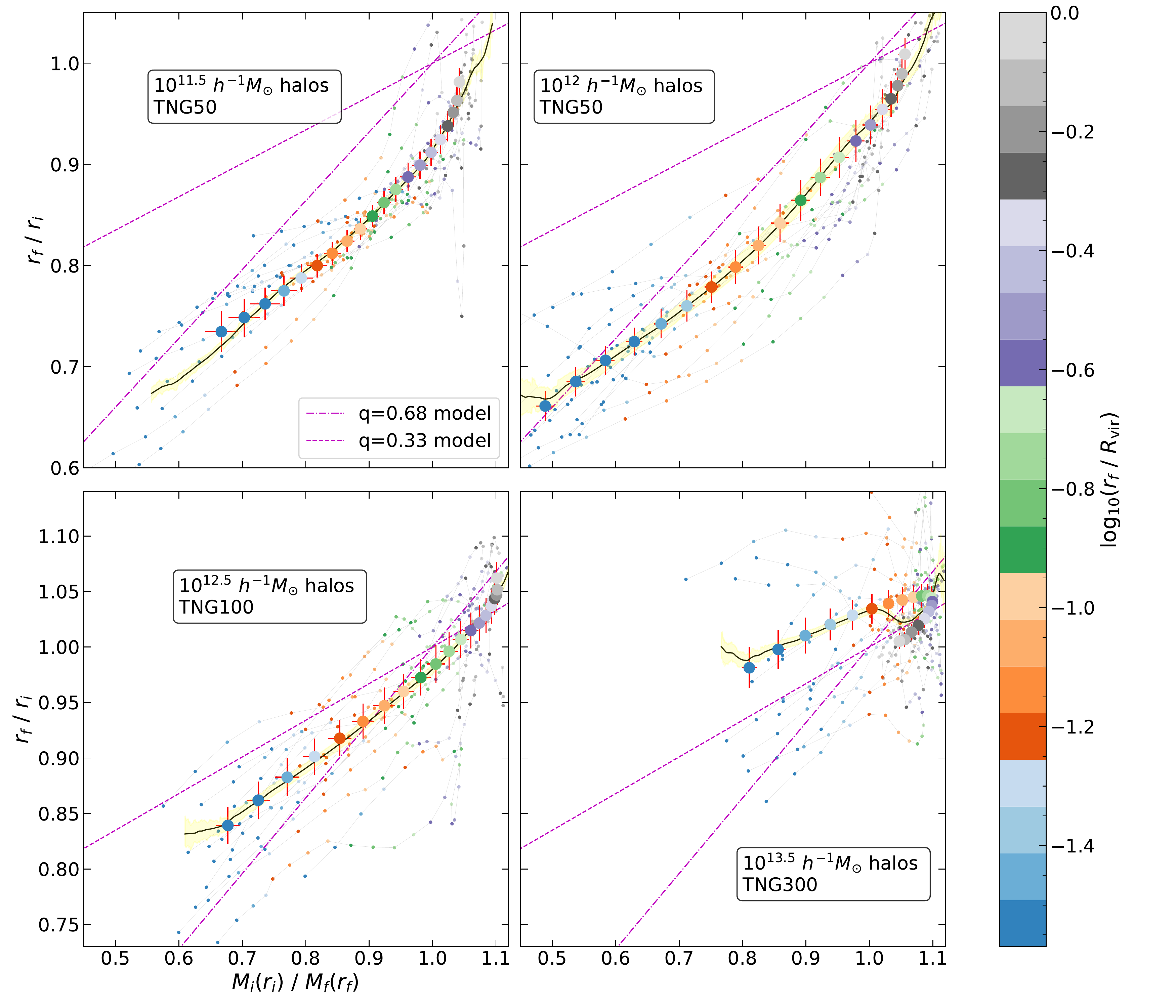}
    \caption{Relaxation relation for 4 different sample haloes selected by mass from the IllustrisTNG simulations. The large coloured circles denote the stacked relaxation ratio and total mass ratio at 20 different shells, whose radii is indicated by the colour. Small coloured markers joined by gray lines show the relaxation relation of a few randomly chosen individual haloes in each of the samples. The black curves denote the radius-independent stack of relaxation relation for each sample (see text). The quasi-adiabatic relaxation model \eqn{eq:chi-linear} with $q=0.68$ and $q=0.33$ are shown by the dot-dashed and dashed purple lines, respectively in each panel.} 
    \label{fig:relx-results-simple}
\end{figure*}

\subsubsection{The relaxation relation} %
\label{sec:methods-relx-reln}
Our focus in this work is to characterise the relaxation relation \eqn{eq:qAR} as a function of halo and galaxy properties over a wide dynamic range; e.g, we would like to ask whether \eqn{eq:chi-linear} is a good description of this relation.
To study this, we must extract this relation for individual haloes in hydrodynamical simulations.
For a given hydrodynamical halo in the matched catalog, we can obtain this relaxation relation by considering its matched halo in the gravity-only run to represent its unrelaxed state. 
We find it convenient to work with $r_f$ as a control variable. In this case, the values of $r_i$, $M_i(r_i)$ and $M_f(r_f)$ must be obtained from the matched halo pair, which can be done as follows.

For a dark matter shell at radius $r_f$ in the relaxed halo enclosing a dark matter mass of $M_f^d(r_f)$, its unrelaxed radius $r_i$ can be obtained by applying \eqn{eq:DMmass} and inverting the mass profile $M_i^d(r)=(1-f_b) M_i(r)$ of the gravity-only halo, where $f_b$ is the cosmic baryon fraction, to obtain 
\begin{align}
\label{eq:inv-mass}
r_i = {M_i}^{-1} \left( \frac{M_f^d(r_f)}{(1-f_b)} \right)\,.
\end{align}
This is because each `particle' in the gravity-only halo consists of collisionless baryons and dark matter in precisely the proportion $f_{b}$. 
The value of $M_i(r_i)$ then follows from direct mass counting in the unrelaxed (i.e., gravity-only) halo in radius $r_i$, and the value of $M_f(r_f)$ follows from direct mass counting in the relaxed (i.e. hydrodynamical) halo in radius $r_f$, as described in \secref{subsec:massprofiles}. In practice, we first obtain the unrelaxed mass profile $M_i(r_i)$ for a wide range of radii in finely spaced bins, in order to then compute the inverse in \eqn{eq:inv-mass} by interpolation.

Thus, for any shell defined by its relaxed radius $r_f$, we can obtain both the relaxation ratio $r_f/r_i$ and the mass ratio $M_i/M_f$ from its unrelaxed radius computed from mass profiles. Hence we can obtain the relaxation relation by placing multiple concentric shells around the halo all the way to its virial radius $R_{\rm{vir}}$.
In Appendix \ref{appen:Mock}, we have tested this algorithm on mock halo + galaxy systems generated with fixed known relaxation relations.

\section{Results}
\label{sec:results}

\subsection{Relaxation of haloes in the IllustrisTNG}
\label{sec:results-1}

We find that the relaxation relation estimated as described in \secref{sec:method} varies widely across haloes in the matched catalogue. In \figref{fig:relx-results-simple}, we show the relaxation relation for four different samples of haloes selected by their unrelaxed mass from the IllustrisTNG simulations. %
The first two samples are from TNG50, with masses $M \sim 10^{11.5} \Mh$ and $10^{12} \Mh$, respectively. Similarly the other two samples are from TNG100 and TNG300 with masses $M \sim 10^{12.5} \Mh$ and $10^{13.5} \Mh$, respectively. 
The relaxation relations of few individual randomly chosen haloes from each sample are shown by grey lines; we also show stacked relaxation relations for each of the sample (see below for measurement details). The quasi-adiabatic relaxation model \eqn{eq:chi-linear} with $q=0.68$ and $q=0.33$ is shown by the dot-dashed and dashed purple lines, respectively, in each panel. The value $q=0.68$ was proposed by \citet{2015JCAP...12..049S} as being a reasonable description of cluster-sized haloes, while \citet{2021MNRAS.507..632P} argued that $q=0.33$ leads to a good description of the radial acceleration relation of Milky Way-sized spiral galaxies (see their Appendix A1). We will use these two models as reference points in the comparisons below.
Since the samples shown are representative of the haloes in IllustrisTNG over a large mass range, it is clear that \emph{\eqn{eq:chi-linear} with a constant $q$ does not work for the majority of haloes in IllustrisTNG.} Similar results hold for EAGLE haloes as well. This motivates a systematic study of the relaxation relation as a function of halo mass and other properties.

For each of the four samples selected by halo mass, we compute the relaxation ratio $r_f/r_i$ and the enclosed total mass ratio $M_i/M_f$ at 20 concentric spherical shells for all haloes in the sample. We take the largest shell at the relaxed virial radius $r_f=R_{\rm{vir}}$, while the remaining 19 shells are taken at fixed values of $r_f/R_{\rm{vir}}$ for each of the halo. This allows us to stack the relaxation relation by simply taking the mean and standard deviation of the relaxation ratio and mass ratio at each of the 20 shells. While the physical size of the shell differs from halo to halo, we ensure that the smallest shell has a radius of at least 10 times the force smoothing length of the simulation. In \figref{fig:relx-results-simple}, we show this stacked relaxation relation in large coloured markers, where the colour denotes the relaxed radius of the shell; and the error bar shown in red corresponds to the statistical error in the estimate of the mean value. 
By comparing with the small markers of same colour, we can see that there is a significant scatter not only in the relaxation ratio but also in the mass ratio at fixed $r_f/R_{\rm{vir}}$ across haloes in each sample. 
To assess the level of systematic error introduced by our default choice of stacking technique, we also tested an alternate stacking definition, wherein we interpolate the relaxation relation of individual haloes to obtain the relaxation ratio at fixed values of mass ratios and stack them 
by ignoring the value of corresponding relaxed radii.
However, this stacking method ignores radius information completely; we discuss the consequence of this later in \secref{sec:results-rad-dep-qadiab}.

\subsection{Trend in relaxation relation with halo mass}
\label{sec:results-mass}

\begin{figure}
    \centering
    \includegraphics[width=0.99\linewidth]{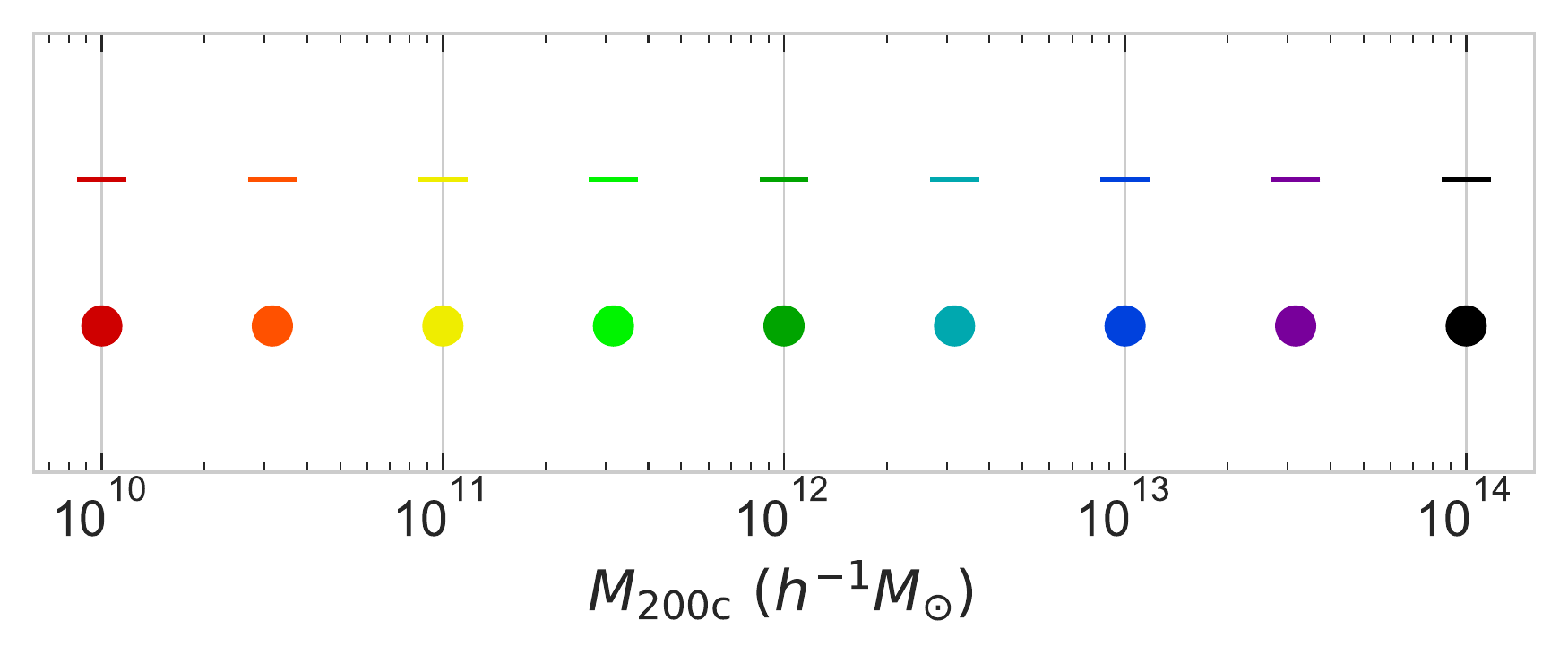}
    \caption{Representative colours we use to denote each of the halo mass bins.}
    \label{fig:mass_bin_label}
\end{figure}

As can be already noted in \figref{fig:relx-results-simple}, the relaxation relation shows very different behaviour at different mass scales. In this section, we focus on the stacked relation (using our default stacking definition) and study how it varies as a function of unrelaxed halo mass. For this, we consider nine mass bins starting from $\log (M/\Mh) = 10$ to $14$ in steps of $0.5$ dex. We list the colour labels used for these mass bins in \figref{fig:mass_bin_label}; this colour-coding will be used in all subsequent plots. None of the five simulations considered, simultaneously provides a sufficiently large sample of cluster-scale haloes and well-resolved low-mass haloes.
In the IllustrisTNG suite, we use the smallest box TNG50 to study haloes with mass $10^{10} \Mh < M < 10^{12} \Mh$, whereas we use TNG100 and TNG300 to study haloes with mass $10^{11} \Mh < M < 10^{12.5} \Mh$ and $10^{12} \Mh < M < 10^{14} \Mh$ respectively. At those mass bins where multiple IllustrisTNG boxes provide halo samples, the smaller box provides a smaller sample but with better resolution. For computational ease, we limit the size of each sample to be $\leq500$ haloes, as we find that the statistics are well-converged with this number.

\begin{figure*}
\centering
\includegraphics[width=0.49\linewidth]{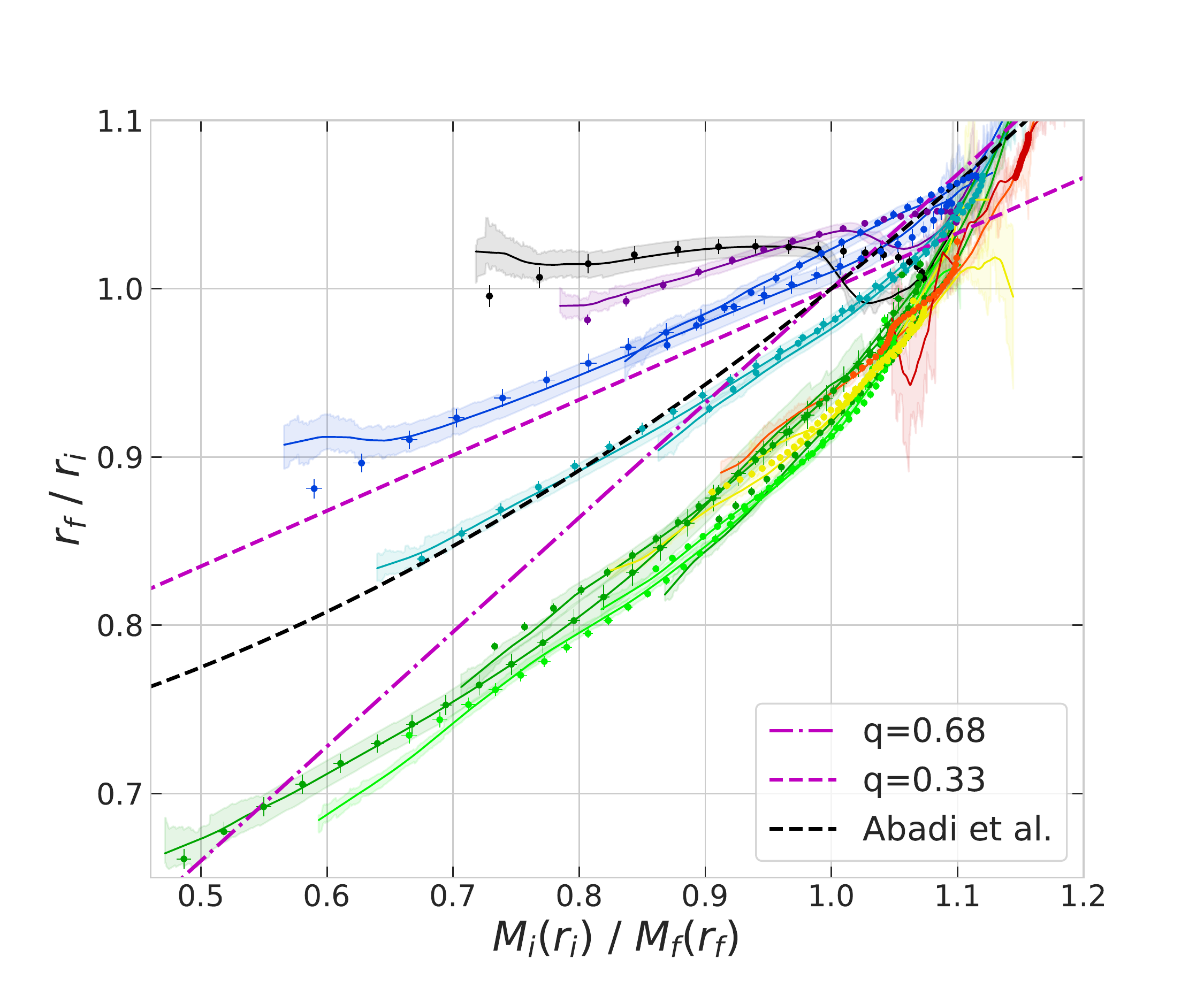}
\includegraphics[width=0.49\linewidth]{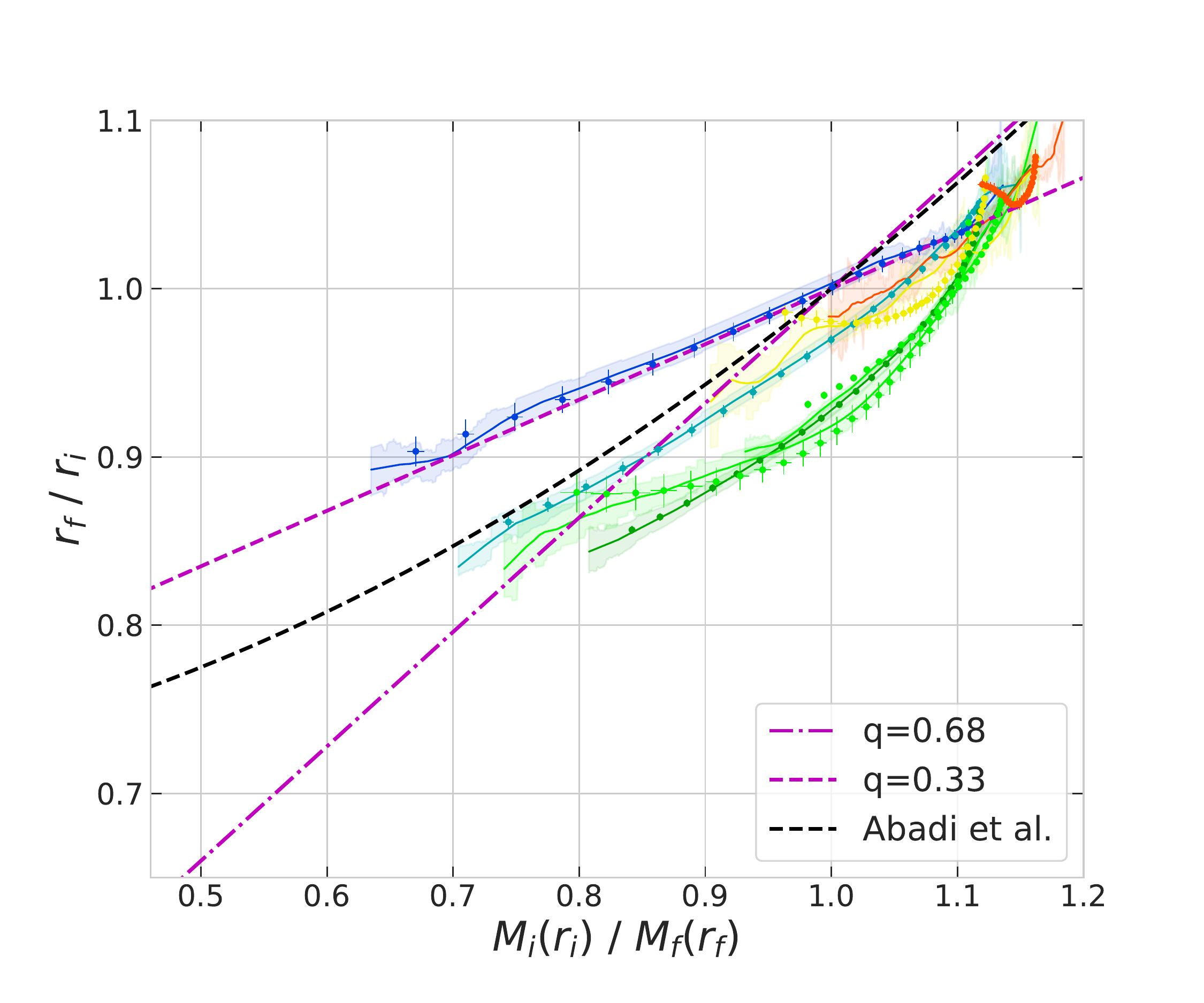}
\caption{The stacked relation between relaxation ratio and mass ratio as a function of halo mass in IllustrisTNG (left panel) and EAGLE (right panel) simulations. Here the points and solid lines represent two different stacking methods as in Fig.~\ref{fig:relx-results-simple}. The colour-coding follows Fig.~\ref{fig:mass_bin_label}.} 
\label{fig:fit-view-mass-indep}
\end{figure*}

By repeating the procedure described in \secref{sec:results-1}, we obtain both the fixed-radius (default) stack and radius-independent (alternate) stack of the relaxation relation for each of the halo samples taken from IllustrisTNG at these nine mass bins (see \emph{left panel} of \figref{fig:fit-view-mass-indep}). 
For reference, note that the case of no relaxation would correspond to a horizontal line at unity in this plot.
Relaxation is strongest for Milky Way-scale haloes, as indicated by the small values of the relaxation ratio for $M\sim10^{12}\Mh$; we discuss the physical implications of this result later. 
Note that the simple quasi-adiabatic relaxation model \eqn{eq:chi-linear} with $q=0.68$ used in \cite{2015JCAP...12..049S} fails to explain the relaxation relation for any of the halo masses considered; however this model with $q=0.33$ is reasonably close to the relaxation relation at %
$M\sim 10^{13} \Mh$. 
And while the quadratic model proposed by \cite{2010MNRAS.407..435A} matches with the relaxation relation of $10^{12.5} \Mh$ haloes in IllustrisTNG, this is possibly a coincidence given that this model was built using zoom simulation of haloes in the mass range $10^{11.5} $-$ 10^{12} \Mh$, which show a very different relaxation relation in IllustrisTNG.\footnote{The \cite{2010MNRAS.407..435A} simulation also suffered from overcooling due to the lack of feedback effects, so that the mass ratios attained much smaller value for shells at the same radii as compared to IllustrisTNG.}

We also take six samples of haloes from the EAGLE simulation, in mass bins $\log (M/\Mh) = 10.5, 11,11.5$  from the small, high-resolution L25 box and in mass bins $12, 12.5, 13$ from the main L100 box. Here too, the $q=0.68$ model fails for all masses, but $q=0.33$ model works reasonably for  $M\sim10^{13} \Mh$ haloes (see the \emph{right panel} of \figref{fig:fit-view-mass-indep}). We find that, despite having a different galaxy formation model, the relaxation relation for haloes found in the primary EAGLE run L100 is consistent with the results from IllustrisTNG. 
IllustrisTNG samples reach lower values of the relaxation ratio and mass ratio than EAGLE because of the better resolution available. For $M_{200}=10^{12} \Mh$, the mean relaxation relation shown in \figref{fig:fit-view-mass-indep}, does not seem to be very different between IllustrisTNG and EAGLE $L100$, atleast not anymore than the difference between different boxes of the IllustrisTNG. However, the haloes from EAGLE $L25$ simulation shows a unique behaviour where the relaxation ratio increases with decrease in mass ratio in the innermost regions. We expect that this might be due to the fact that the EAGLE reference model required recalibration at this higher resolution. In a future work we will explore how the dark matter response depends on such variations in the baryonic prescription.

\subsection{Parametrised model of quasi-adiabatic relaxation}
\label{sec:results-rad-dep-qadiab}
In this section, we model the relaxation relations discussed above, with a focus on conveniently quantifying this response across a wide range of halo masses.

\begin{figure*}
    \centering
    \includegraphics[width=0.48\linewidth]{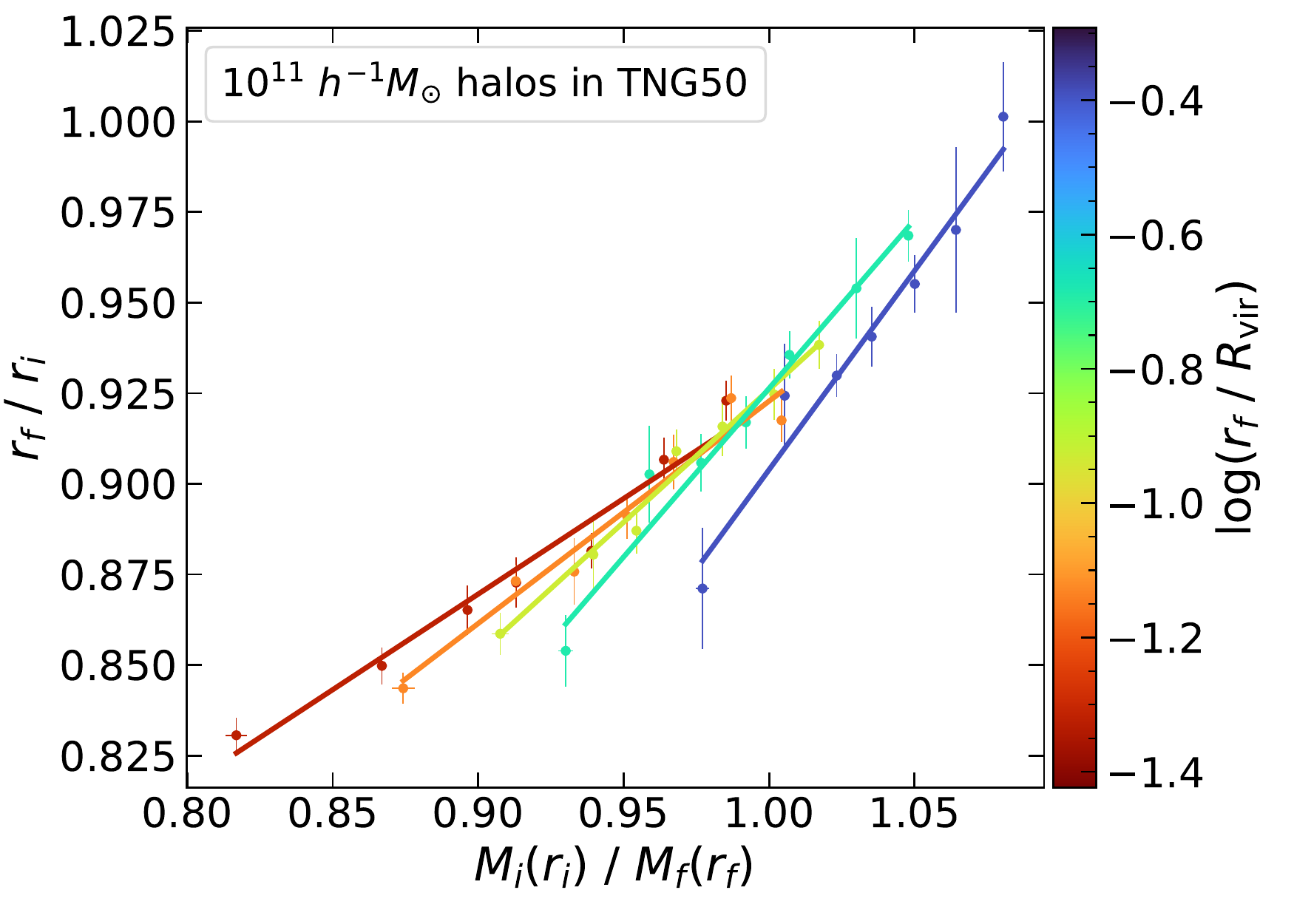}
    \includegraphics[width=0.48\linewidth]{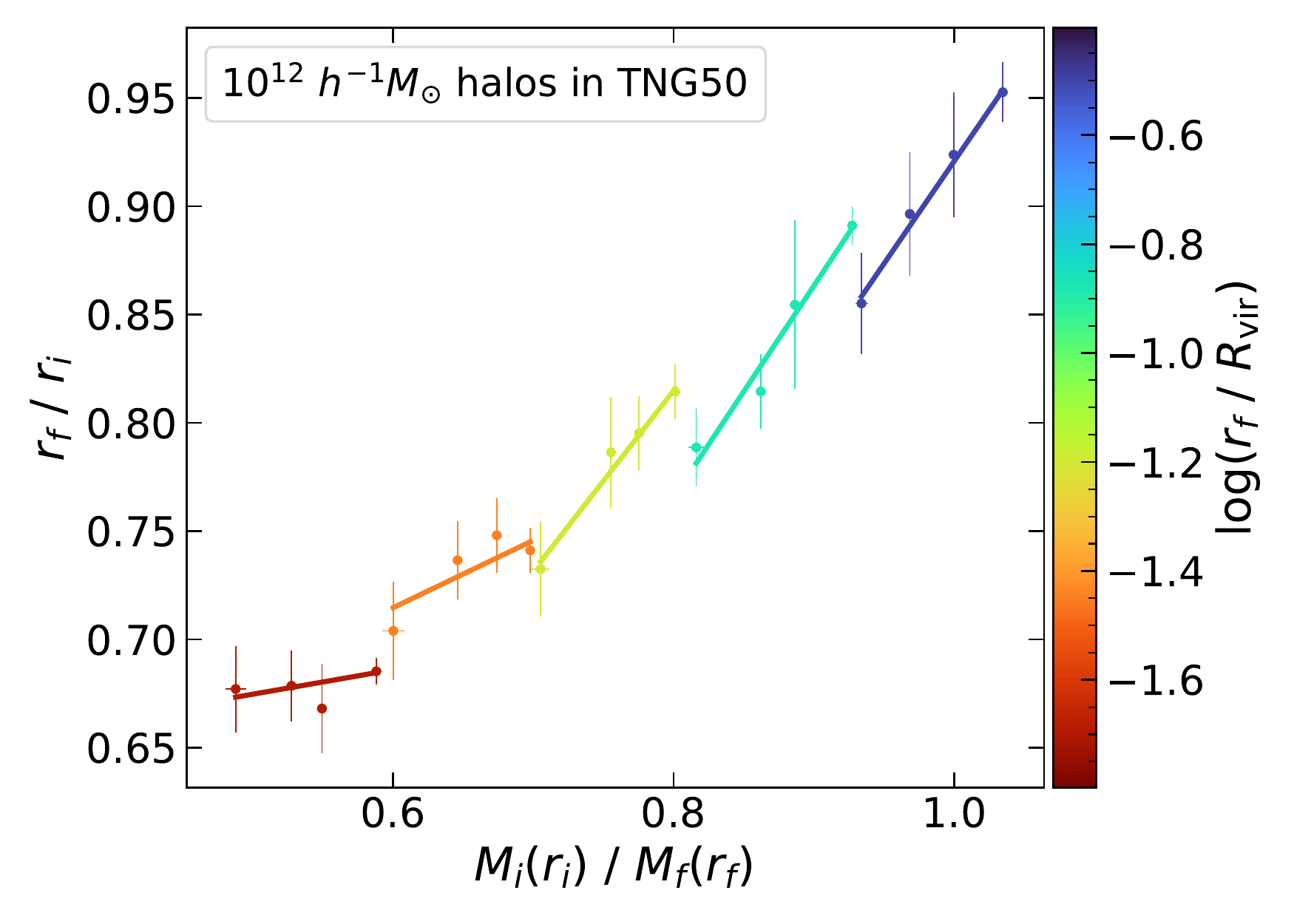}
    \includegraphics[width=0.48\linewidth]{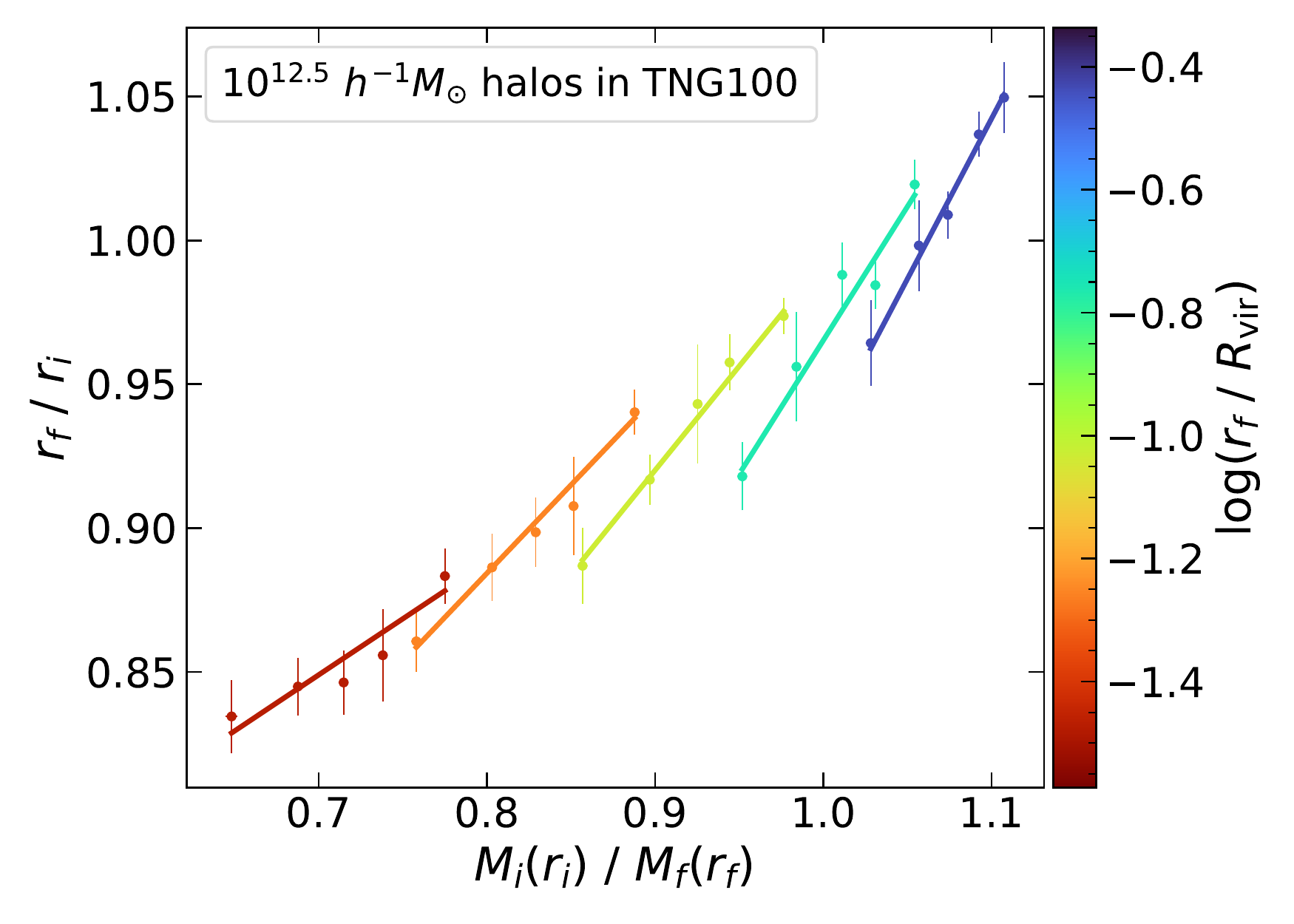}
    \includegraphics[width=0.48\linewidth]{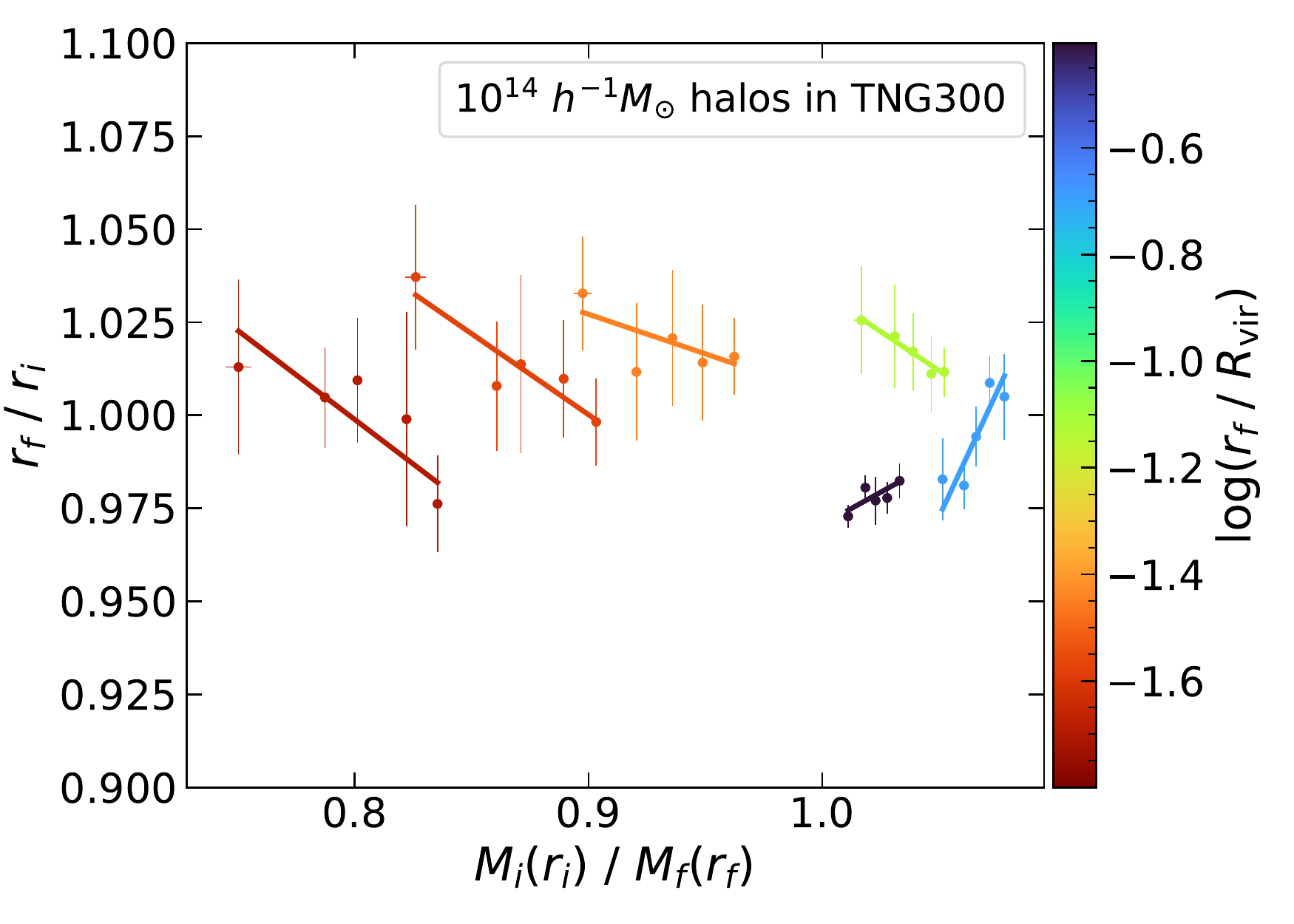}
    \includegraphics[width=0.48\linewidth]{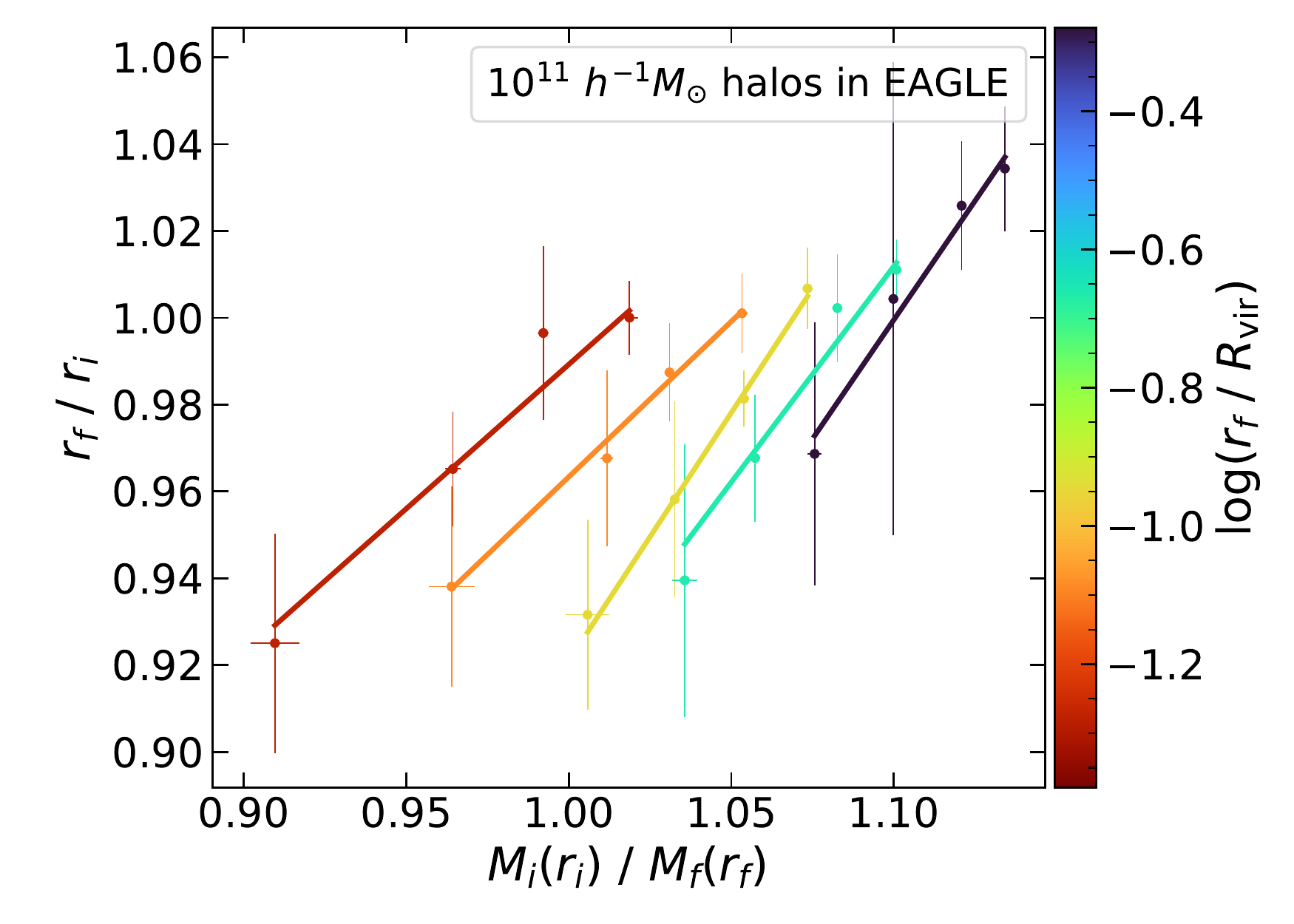}
    \includegraphics[width=0.48\linewidth]{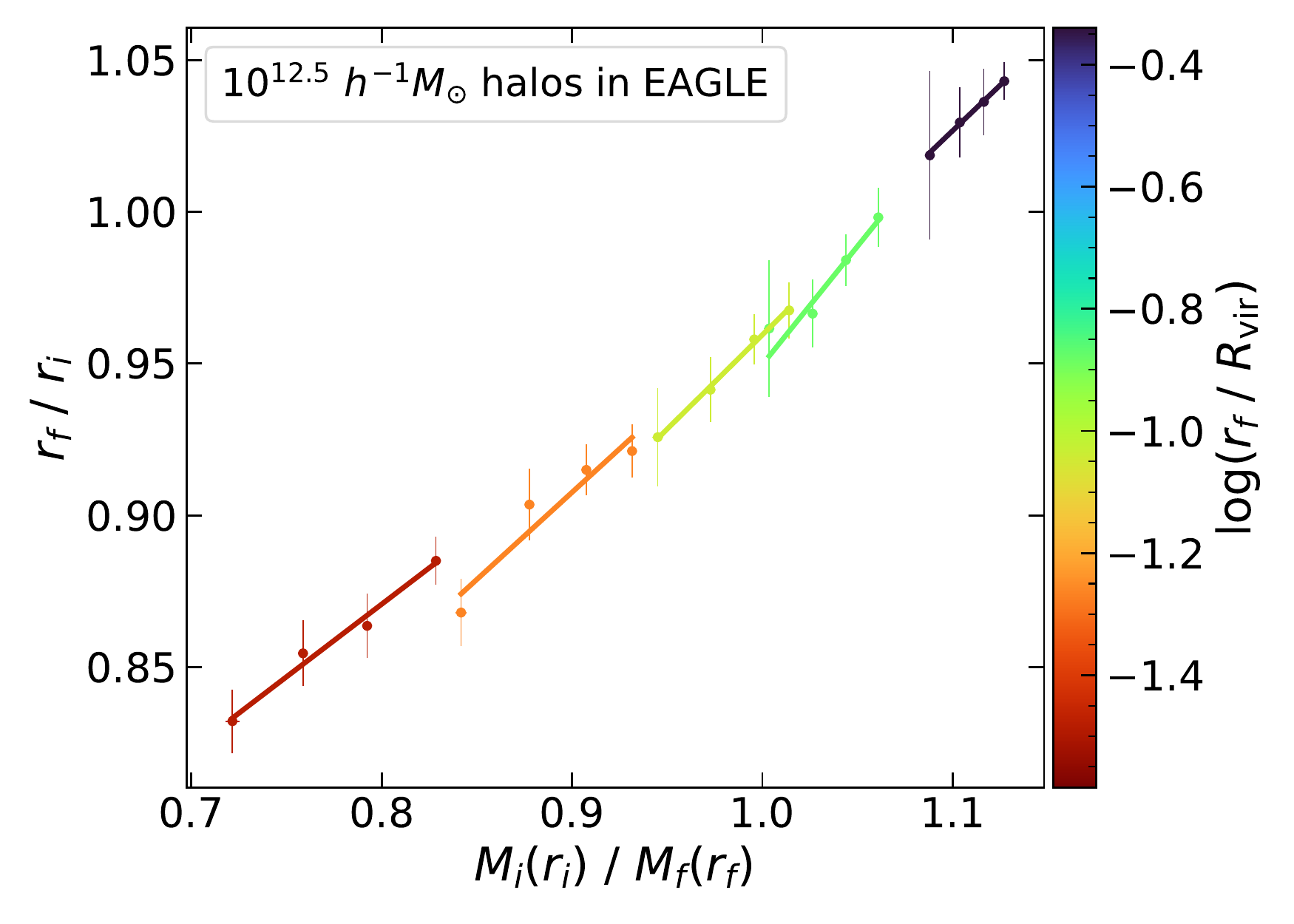}
    \caption{Relaxation relation stacked separately at 5 different radii indicated by color in six samples of haloes selected by mass from IllustrisTNG and EAGLE simulations. We also show linear polynomial fit to this relation, following \eqn{eq:chi-linear-q0} with the best-fit values for the parameters $q_0$ and $q_1$ at each of the selected radii for each of the six halo samples.} %
    \label{fig:rf-fit-show}
\end{figure*}

\subsubsection{Expectations from simulation measurements}
\label{subsubsec:sim-relax}
In both IllustrisTNG and EAGLE, for all masses other than $10^{13} \Mh$, the simple quasi-adiabatic relaxation model \eqn{eq:chi-linear} fails to explain the measured relation with any value of $q$. 
As seen in Fig.~\ref{fig:fit-view-mass-indep}, an important aspect of this mismatch is caused by the model's requirement that shells which hold their baryonic mass fixed (i.e., for which $M_i/M_f=1$) must necessarily also hold their radius fixed ($r_f/r_i=1$), and vice-versa. The measurements, however, show substantial offsets in the relaxation ratio from unity for shells with $M_i/M_f = 1$, and also substantial offsets in the mass ratio from unity for shells with $r_f/r_i=1$, across nearly the entire range of halo mass. One way of understanding this effect physically is due to feedback-related baryonic outflows: a particular shell which maintains its radius after relaxation ($r_f/r_i=1$), could still lose its baryonic mass due to outflows, resulting in $M_i/M_f > 1$ \citep[][]{2022MNRAS.511.3910F}. Alternatively, the interplay between cooling-related condensation (which increases baryonic mass in a given shell) and feedback-related outflows (which decrease baryonic mass) could result in a situation where the baryonic mass after relaxation retains its initial value despite an overall relaxation, e.g. due to approximate angular momentum conservation, leading to $r_f/r_i < 1$ while $M_i/M_f=1$. These trends are visible  in  Fig.~\ref{fig:fit-view-mass-indep}  for haloes with $M<10^{13}\Mh$. Fig.~\ref{fig:relx-results-simple} shows that the former trend ($M_i/M_f > 1$ when $r_f/r_i=1$) occurs in the halo outskirts ($r_f\sim R_{\rm vir}$) and the latter ($r_f/r_i < 1$ when $M_i/M_f=1$) in the inner halo ($r_f\lesssim0.3\,R_{\rm vir}$), for $M<10^{13}\Mh$. 
On the other hand, more massive haloes show little to no relaxation in both inner halo (where there is net baryonic inflow, $M_i/M_f < 1$) and outer halo (where there is net baryonic outflow, $M_i/M_f > 1$).

To account for such effects, we expand the simple quasi-adiabatic relaxation model \eqn{eq:chi-linear} by adding 
a null offset parameter $q_0$:
\begin{align}
    \label{eq:chi-linear-q0}
    \frac{r_f}{r_i} - 1 &= q_1 \left( \frac{M_i(r_i)}{M_f(r_f)} - 1 \right) + q_0\,.
\end{align}
With this model, the ratio of angular momenta of the dark matter particles in approximately circular orbits before and after relaxation can be expressed simply as follows (with $L_i$ and $L_f$ denoting the angular momenta of the unrelaxed and relaxed shell, respectively),
\begin{align}
\left( \frac{L_f}{L_i} \right)^2 &= \frac{M_f}{M_i} \frac{r_f}{r_i}\\
&= \frac{M_f}{M_i} \left[ q_1 \left( \frac{M_i}{M_f} - 1 \right) + q_0 + 1 \right]\\
\label{eq:Lf-Li-ratio}
&= (1 + q_0 - q_1) \frac{M_f}{M_i} + q_1
\end{align}
For example, the special case $q_0=-(1-q_1)$ can be thought of as a natural generalisation of the original adiabatic relaxation model, because in this case we have $L_f/L_i = \sqrt{q_1}$, relating $q_1$ directly to angular momentum loss or gain.
Below, however, we will see that there is no simple relation between $q_1$ and $q_0$ for generic measurements in the simulations. In general, then, one can only say that a particular shell has gained or lost angular momentum when the value of $(1 + q_0 - q_1) (M_f/M_i)$ is, respectively, larger or smaller than $1-q_1$.

However, the above holds true only when the dark matter particles are in circular orbits. When galactic processes lead to changes in the baryonic mass profile, even the dark matter particles in circular orbit can go into elliptical orbits \citep[see, e.g.][]{2005ApJ...634...70S}. For example, when there is a sudden expulsion of gas due to feedback events, the total mass enclosed decreases and the particles start moving radially outward. During this period the mass ratio $M_i/M_f$ can become greater than one and still have no relaxation (i.e. $r_f/r_i=1$) as discussed in the start of this section.

While this extended linear model can describe the relaxation relation at few other halo masses (see for example $10^{12.5}\Mh$ halos in both left and right panel of \figref{fig:fit-view-mass-indep}), even this model fails at many halo masses.
Moreover, we have checked that there is no simple polynomial model favoured by standard information criteria such as AICC \citep[][]{2007MNRAS.377L..74L}
to describe the relaxation relation at all masses.
Rather,  we find that, if we 
simply elevate $q_0$ and $q_1$ in \eqn{eq:chi-linear-q0} to functions of $r_f/R_{\rm vir}$, then this model
can be applied at all the mass scales that we consider. For this, we need the relaxation relation at fixed relaxed radius, which we obtain as follows. We measure the relaxation ratio and mass ratio of shells having fixed $r_f/R_{\rm vir}$ for all haloes in a selected sample, then stack them in bins
of mass ratio at each spherical shell separately. 
We find that \eqn{eq:chi-linear-q0}
is consistent with the relaxation relation of all halo masses considered, where we infer the values of $q_0$ and $q_1$ at each $r_f$ using standard least squares fitting (the reduced $\chi^2$ values are always close to unity, for all masses and radial shells).

In \figref{fig:rf-fit-show} we show the measured relaxation relation for six different sample of haloes at shells of selected radii, compared with the best-fit model \eqn{eq:chi-linear-q0} for each case; the model clearly describes these measurements extremely well. We already noted from \figref{fig:fit-view-mass-indep}, that the haloes in the small volume EAGLE simulation with the reference model shows a different relaxation behaviour. This is also apparent in \figref{fig:rf-fit-show}, between the $10^{11} \Mh$ haloes from that L25 simulation (last row, left panel) and the haloes of the same mass from IllustrisTNG (first row, left panel); however they both follow the linear relaxation relation at fixed radii. An interesting feature to note is the dramatic change in slope $q_1$ for the most massive haloes (middle row, right panel), with $q_1$ changing sign as one moves outwards through the halo. This also qualitatively explains the non-monotonicity and multi-valued nature of the default stacks in the lower right panel of Fig.~\ref{fig:relx-results-simple}.

\begin{figure}
    \centering
    \includegraphics[width=0.99\linewidth]{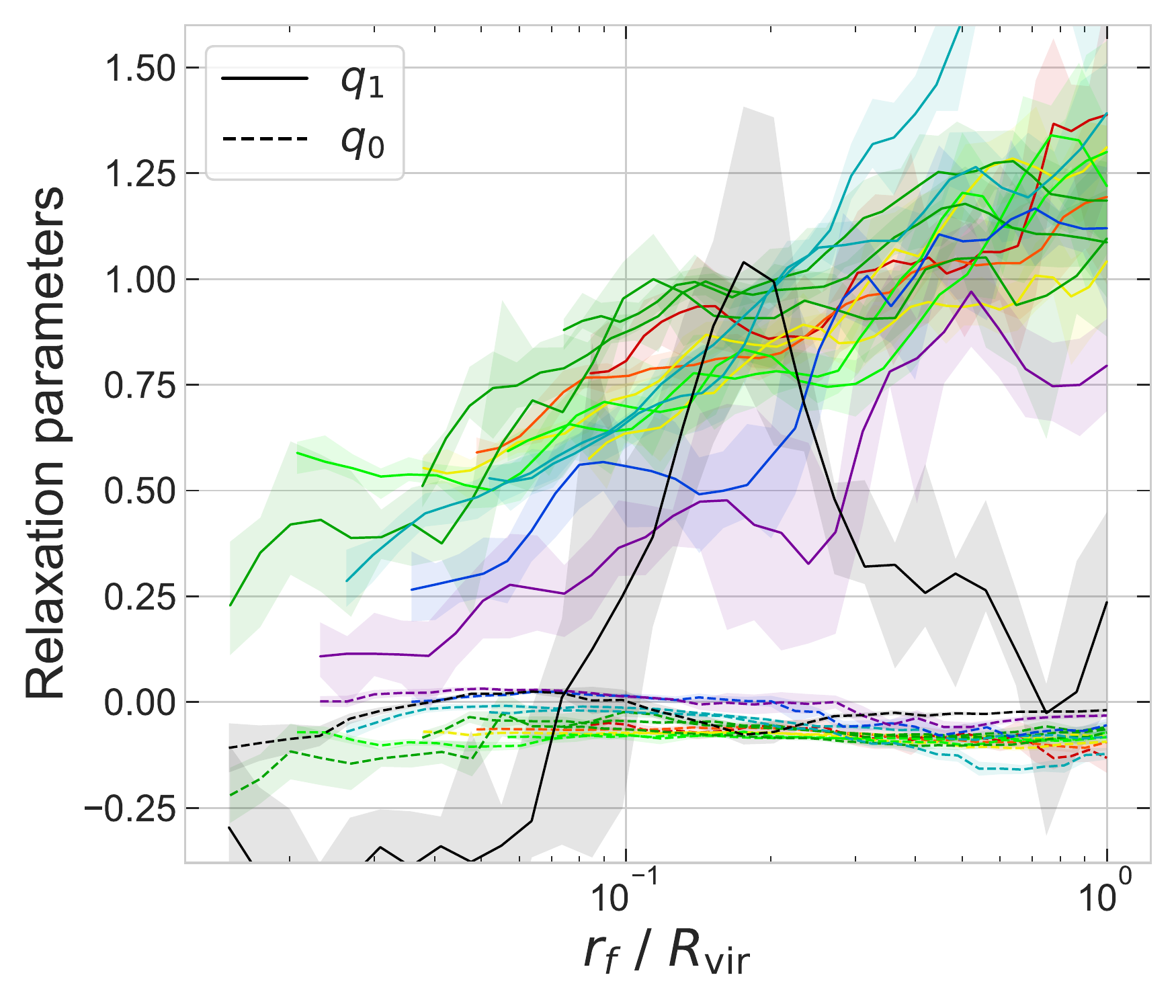}
    \includegraphics[width=0.99\linewidth]{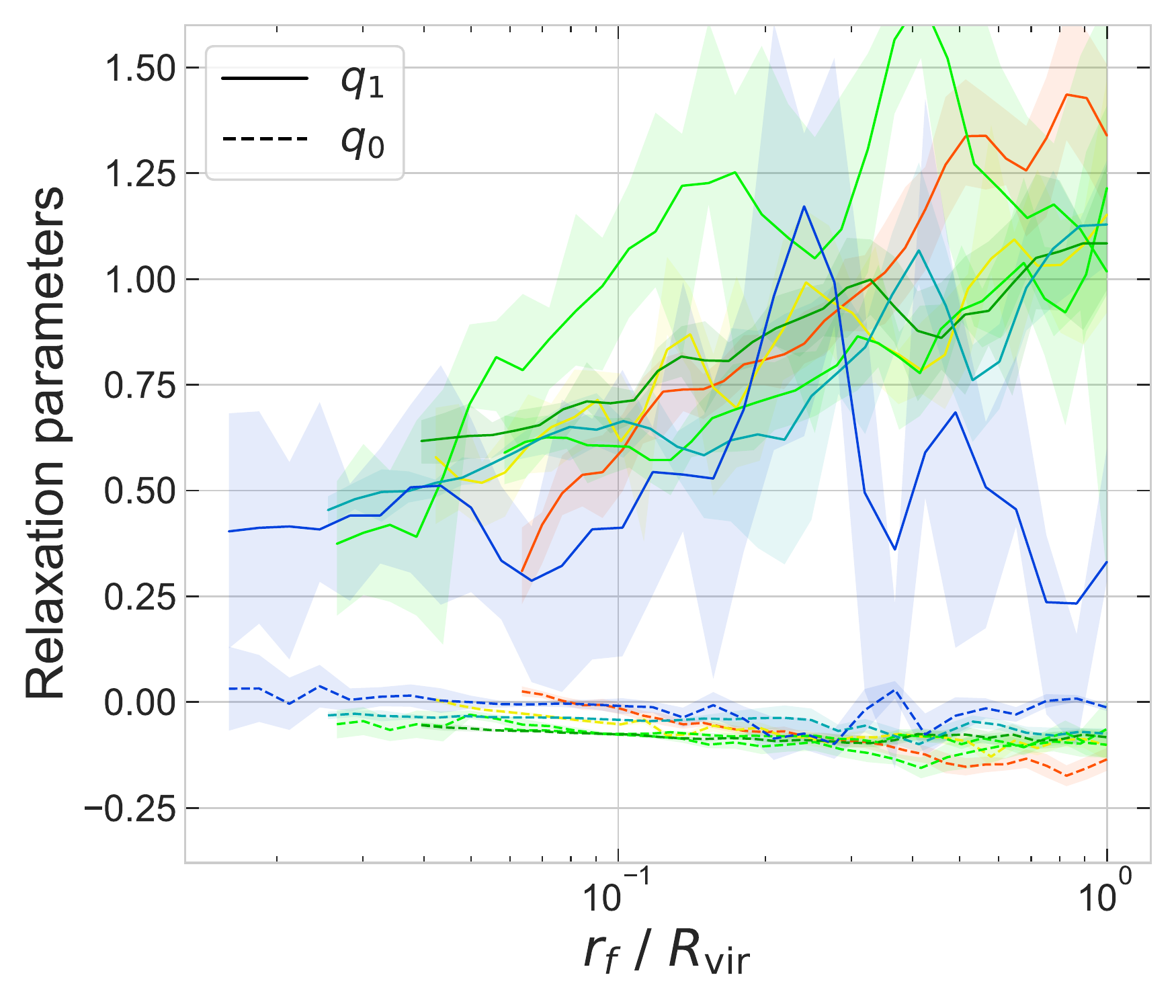}
    \caption{Linear quasi-adiabatic relaxation model parameters as a function of the radius of relaxed halo at different halo masses in IllustrisTNG \emph{(upper panel)} and EAGLE \emph{(lower panel)}. The colour-coding follows Fig.~\ref{fig:mass_bin_label}. See text for details.}
    \label{fig:rf-fit-params}
\end{figure}

\subsubsection{Modelling the radial dependence of the relaxation relation}
As described above, we obtained the best fit parameters $q_0$ and $q_1$ of the relaxation relation at each relaxed radius $r_f$, this is shown in \figref{fig:rf-fit-params} as a function of $r_f/R_{\rm vir}$. 

For haloes of mass $M<10^{13}\Mh$, the $q_1$ parameter increases monotonically with radius and this dependence can be modelled as 
\begin{align}
q_1 (r_f) = q_{10} + q_{11} \log \left( r_f/R_{\rm vir} \right) \,,
\label{eq:q1(r_f)}
\end{align}
where $q_{10}$ and $q_{11}$ are constants.
The parameter $q_0$, on the other hands, remains relatively constant at small negative values for each halo sample.
This means the factor $(1 + q_0 - q_1)$ starts with a positive value in the inner halo and becomes negative in the outer halo, inverting the relationship between change in angular momentum and mass ratio (see equation~\ref{eq:Lf-Li-ratio}). And due to $q_0$ being small in magnitude, the radius at which this transition happens roughly satisfies the condition $L_f/L_i=1$. 

For cluster-scale haloes, this simple monotonic dependence of $q_1(r_f)$ is replaced with oscillatory behaviour. In fact, some of the peaks in $q_1$ correspond to $q_1\approx1$; combined with $|q_0|\ll1$, this indicates that these peaks are shells which nearly perfectly conserve angular momentum. (E.g., this happens at $r_f/R_{\rm vir}\simeq0.5\,(0.2)$ for $M=10^{13.5}\,(10^{14})\Mh$.)
This is in strong contrast to \figref{fig:fit-view-mass-indep} where the slope of the relaxation relation represented by a globally defined $q_1$ (e.g., equation~\ref{eq:chi-linear-q0} without explicit $r_f$ dependence in the parameters) is close to zero for these haloes, indicating maximum deviation from the adiabatic relaxation. This complicated behaviour in the cluster-scale haloes could be due to the presence of substructures. The $q_0$ parameter now also shows interesting behaviour; it is only slightly negative in the outer halo but becomes close to zero in the inner halo for these haloes ($q_0$ was relatively constant with more negative value for less massive haloes).

\begin{figure}
    \centering
    \includegraphics[width=\linewidth]{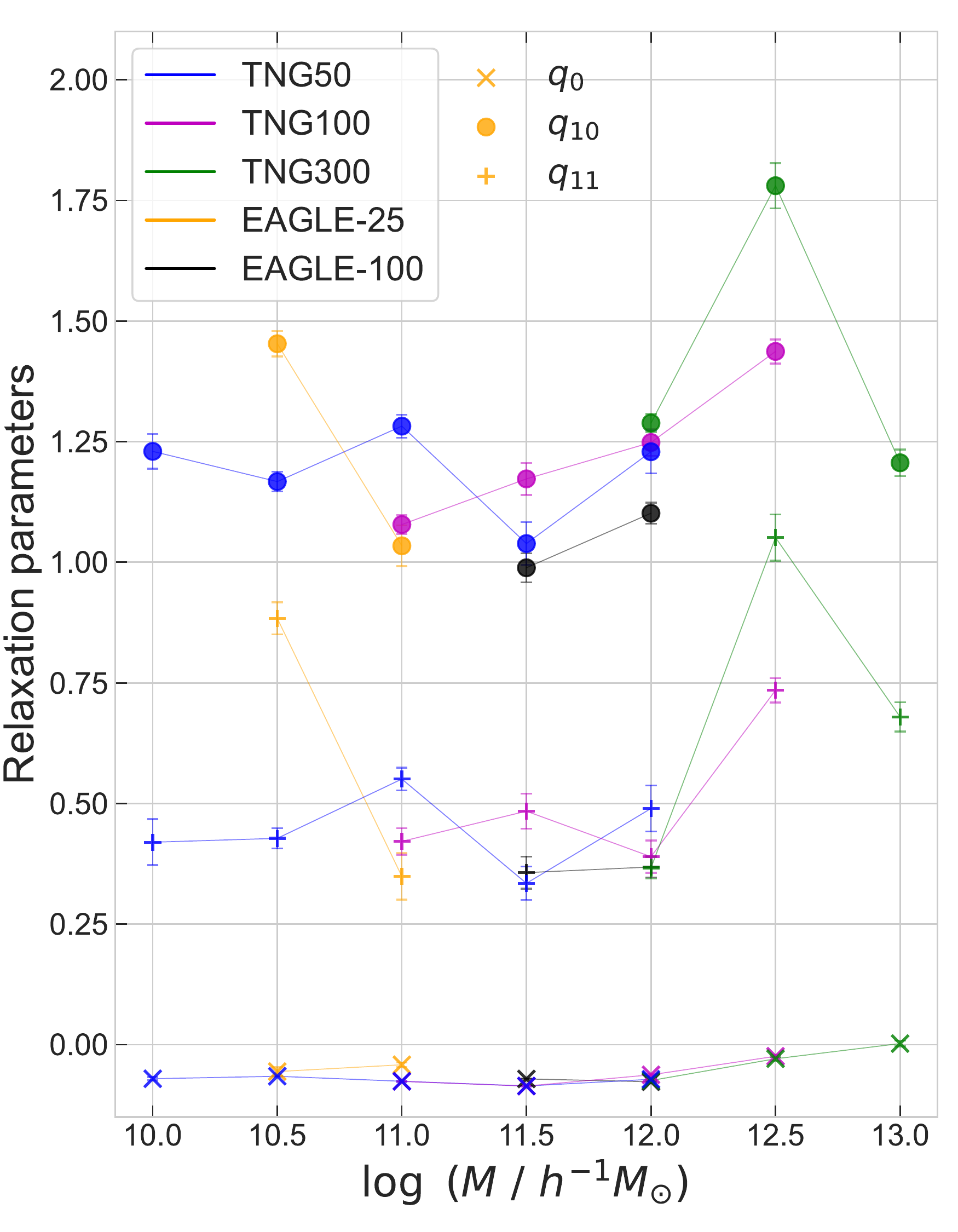}
    \caption{Fitting values for the three parameters namely $q_{0}$, $q_{10}$ and $q_{11}$ in radially dependent quasi-adiabatic relaxation model described by equation~\ref{eq:q3-model} as a function of halo mass in the three IllustrisTNG simulations and the two EAGLE simulations as listed in \secref{sec:sim-and-teq}.}
    \label{fig:3-param-mass-only}
\end{figure}

Excluding those cluster scale haloes, we propose a three parameter model as an extension to the quasi-adiabatic relaxation model, where the relaxation ratio depends linearly on the mass ratio as in \eqn{eq:chi-linear-q0}, however the slope of this relationship has explicit logarithmic dependence on the radius:
\begin{align}
\label{eq:q3-model}
\frac{r_f}{r_i} - 1 &=  \left[ q_{10} + q_{11} \log \left( \frac{r_f}{R_{\rm vir}} \right) \right] \left( \frac{M_i(r_i)}{M_f(r_f)} - 1 \right) + q_0
\end{align}
Using this model we can quantify the response of these haloes to galaxy formation;
in \figref{fig:3-param-mass-only}, we show these 3 parameters estimated as a function of halo mass for IllustrisTNG and EAGLE. We see that each of the parameters is nearly mass-independent for $M\lesssim10^{12}\Mh$, showing significant trends with mass only above $M\gtrsim10^{12.5}\Mh$ (with the exception of $10^{10.5} \Mh$ in EAGLE). This simplified but accurate relaxation model can be of great use in modelling the rotation curves of low-mass and Milky Way-like galaxies \citep{2021MNRAS.503.4147P,2021MNRAS.507..632P}, which we will explore in future work.

\subsection{Effect of properties beyond halo mass}
\label{sec:dep-on-hal-gal-props}
\begin{figure}
    \centering
    \includegraphics[clip,trim={0.2cm 1.18cm 0.25cm 0.3cm},width=0.49\linewidth]{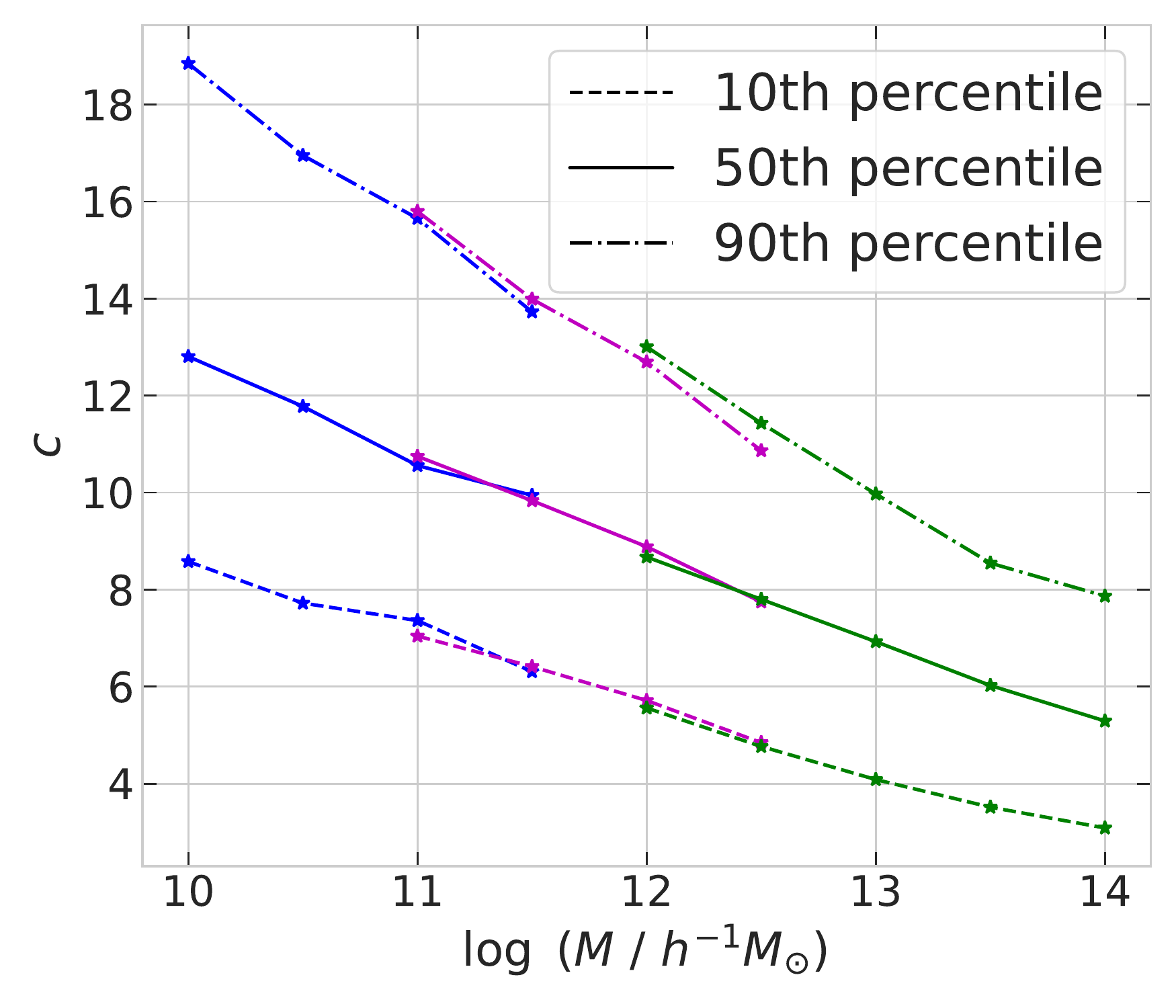}
    \includegraphics[clip,trim={0.2cm 1.18cm 0.25cm 0.3cm},width=0.495\linewidth]{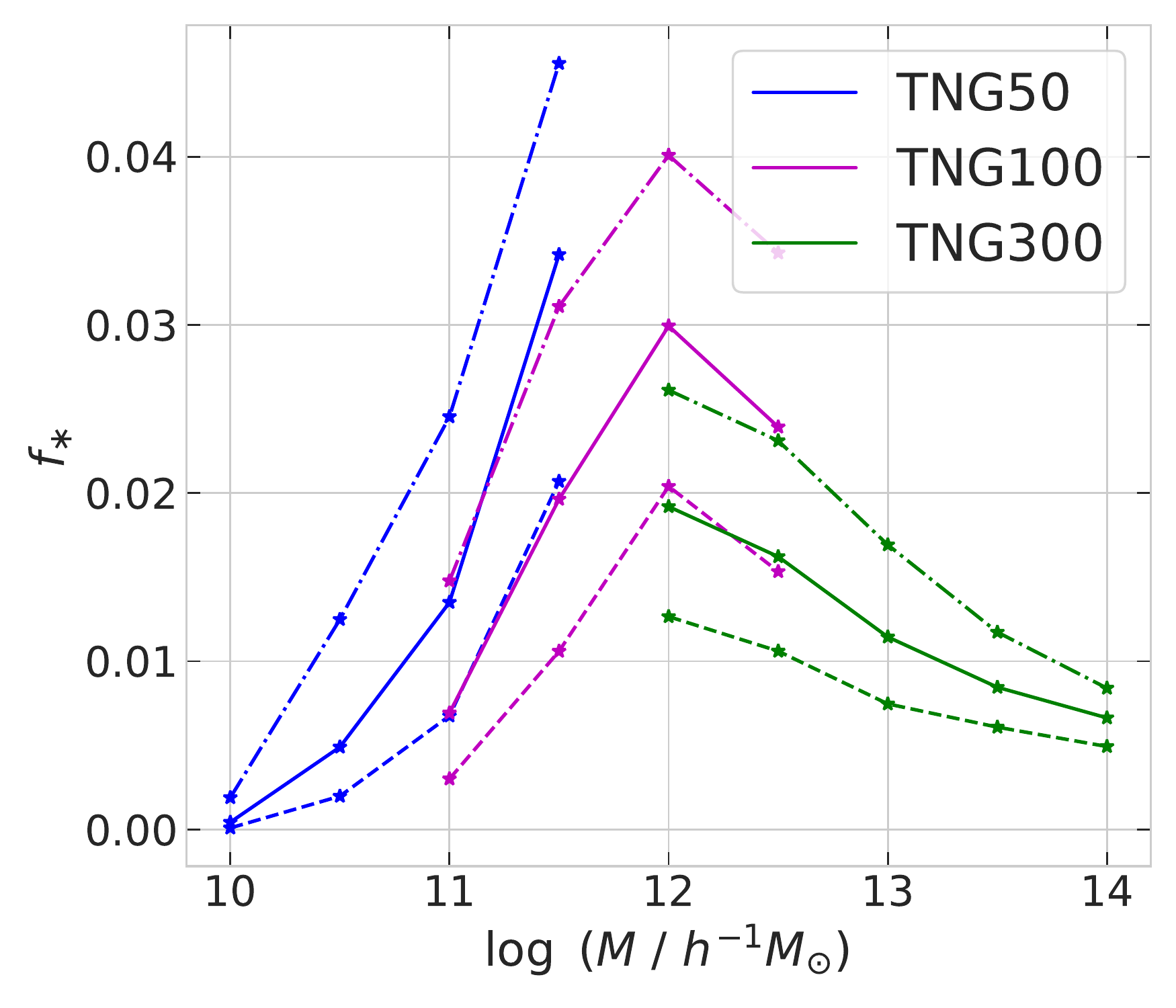}
    \includegraphics[clip,trim={0.2cm 0cm 0.25cm 0.3cm},width=0.49\linewidth]{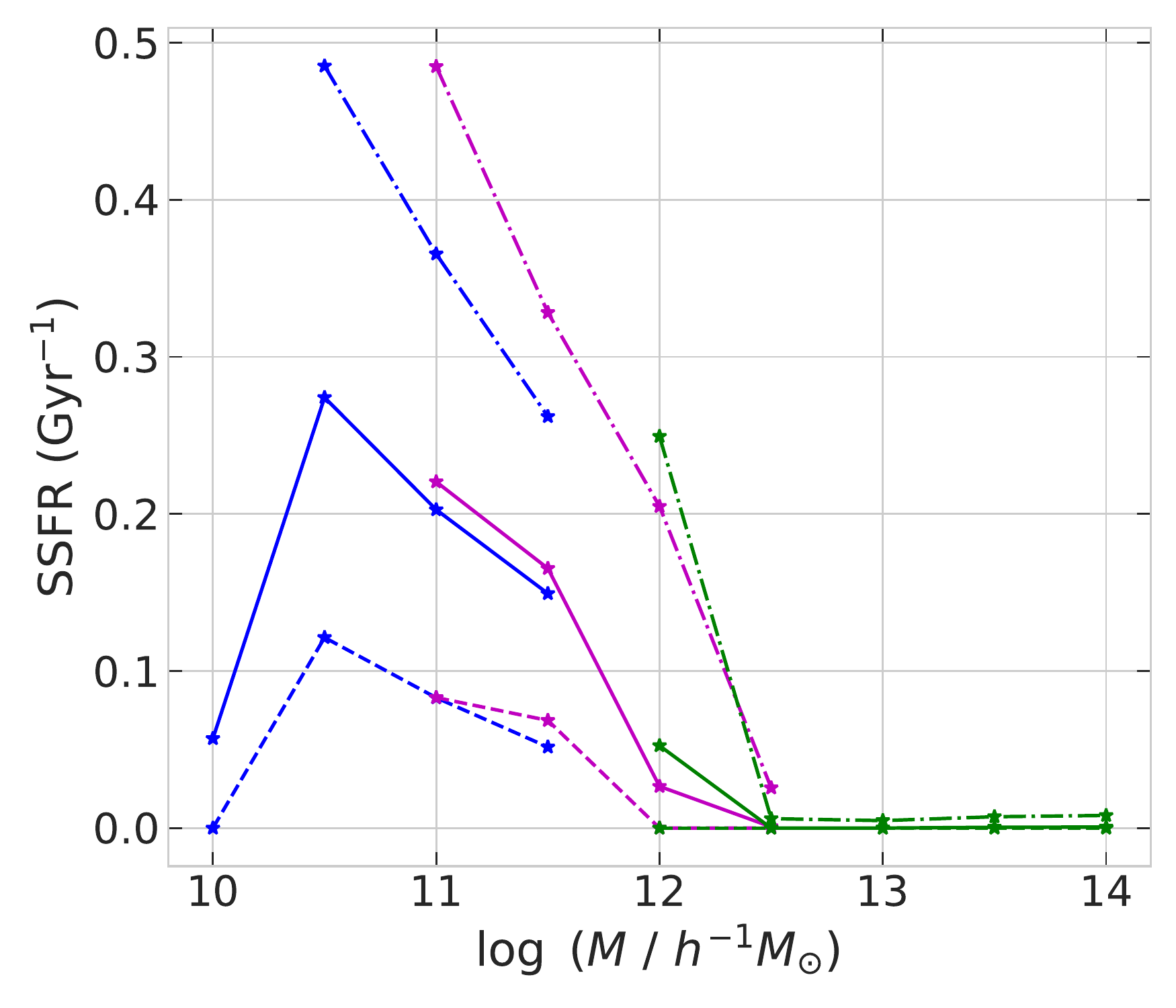}
    \includegraphics[clip,trim={0.2cm 0cm 0.25cm 0.3cm},width=0.495\linewidth]{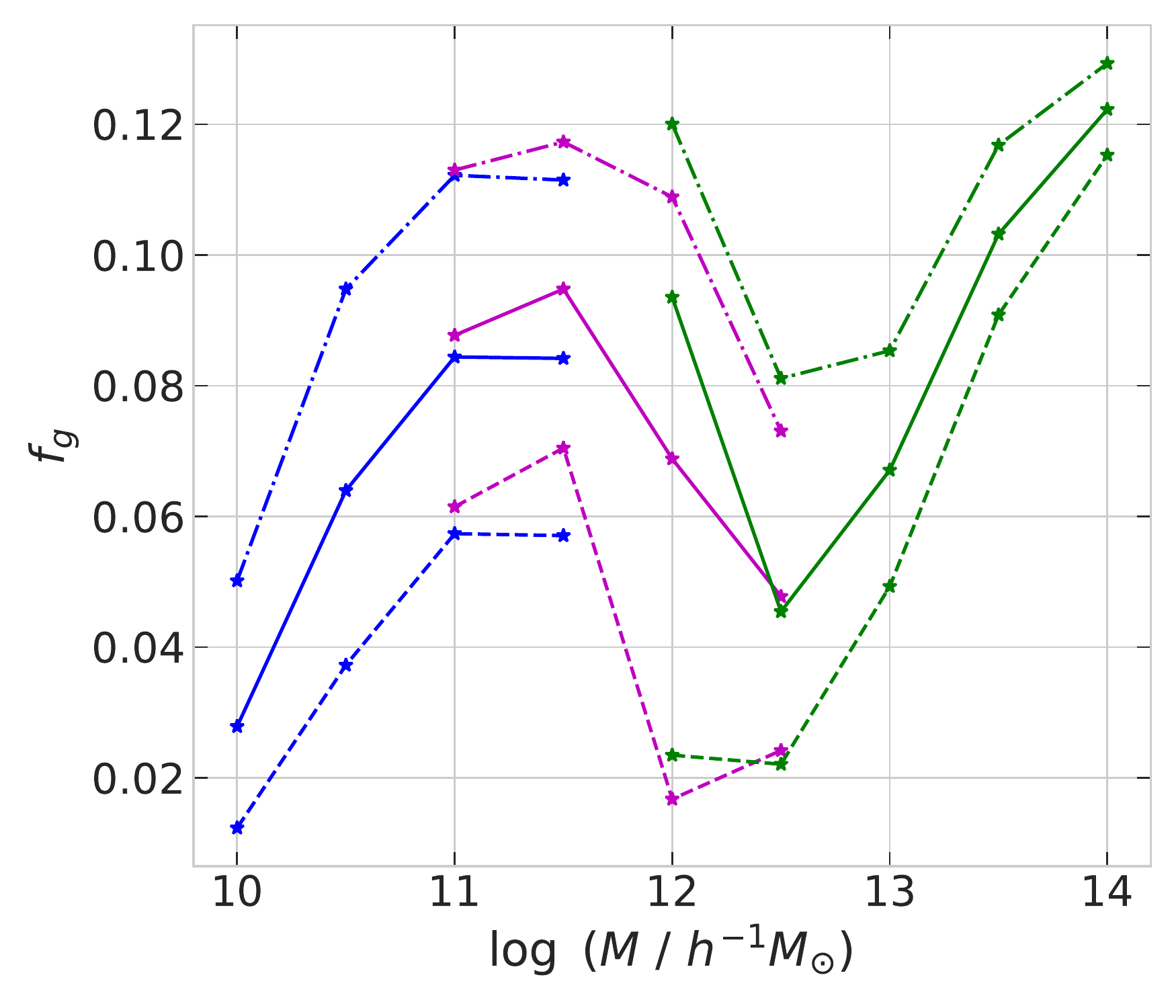}
    \caption{The 10th, 50th and 90th percentile of the four different halo properties in each of the sample selected by mass from three different cosmological boxes of the IllustrisTNG.
    This includes the concentration ($c$) of the unrelaxed halo, and the stellar fraction ($f^{\ast}$), specific star formation rate (SSFR) and gas fraction ($f_g$) of the hydrodynamical halo, all of which are defined in \secref{sec:halo-props}.} %
    \label{fig:halo-prop-pers-data}
\end{figure}

In this section, we study our parametrised model of the response of dark matter in the halo as a function of halo and galaxy properties beyond halo mass. This is motivated by the fact that there is a significant scatter in the relaxation relation (see Fig.~\ref{fig:relx-results-simple}), even within halo samples selected by mass. Such a study would also have implications for halo assembly bias and similar environmental correlations predicted in the $\Lambda$CDM framework  \citep[see, e.g., the discussion in][]{2021arXiv211200026P}. We focus on the four halo properties defined in \secref{sec:halo-props}; while these halo properties show overall trend with halo mass, they take a wide range of values even at fixed mass scale (see Fig.~\ref{fig:halo-prop-pers-data}). 

For all subsamples selected by a secondary halo/galaxy property at fixed halo mass, we use the 3-parameter model discussed above in the mass range $10^{10}$ to $10^{12.5}$ $\Mh$, 
while for cluster-scale haloes, where our 3-parameter model fails, we directly compare the linear quasi-adiabatic relaxation model parameters $q_0$ and $q_1$ as a function of the scaled halo-centric radius $r_f/R_{\rm vir}$. The results for low-mass (massive) haloes are shown in Fig.~\ref{fig:fit-fit-func-q} (Fig.~\ref{fig:fit-func-rf-13514}).

For reference, the upper panels of Fig.~\ref{fig:fit-fit-func-q} show the best-fit values of $q_0$, $q_{10}$ and $q_{11}$ as a function of halo mass alone for haloes with $M\leq10^{13}\Mh$,
which repeat the corresponding curves in Fig.~\ref{fig:3-param-mass-only}. The upper panels of Fig.~\ref{fig:fit-func-rf-13514} similarly repeat the results for $q_0(r_f)$ and $q_1(r_f)$ for the mass bins $M=10^{13},10^{13.5},10^{14}\Mh$ from Fig.~\ref{fig:rf-fit-params}. By displaying both, the full radial dependence as well as the 3-parameter description for the mass bin $10^{13}\Mh$, we can assess the reliability of the latter around the mass scale where it begins to fail. We repeat this for subsamples split by secondary halo/galaxy properties below.

\subsubsection{Dependence on unrelaxed halo concentration}
Unrelaxed haloes at fixed mass, as found in gravity-only simulations are known to have universal mass profiles characterised by their concentration alone (defined in section \ref{sec:halo-props}) together with their mass.
As can already be noted in \figref{fig:halo-prop-pers-data}, this NFW concentration is correlated with the halo mass \citep[see e.g. ][]{2006ApJ...652...71W,2007MNRAS.378...55M,2015ApJ...799..108D,2017MNRAS.468.2984P}.
In order to isolate the effect of concentration on the response, we define \textbf{concentration significance} $(c_s)$.
\begin{align}
c_s = \left(\log c- \log \bar{c}(M)\right)/\sigma\,. \nonumber
\end{align}
Here we use the median $\bar{c}(M)$ and scatter $\sigma$ of the concentration-mass relation as given by \citet{2019ApJ...871..168D} and computed with the COLOSSUS code \citep[][]{2018ApJS..239...35D}.

\begin{figure*}
    \centering
    \includegraphics[width=0.32\linewidth]{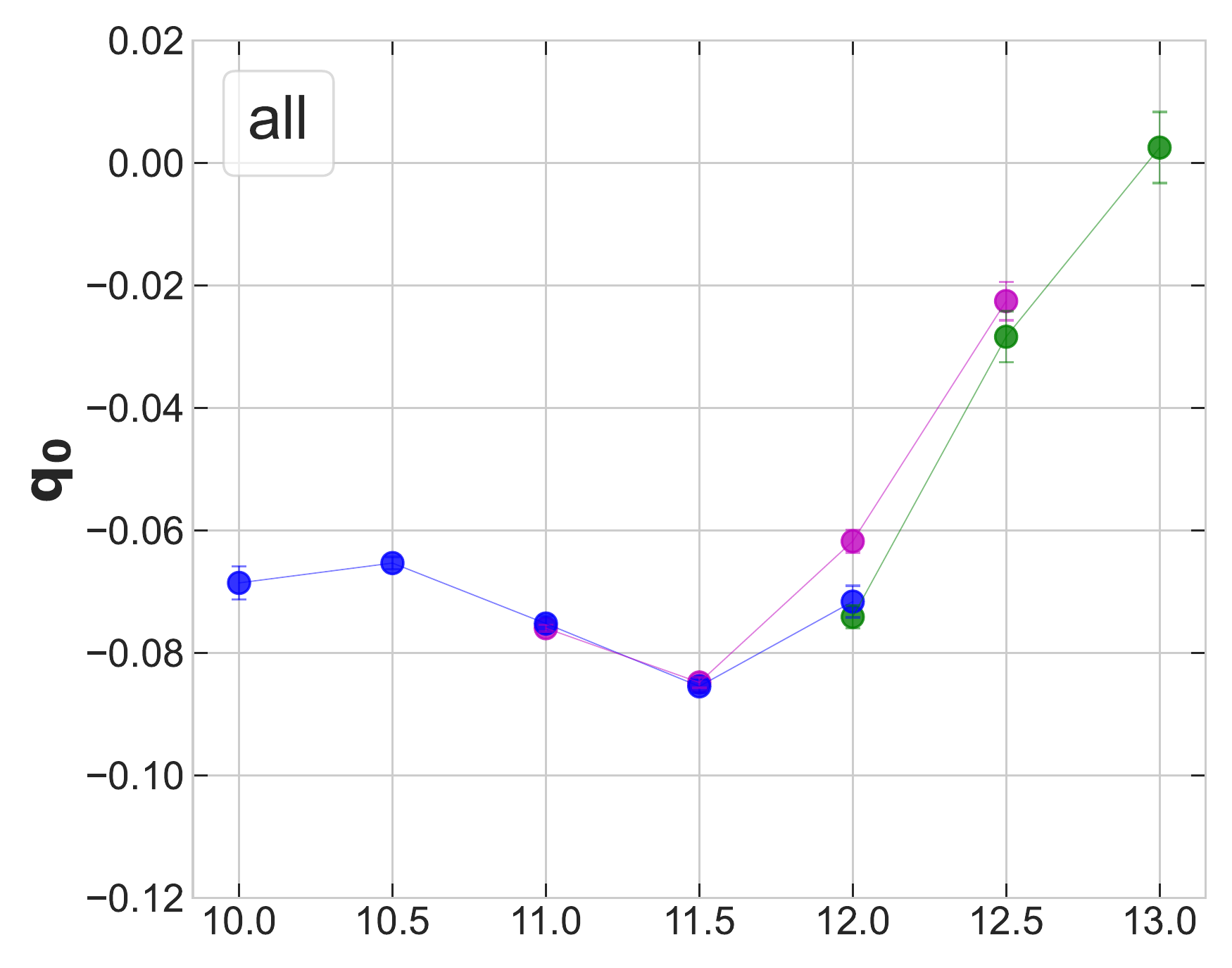}
    \includegraphics[width=0.32\linewidth]{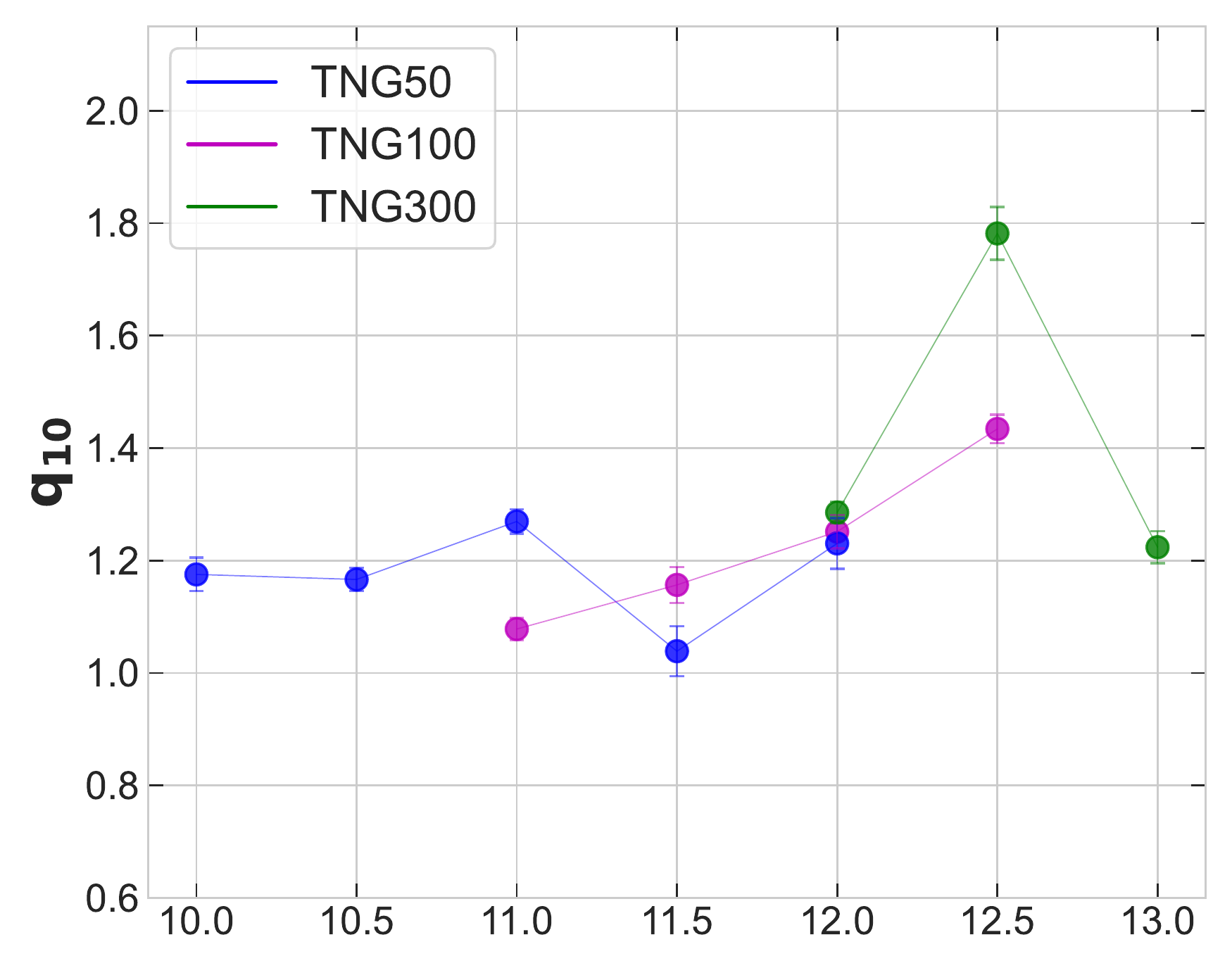}
    \includegraphics[width=0.32\linewidth]{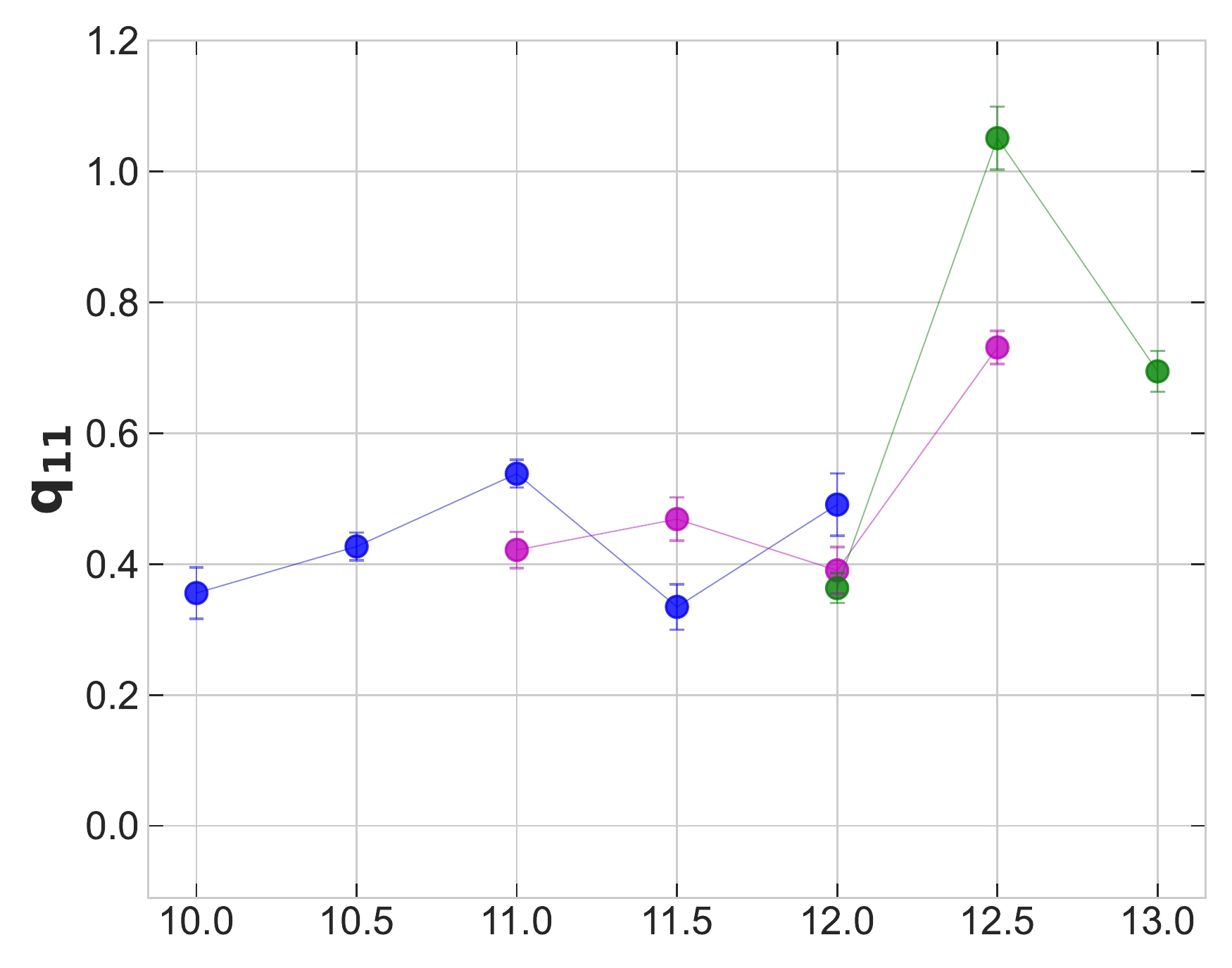}
    \includegraphics[width=0.32\linewidth]{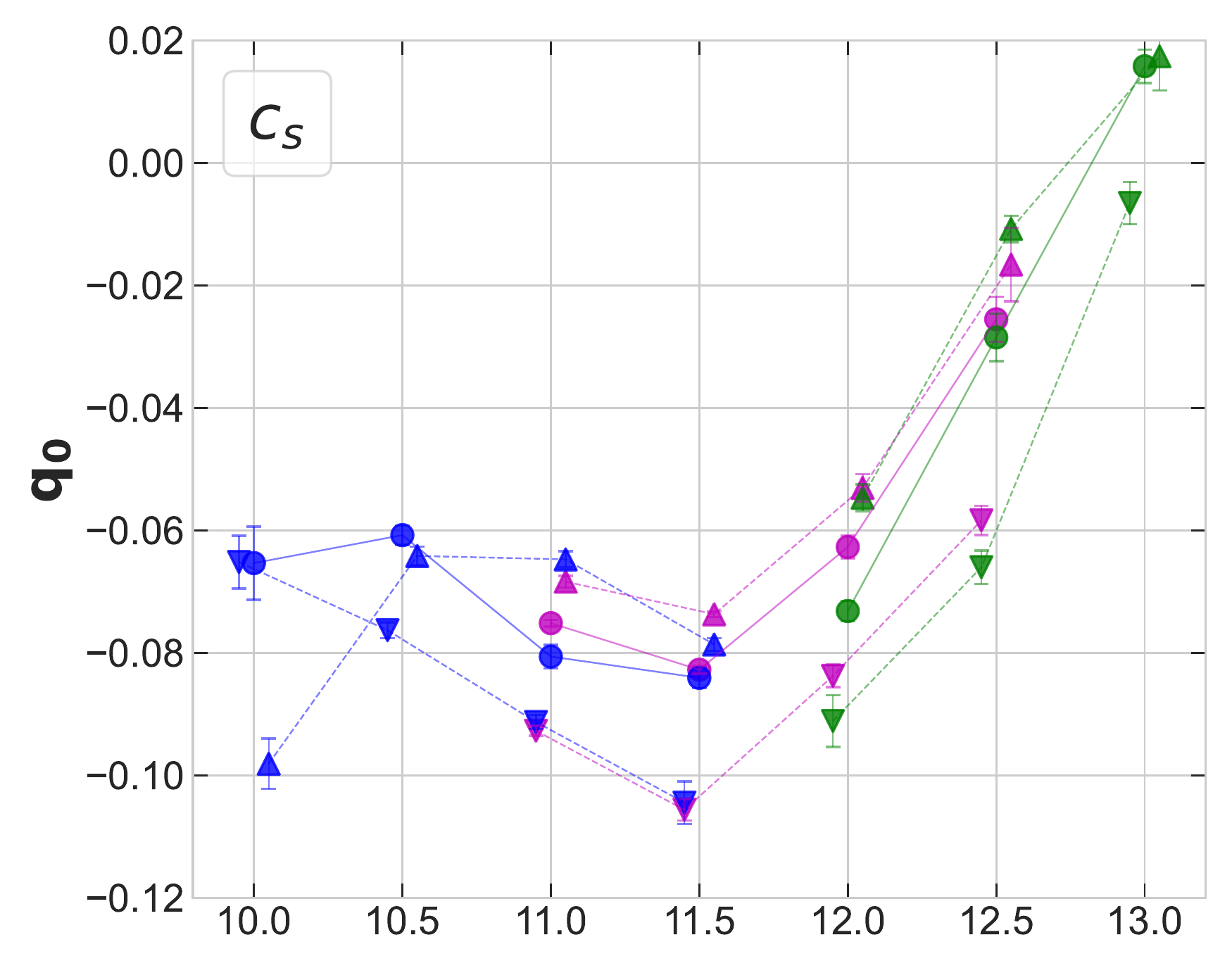}
    \includegraphics[width=0.32\linewidth]{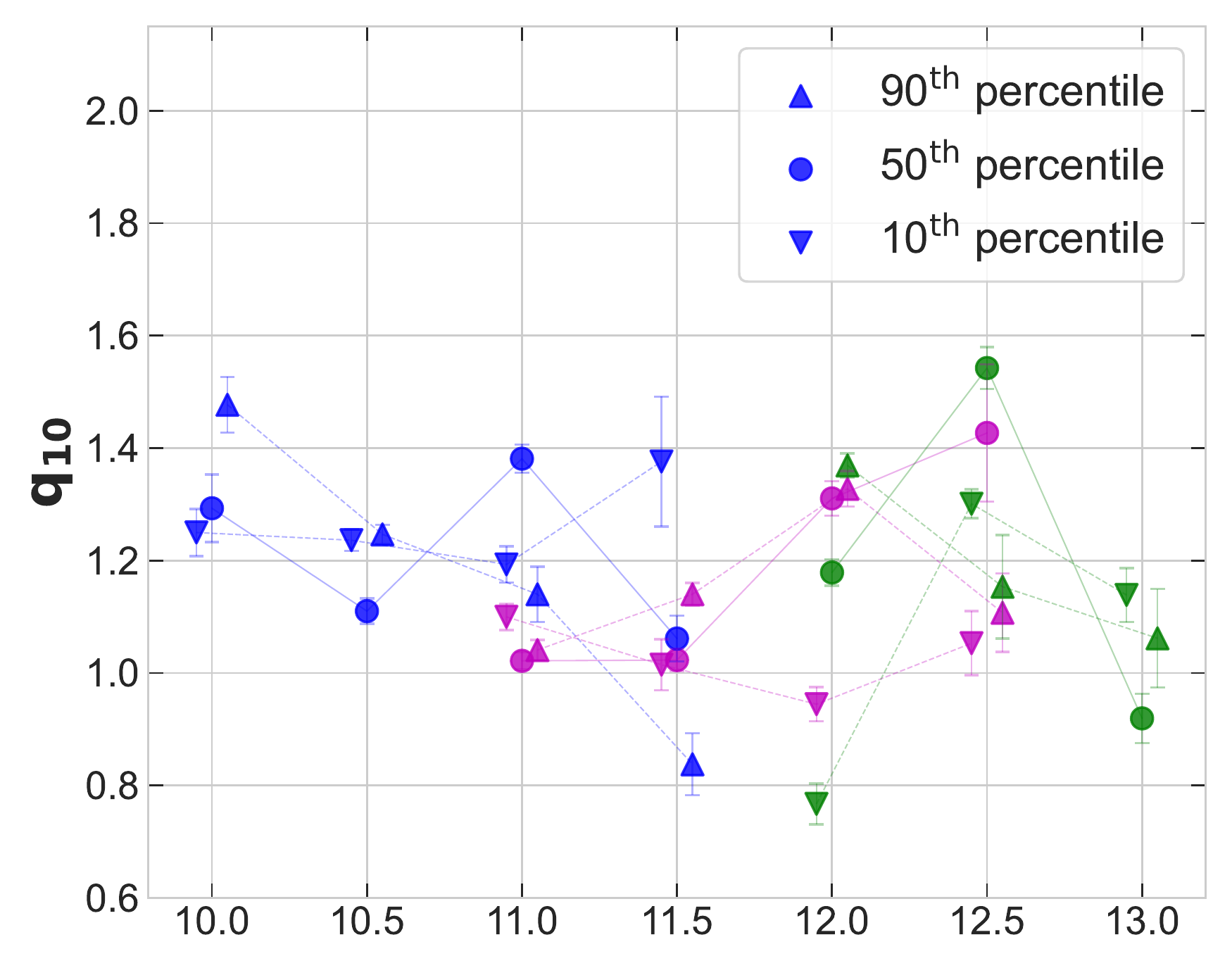}
    \includegraphics[width=0.32\linewidth]{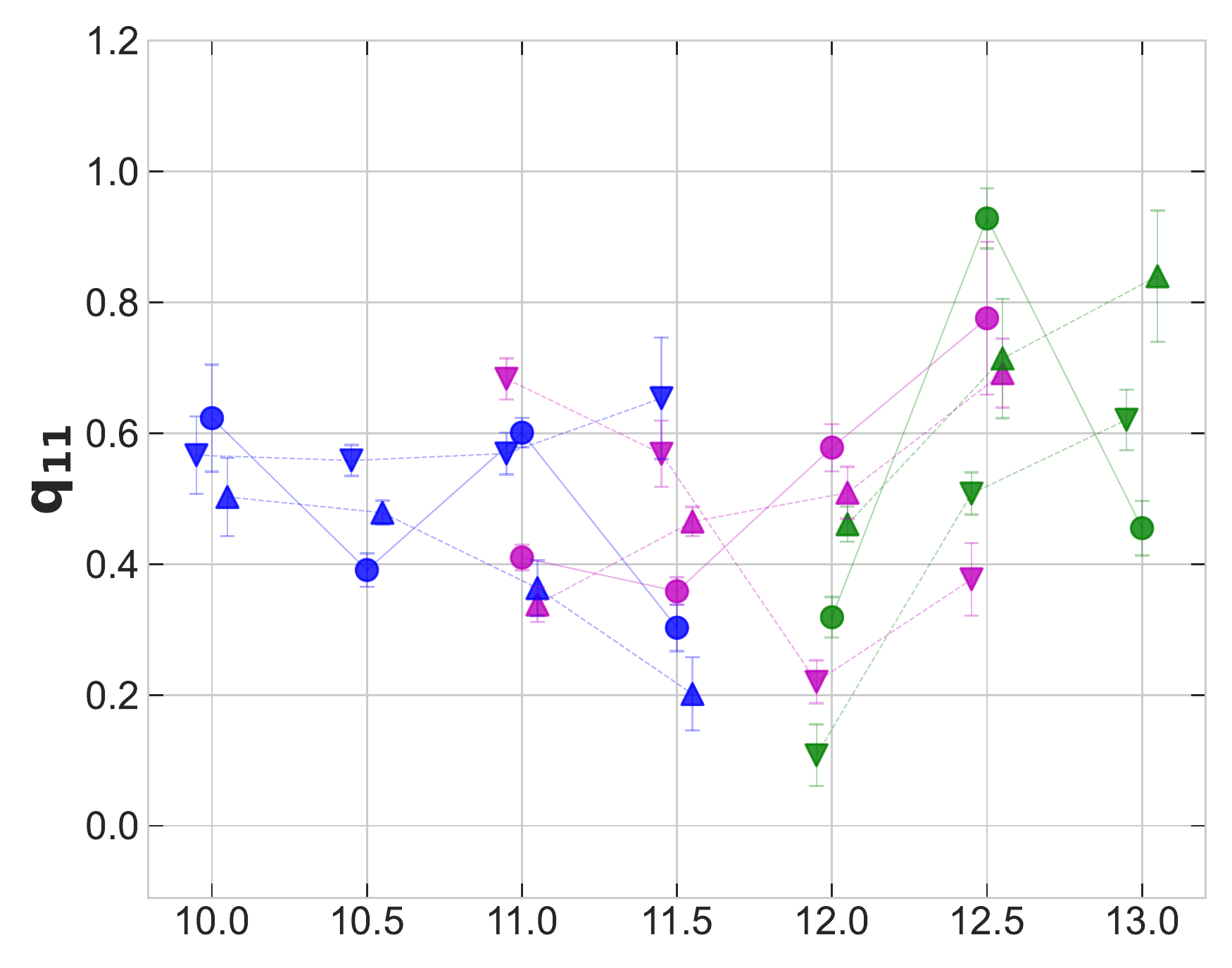}
    
    \includegraphics[width=0.32\linewidth]{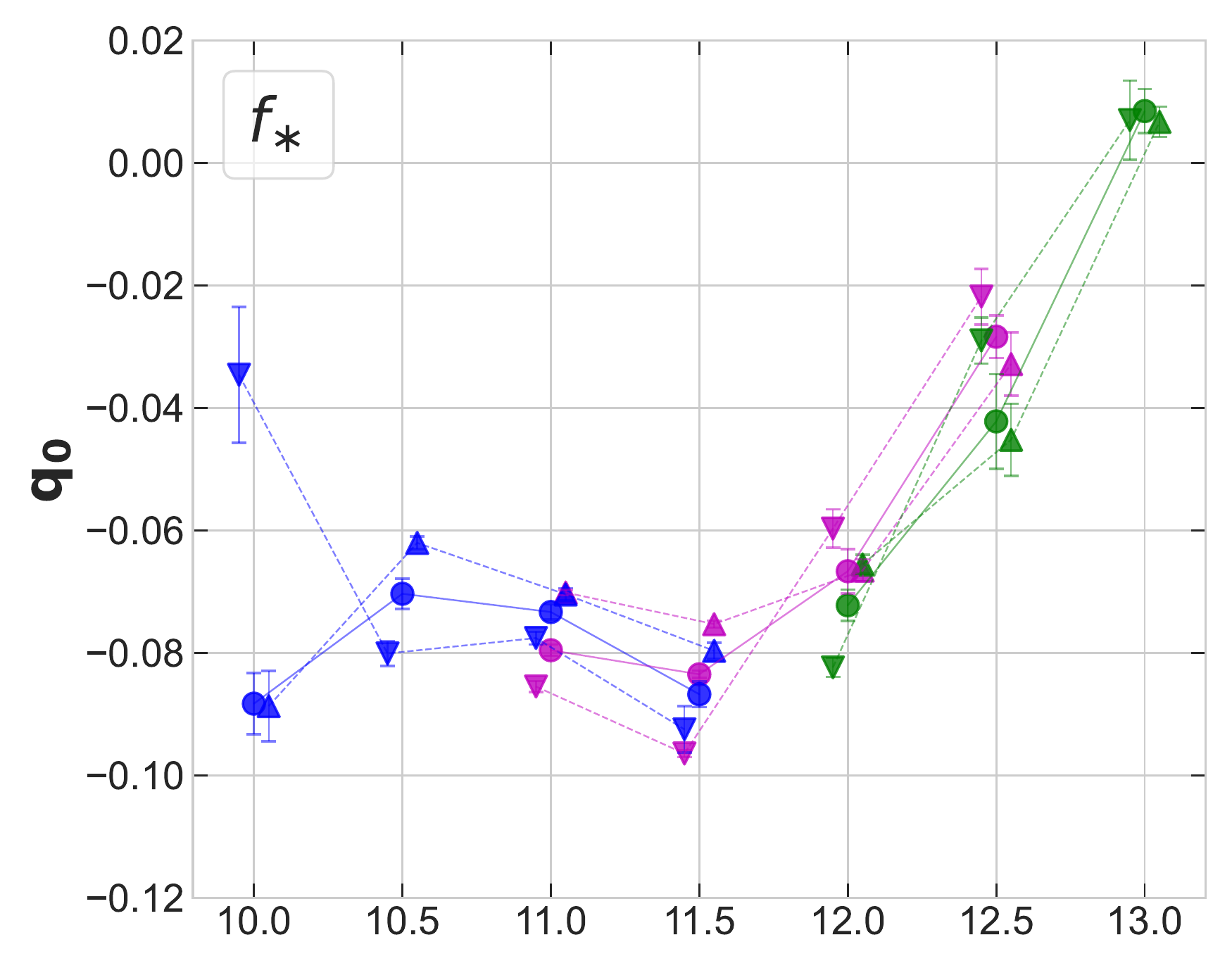}
    \includegraphics[width=0.32\linewidth]{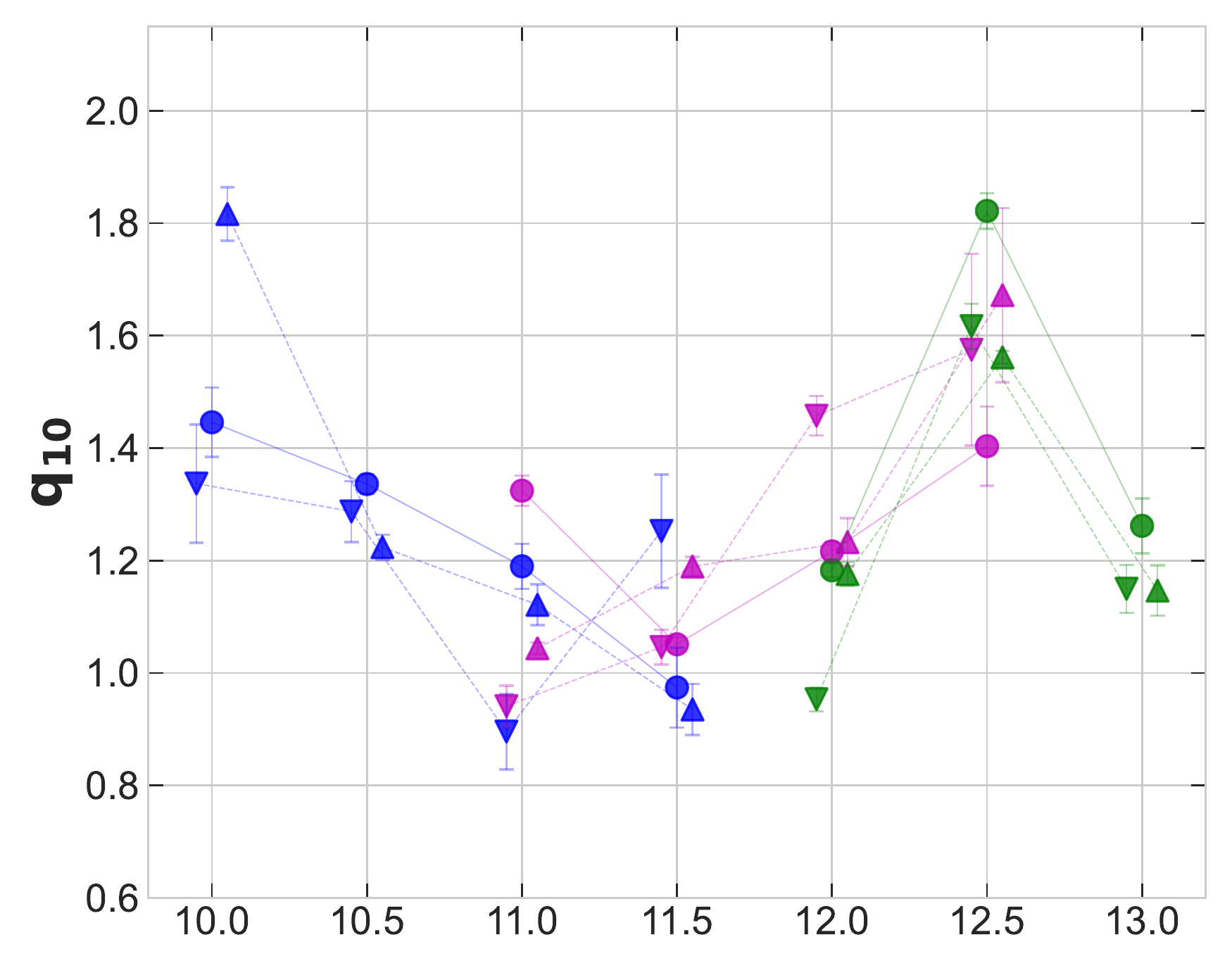}
    \includegraphics[width=0.32\linewidth]{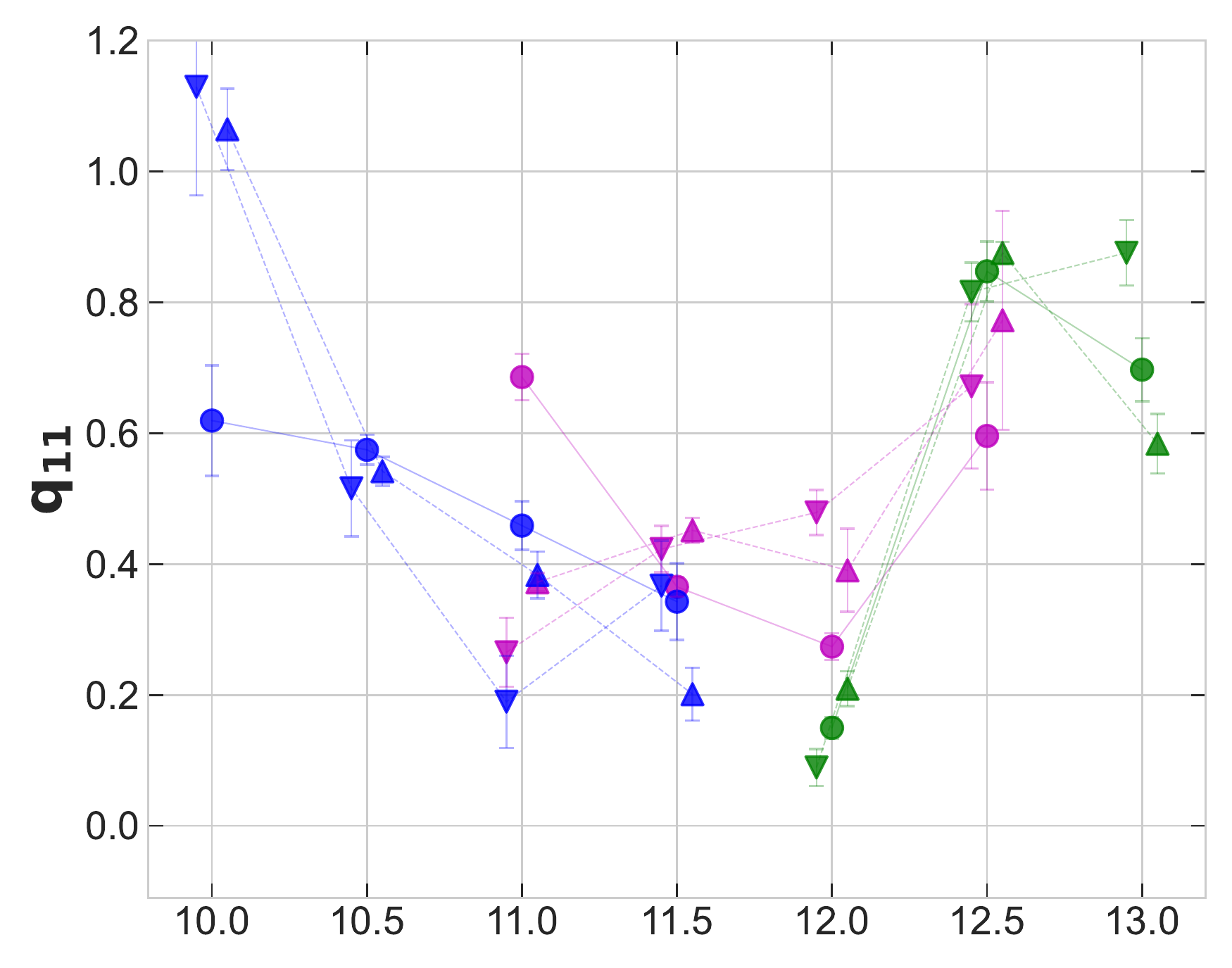}
    
    \includegraphics[width=0.32\linewidth]{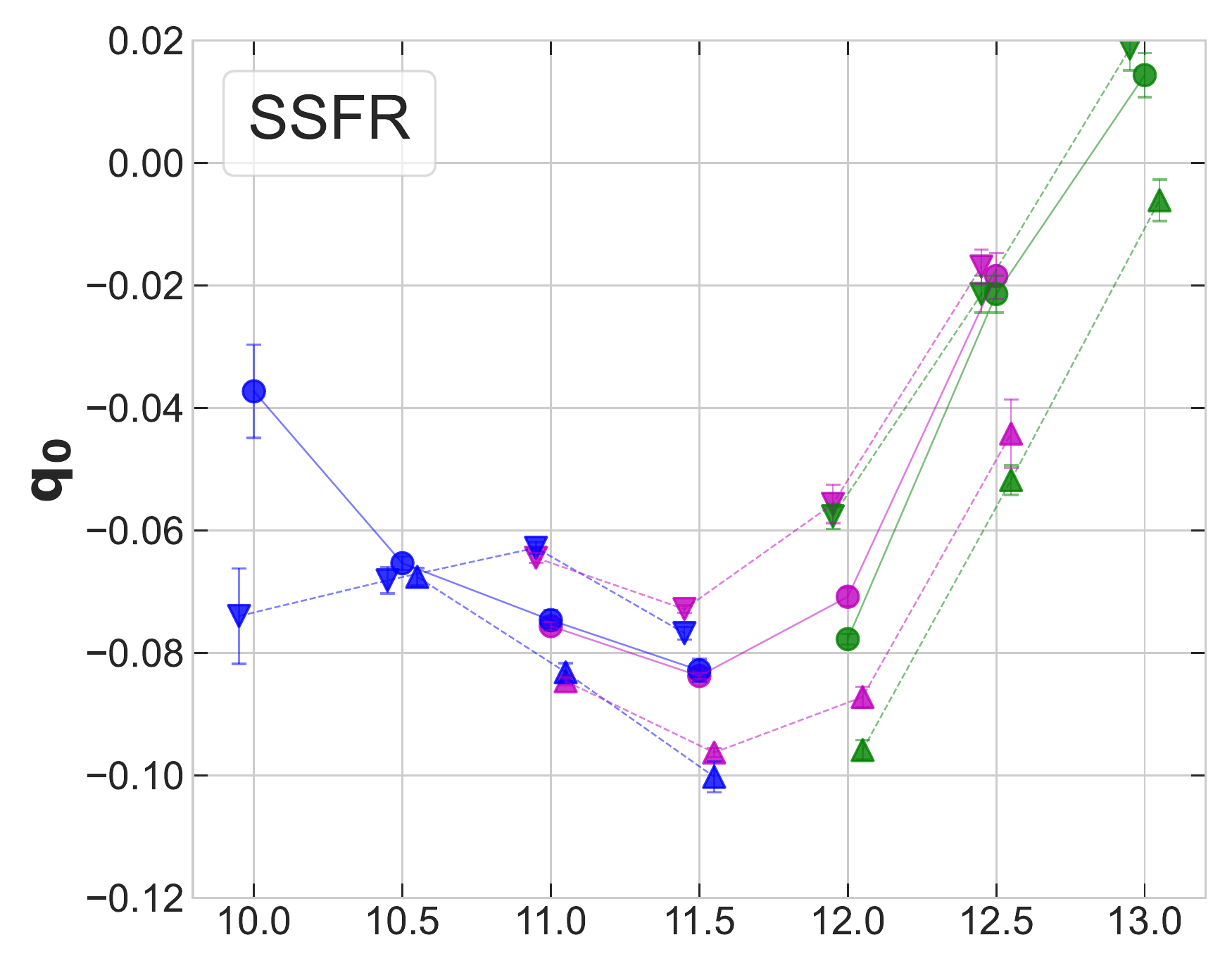}
    \includegraphics[width=0.32\linewidth]{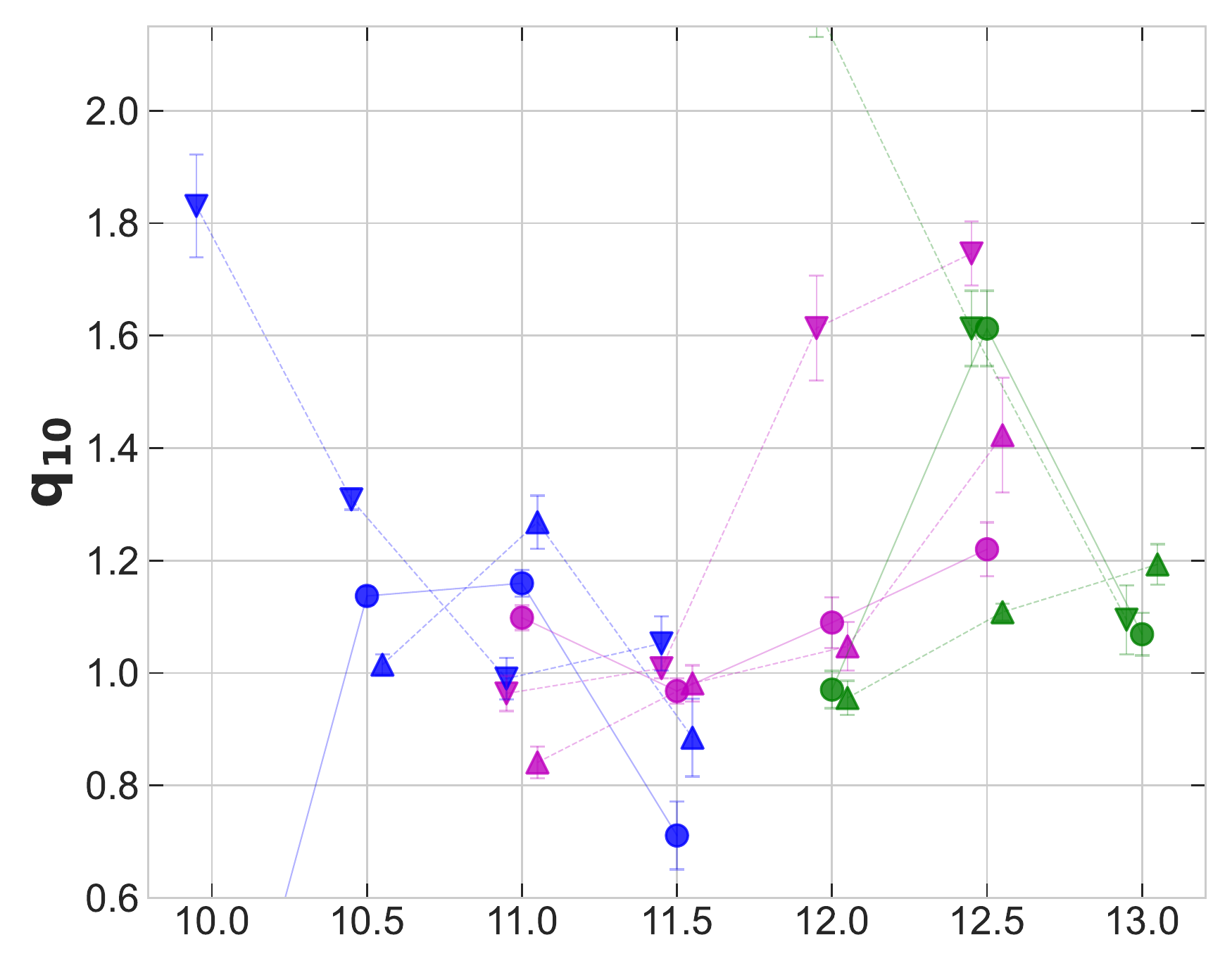}
    \includegraphics[width=0.32\linewidth]{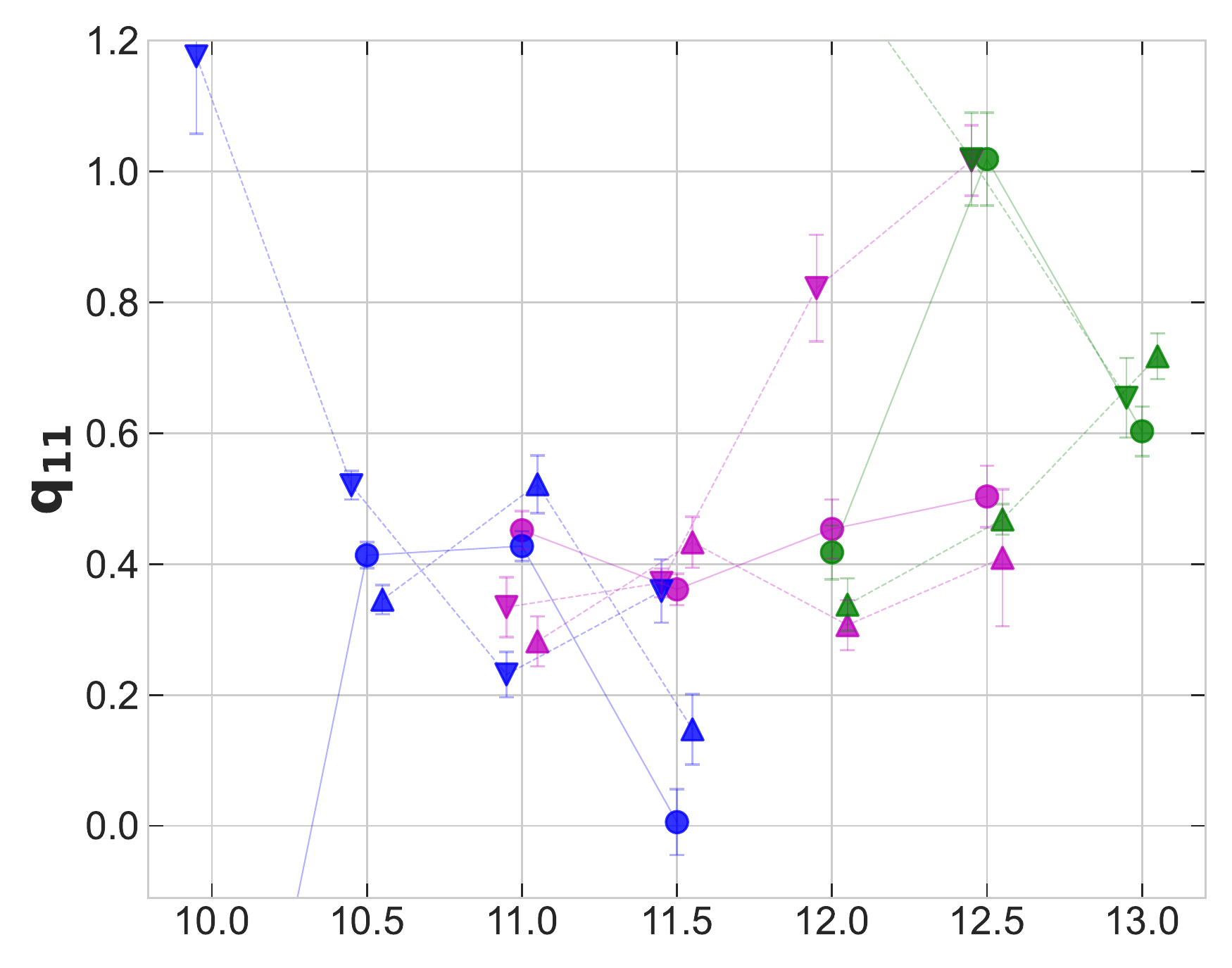}
    
    \includegraphics[width=0.32\linewidth]{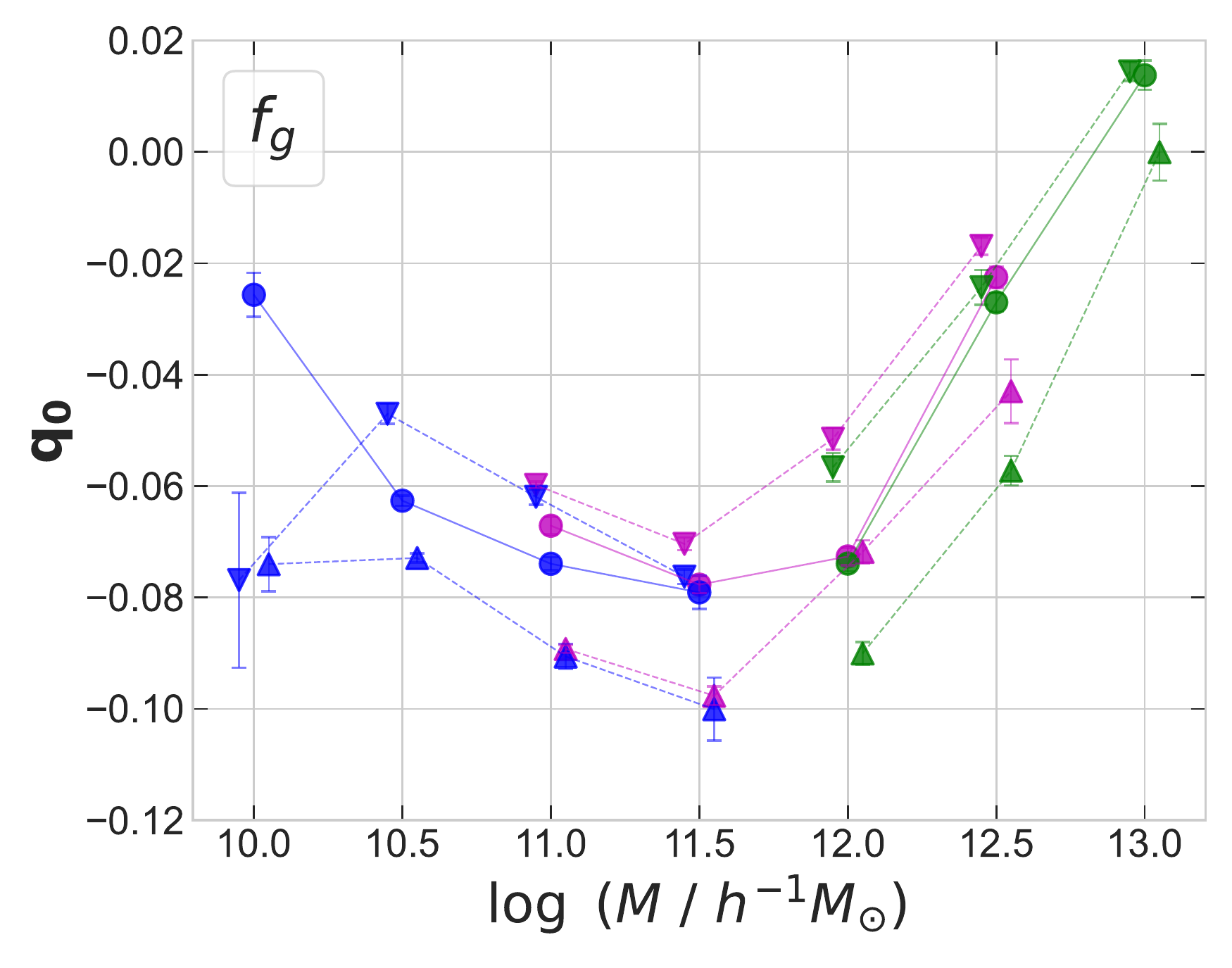}
    \includegraphics[width=0.32\linewidth]{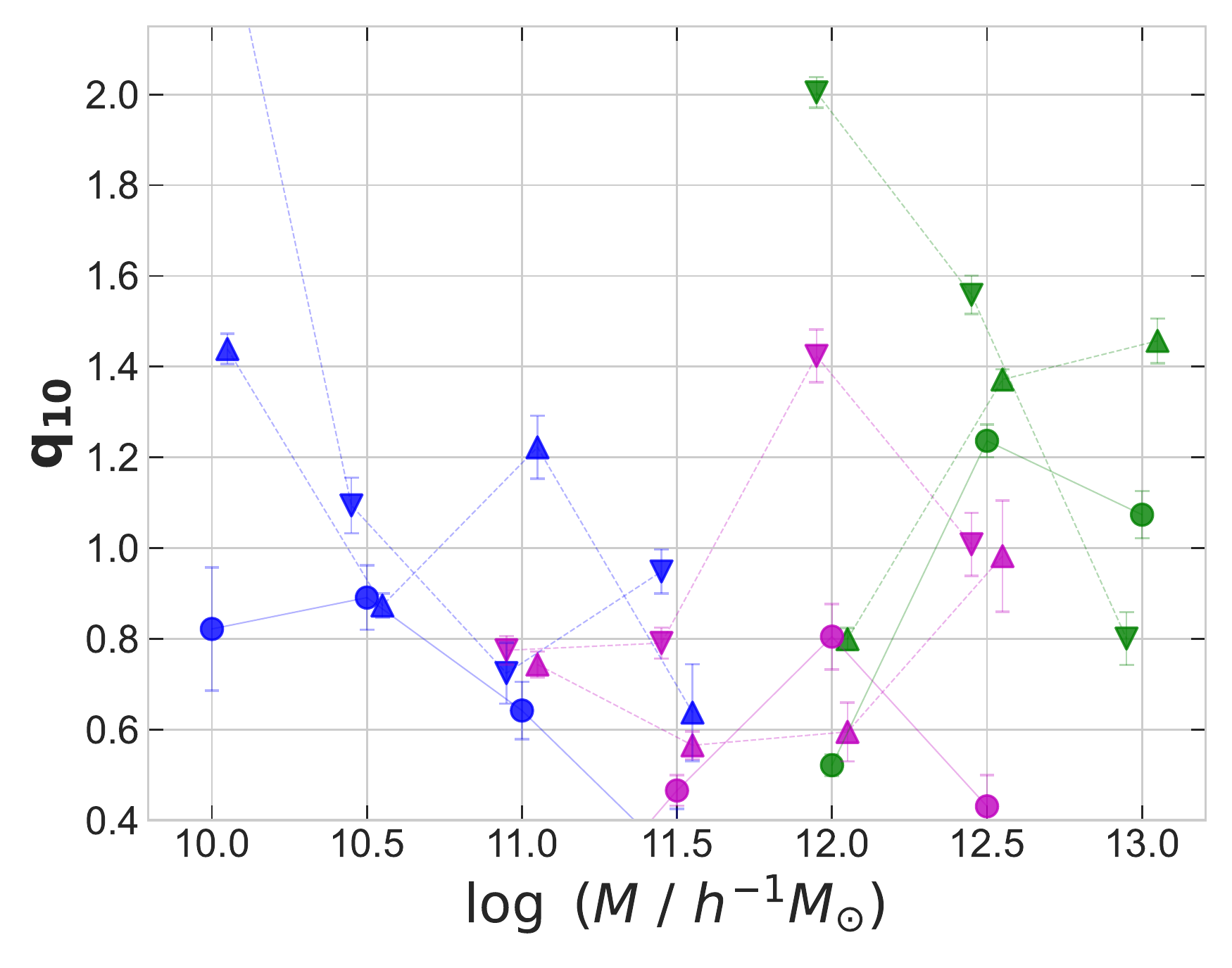}
    \includegraphics[width=0.32\linewidth]{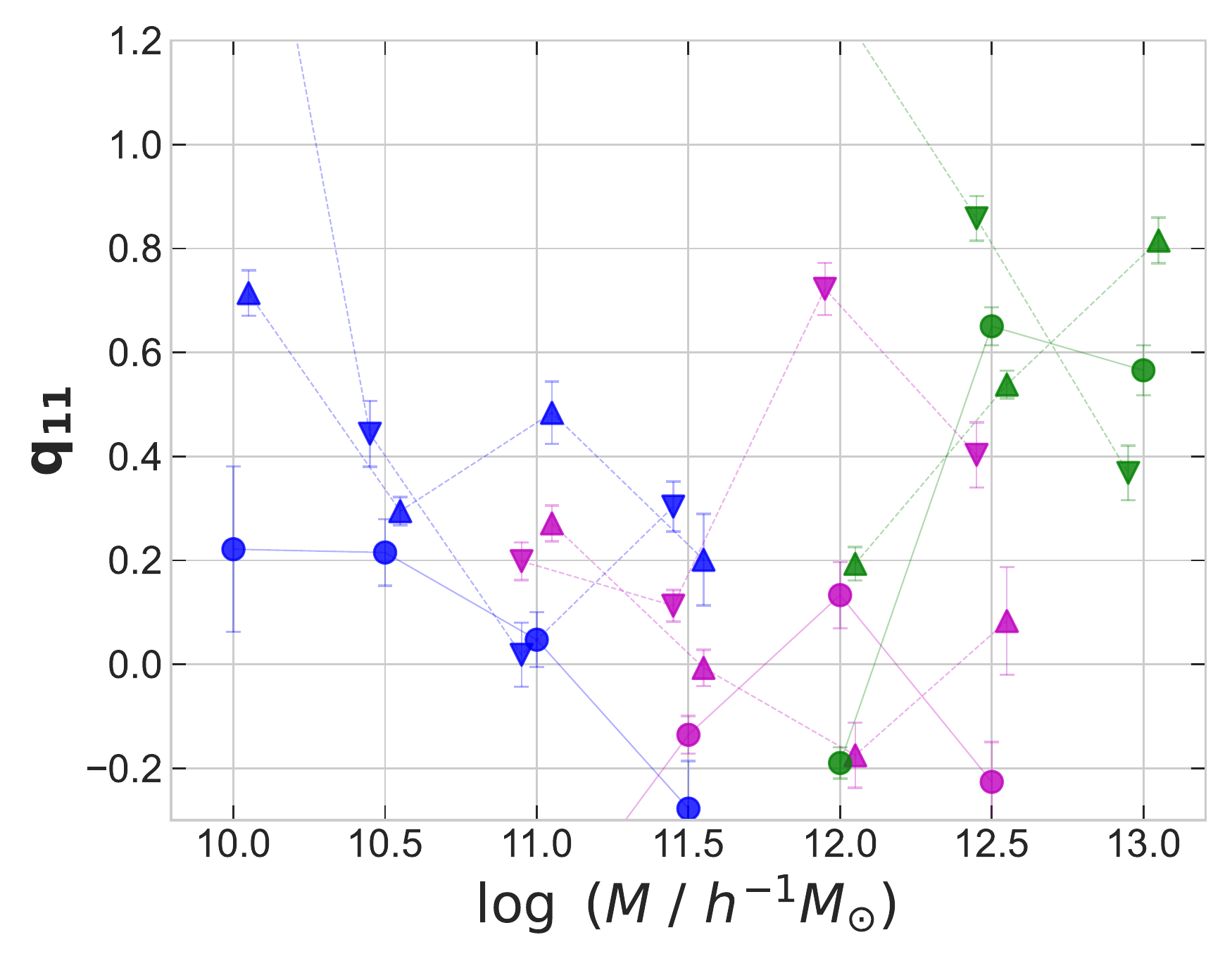}
    \caption{Radially dependent quasi-adiabatic relaxation model parameters $q_{0}$, $q_{10}$ and $q_{11}$ estimated as a function of halo properties in IllustrisTNG simulations. In the top row panels, only halo mass dependence is shown; whereas in the next three rows, we show the dependence on halo concentration, stellar mass fraction, specific star formation rate and gas fraction in terms of percentiles respectively.} %
    \label{fig:fit-fit-func-q}
\end{figure*}

Then we select haloes from each of the mass bins in three separated $c_s$ percentile bins $(10\pm10, ~50\pm10 ~\&~ 90\pm10)$  and compute the relaxation relation as described in previous sections for each of those samples. 
We find that $q_0$ shows strong dependence on the concentration, with more concentrated haloes having higher value of $q_0$ at most halo masses with $M \leq 10^{13} \Mh$ (see second row in \figref{fig:fit-fit-func-q} and second row, first column in \figref{fig:fit-func-rf-13514}). This can be understood in terms of the formation time of the halo, since concentration is correlated with the formation time. More concentrated haloes, that have formed earlier might have had enough time for the dark matter to respond to the baryonic feedback, and hence there is less offset. On the other hand, $q_{10}$ or $q_{11}$ display a more complex dependence at low mass, with no clear monotonic trend. Meanwhile, for cluster-scale haloes, $q_1$ shows a very different behaviour as a function of $r_f$ at different halo concentrations (see second row of \figref{fig:fit-func-rf-13514}). We leave a fuller exploration of these trends, particularly their dependence on substructure properties, to future work.

\subsubsection{Dependence on baryonic halo properties}
We now shift our focus to hydrodynamical halo properties. In this regard, we study the response as a function of the three halo properties defined in \secref{sec:halo-props}; namely the specific star formation rate (SSFR) at current redshift $z=0$ and the total stellar mass fraction ($f_{\ast}$) and gas fraction ($f_g$) at redshift $z=0$ which respectively represent the integrated star formation activity and gas content of the central subhalo. At each halo mass bin, we take three subsamples selected by bins of percentiles in $f_{\ast}$, SSFR and $f_g$, in a similar fashion as with concentration significance. %

From the third column of \figref{fig:fit-fit-func-q}, we note that $f_\ast$ does not affect the relaxation response significantly; in particular the $q_0$ parameter is relatively least dependence on $f_{\ast}$, compared to other halo properties. This is consistent with the fact that the $q_0$ parameter clearly converges between three TNG boxes (see upper panel of \figref{fig:fit-fit-func-q}), despite large differences in $f_\ast$ with resolution (see upper right panel of \figref{fig:halo-prop-pers-data}).
On the other hand, we can see a clear trend  in $q_0$ parameter with SSFR (see the first column in the third row of \figref{fig:fit-fit-func-q}); at a given halo mass, the $q_0$ value is closer to zero when the star formation activity is lower. To recall, $q_0 \simeq 0$ would mean no offset in the relaxation relation, and in that case for shells having no relaxation, the mass ratio is unity indicating that the enclosed baryonic mass also remains same. This result is consistent with our argument in \secref{sec:results-rad-dep-qadiab} that the $q_0<0$ is caused by the recent baryonic outflows due to feedback which is lower in these low mass haloes when SSFR is low.
From the first panel in the last row of \figref{fig:fit-fit-func-q}, we can see a similar trend in $q_0$ with gas fraction $f_g$ of the halo; this is likely due to the fact that the FOF haloes with more gas have relatively higher active star formation with larger recent baryonic outflows. However, even the halos with similar SSFR and different gas fraction may show different relaxation behaviour. In a future work, we will study such effects using hydrodynamic simulations with different baryonic prescriptions, that produces haloes with same $f_g$ but very different SSFR and vice versa.
On the other hand, for the cluster-scale haloes shown in \figref{fig:fit-func-rf-13514}, the $q_0$ parameter does not vary significantly with any of the halo property that we considered.

Turning to $q_1$, while we see no clear dependence of its constituent parameters $q_{10}$ and $q_{11}$ on the hydrodynamical halo properties of low-mass haloes in \figref{fig:fit-fit-func-q}, for cluster-scale haloes we do see a strong, albeit complex, dependence of $q_1$ on SSFR (see the third row of \figref{fig:fit-func-rf-13514}). Like the case of the concentration significance, it will be interesting in future work to understand the physical mechanisms driving some of the stronger correlations of the halo response with properties such as SSFR and $f_g$ seen above.

\begin{figure*}
    \centering
    \includegraphics[width=0.84\textwidth]{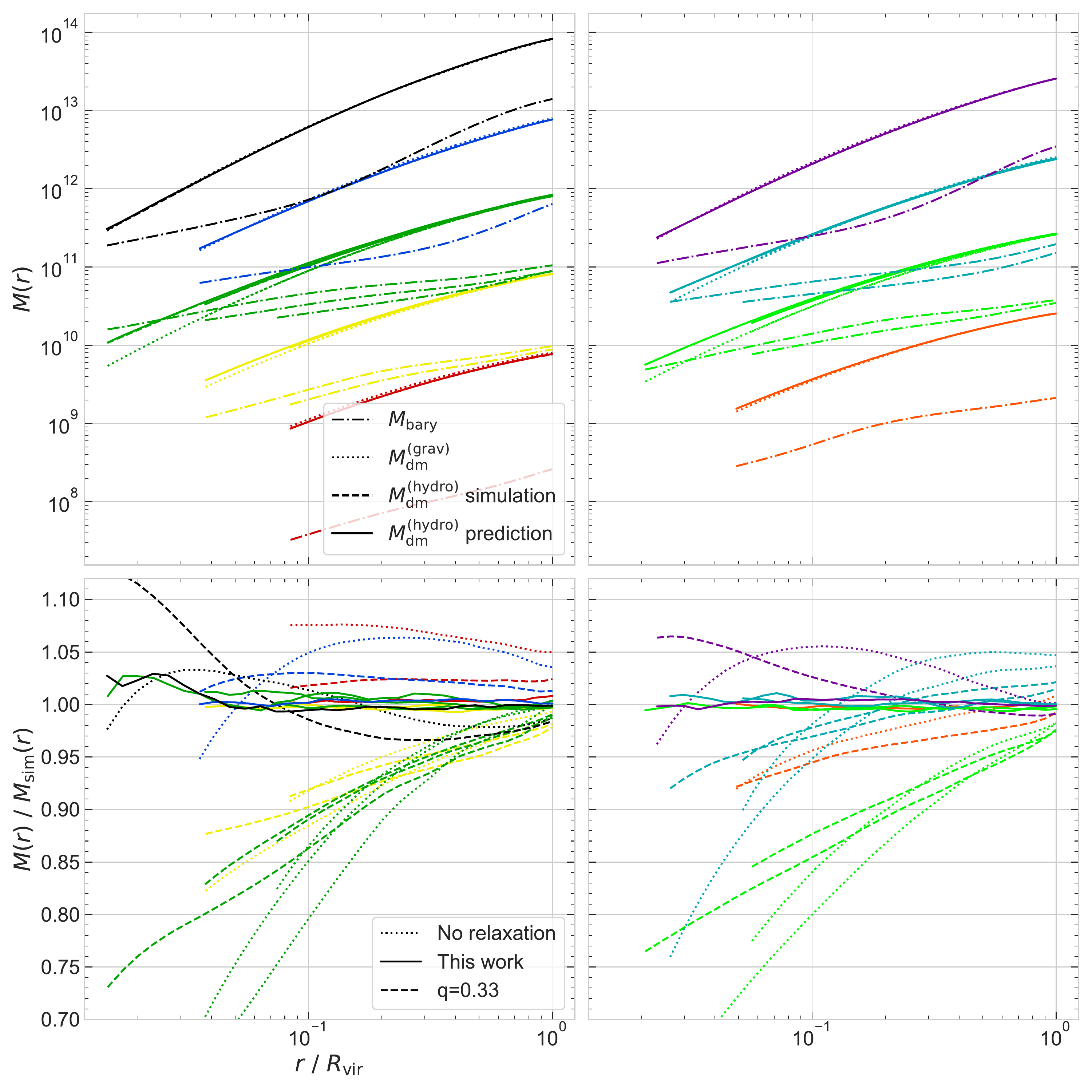}
    \caption{\emph{(Top row:)} For the haloes in IllustrisTNG simulations, the mean radial mass profiles are shown in bins of halo mass for the baryonic component (dash-dotted curves) and dark matter component in hydrodynamic (dashed curves) and gravity-only (dotted curves) runs.  The relaxed dark matter mass profile predicted as described in \secref{sec:mass-prof-demo} is shown by solid curves. The colour-coding follows Fig.~\ref{fig:mass_bin_label}; for clarity we use two panels to show the averaged mass profiles for the nine mass bins. 
    \emph{(Bottom row:)} The ratio of the relaxed dark matter mass profile predicted by our model to that from the hydrodynamic simulation is shown by solid curves. For comparison, the corresponding ratio for quasi-adiabatic relaxation model with $q=0.33$ is shown by dashed curves and the ratio of dark matter mass profile between gravity-only simulation to the full hydrodynamic simulation is shown by dotted curves, representing the case of no relaxation. } 
    \label{fig:demo-fit}
\end{figure*}

\section{Applications}
\label{sec:applic}
In this section, we briefly discuss a few (potential) applications of our analysis.

\subsection{Baryonification schemes}

The results above show that the response of a halo's dark matter content to the galaxy and gas evolving in it depends not only on the integrated properties of the halo and galaxy (such as mass, concentration, etc.) but also on halo-centric distance, even at fixed mass ratio. This is in stark contrast to analytical approximations employed in the literature which typically use simplified  relations between the relaxation ratio and mass ratio, ignoring the radial dependence. These analytical approximations are now commonly employed in baryonification schemes to predict the total matter power spectrum for a given cosmological model using only the results of gravity-only $N$-body simulations \citep{2015JCAP...12..049S,2018MNRAS.480.3962C,2021MNRAS.503.3596A}. Our results above can directly impact such predictions by modifying the small-scale (deep 1-halo regime) behaviour of the power spectrum. 

For example, to model the effect of baryons in low- and intermediate-mass haloes ($\lesssim10^{13}\Mh$), we advocate the use of our fitting function \eqn{eq:q3-model} for the relaxation relation, with parameters set to $q_0\simeq-0.05$, $q_{10}\simeq1.1$ and $q_{11}\simeq0.5$,\footnote{We have tested that the relaxed mass profile predicted by such a generic model agrees reasonably with the simulation; see \secref{sec:mass-prof-demo} and Appendix \ref{sec:apndx-demo} for a more accurate prediction} which gives a good description of the results of both IllustrisTNG and EAGLE haloes (see Fig.~\ref{fig:3-param-mass-only}). For larger (cluster-sized) haloes, the response is still accurately described by the relation \eqn{eq:chi-linear-q0}, but with more complex behaviour for the parameters $q_1(r_f)$ and $q_0(r_f)$, which presently needs to be accounted for numerically (see, e.g., Fig.~\ref{fig:fit-func-rf-13514}). We discuss this further in \secref{sec:conclusion}.

\subsection{Mass profiles}
\label{sec:mass-prof-demo}
The primary utility of an analytical model (or fitting function) such as \eqn{eq:chi-linear-q0} for the relaxation relation is to be able to predict the relaxed dark matter profile $M_f^d(r_f)$ of a halo which has responded to its baryonic content. The procedure for obtaining this profile is straightforward \citep[see, e.g., Appendix A of][]{2021MNRAS.503.4147P}: the relaxation relation is solved iteratively using the unrelaxed mass profile and the baryonic mass profile as inputs, until a converged answer for $M_f^d(r_f)$ is achieved.\footnote{In some cases, when these input profiles can be described by simplified analytical forms, a fully analytical expression for the relaxed dark matter profile can also be obtained \citep[see, e.g., Appendix A of][]{2021MNRAS.507..632P}.} In our case, the procedure to obtain the relaxed dark matter profile using \eqn{eq:chi-linear-q0} works identically. The additional radial dependence of $q_1$ and $q_0$ is not an issue, since the radius $r_f$ itself is used as the control variable in solving for the enclosed mass.

As an example, we compare the relaxed profiles predicted by this procedure -- using the unrelaxed and baryonic mass profiles and the fits to the relaxation relation \eqn{eq:chi-linear-q0} as inputs -- with the dark matter profiles actually measured in the hydrodynamical simulations for the same haloes.
For simplicity, in this analysis we ignore the dependence of the dark matter response on halo properties other than the mass;
we use $q_0$ and $q_1$ as a function of $r_f/R_{\rm{vir}}$ as shown in the \figref{fig:rf-fit-params} for each halo mass bin. In the upper panel of Fig.~\ref{fig:demo-fit}, we show this estimated mass profile along with the actual mass profiles found in the IllustrisTNG simulation.
For comparison, we also show the results of replacing the relaxation relation with simpler approximations from the literature (while still using the unrelaxed and baryonic mass profiles from the simulation as inputs in the iterative procedure).
Our model produces significantly better estimates of the relaxed dark matter profile, especially in the inner halo we obtain an order of magnitude better accuracy in comparison to such simple models (see lower panel of Fig.~\ref{fig:demo-fit}). In Appendix \ref{sec:apndx-demo}, we show that even the simple three parameter model gives a reasonably good prediction of the relaxed mass profile, while also being easier to incorporate into the existing procedures that use an adiabatic relaxation model.

\subsection{Rotation curves}
Since our model can predict relaxed mass profiles using unrelaxed and baryonic mass profiles as inputs, it can also predict rotation curves of galaxies using the same inputs, along with some assumptions regarding the geometry of various mass components. 
The interpretation of observed rotation curves and related statistics such as the radial acceleration relation, using data from spatially resolved spectroscopy of nearby galaxies, forms a key aspect of discussions in the literature regarding the nature of gravity at galactic scales \citep[e.g.,][]{lms16b,lmsp17}. 

In the $\Lambda$CDM context, such studies typically model the relaxed dark matter profile using a generalised NFW profile, with or without a core, but unconnected to the baryonic mass \citep[e.g.,][]{llms20}. Previous work has suggested that the use of a parametrised model of dark matter \emph{response}, rather than the relaxed profile itself, should lead to more robust results \citep{2021MNRAS.507..632P,pscs21}. For example, it is known that the use of smooth NFW-like profiles does not produce formally good fits in cases where the observed rotation curve shows oscillatory behaviour. Rather, these oscillations in the rotation curve  correlate with similar oscillations seen in the measured baryonic mass profiles \citep[see, e.g., figs.~4 and~6 of][]{llms20}.\footnote{There could also be additional biases induced by various simplifying modelling assumptions regarding, e.g., circularity of orbits and disk thickness, which must be accounted for especially in the context of cored versus cuspy inner halo profiles \citep[see, e.g., the discussion in][]{roper+22}.} It is then reasonable to speculate that a model which smoothly parametrises the physics of the dark matter response, rather than the profile of dark matter itself, might account for such correlations naturally. More generally, such a model is more physically motivated than one which directly parametrises the dark matter profile itself. 

In future work, we plan to confront observed rotation curves for low-mass systems with the 3-parameter relaxation model presented above. Our specific results for the values of these parameters in IllustrisTNG and EAGLE can then provide useful priors for the statistical comparison with data.

\section{Conclusion}
\label{sec:conclusion}
In this work, we have explored in detail the response of the dark matter content of a halo to the galaxy and gas it hosts. Understanding and accurately modelling this response is important for a number of applications including baryonification schemes for small-scale power spectrum emulation, rotation curve modelling, %
constraining the nature of dark matter using inner halo mass profiles, etc. 

Using haloes and galaxies identified in the IllustrisTNG and EAGLE simulations and matched to their gravity-only counterparts, our analysis demonstrates that the simplified analytical schemes used thus far to model the dark matter response \citep[e.g.,][]{1986ApJ...301...27B,2010MNRAS.407..435A,2015JCAP...12..049S} are inadequate in describing its detailed behaviour across a variety of halo and galaxy types. Specifically, we showed that the dark matter response, or relaxation relation (see equation~\ref{eq:qAR}), which connects the relaxation ratio $r_f/r_i$ to the mass ratio $M_i/M_f$ between unrelaxed (gravity-only) and relaxed (hydrodynamical) haloes, explicitly depends on halo-centric distance $r_f$ in the relaxed halo, in addition to being sensitive to a number of halo and galaxy properties including halo mass, halo concentration, stellar and gas mass fraction, and specific star formation rate. These effects, especially the dependence on halo-centric distance, have been typically neglected by existing quasi-adiabatic relaxation models. 

We presented a simple, physically motivated extension (equation~\ref{eq:chi-linear-q0}) of the existing models which accurately captures the dark matter response over 4 orders of magnitude in halo mass ($10^{10}\lesssim M/(\Mh)\lesssim 10^{14}$) and $\sim2$ orders of magnitude in relative halo-centric distance ($0.02\lesssim r_f/R_{\rm vir}\leq1$). Apart from an explicit radial dependence of the relaxation relation (e.g., equation~\ref{eq:q3-model} for low-mass haloes), a second novelty of our model is the inclusion of a parameter $q_0$ which characterises feedback-induced offsets seen in the relaxation relation measured in IllustrisTNG and EAGLE haloes in which, e.g., shells that do not show an overall change in radius ($r_f/r_i\simeq1$) nevertheless have $M_i/M_f>1$ (indicating loss of baryonic material). The existing quasi-adiabatic relaxation models do not allow for the existence of such shells, which are however captured well by our new null-offset parameter $q_0$ (see \secref{subsubsec:sim-relax} for a detailed discussion).
We argued that our results could have a significant impact on the applications listed above.

Our analysis also raises some interesting questions, which we briefly discuss before concluding. We noted in \secref{sec:results-rad-dep-qadiab} that, unlike low-mass haloes whose relaxation relation is well-described by \eqn{eq:q3-model}, the radial dependence of the relaxation parameters $q_1$ and $q_0$ in \eqn{eq:chi-linear-q0} for haloes with $M\gtrsim10^{13}\Mh$ shows non-trivial features and oscillations that are not easily captured by simple fitting functions (see Fig.~\ref{fig:fit-func-rf-13514}). These features, which typically occur in the halo outskirts, are likely due to the presence of substructure or recent mergers, which would generically lead to a disturbed dynamical state of the halo. In this work we have not attempted to model these features; it will be interesting in the future to systematically study the dependence of these features on substructure fraction, merger history, locations of shocks, etc.

At the other extreme, in the inner halo of low-mass systems, it is very interesting to ask whether the simple quasi-adiabatic relaxation prescription we have calibrated in this work can naturally lead to cored inner dark matter profiles. Previous attempts at coupling the relaxed dark matter profile to baryonic physics using simple prescriptions have focused on introducing a baryonic dependence of the parameters describing the dark matter profile itself \citep[e.g.,][]{2014MNRAS.441.2986D}. Our approach, on the other hand, parametrises the \emph{physics} of the dark matter response, and it will be interesting to see whether this leads to more robust results for cored inner profiles. For such an exercise, it will also be important to understand the dependence of our calibrated parameters on technical choices defining the sub-grid physics models used in the simulations, which can significantly impact the formation of cores \citep{bfln18}. 

Finally, building a more in-depth understanding of our results will need a physical, preferably analytical, model. One possibility is to use the self-similar approximation \citep[][]{fg84,bertschinger85,launagai+15,shi16}
to model the combined dynamical evolution of dark matter, gas and stars in a halo.
We will report the results of such studies in future work.

\section*{Acknowledgments}
We thank Nishant Singh, Nishikanta Khandai and Kandaswamy Subramanian for useful discussions in the early phases of this work.
We thank the anonymous referee for useful comments that improved the clarity of the presentation.
We gratefully acknowledge the use of high performance computing facilities at IUCAA (\url{http://hpc.iucaa.in}). This work made extensive use of the open source computing packages NumPy \citep{vanderwalt-numpy},\footnote{\url{http://www.numpy.org}} SciPy \citep{scipy},\footnote{\url{http://www.scipy.org}} Matplotlib \citep{hunter07_matplotlib},\footnote{\url{https://matplotlib.org/}} Pandas \citep[][]{reback2020pandas},\footnote{\url{https://pandas.pydata.org/about/}} Schwimmbad \citep{schwimmbad},\footnote{\url{https://joss.theoj.org/papers/10.21105/joss.00357}} H5py,\footnote{\url{https://www.h5py.org/}} Colossus \citep{colossus},\footnote{\url{http://www.benediktdiemer.com/code/colossus/}}  Jupyter Notebook\footnote{\url{https://jupyter.org}} and Code-OSS.\footnote{\url{https://github.com/microsoft/vscode}}

\section*{Data availability}
The IllustrisTNG simulations are publicly available at \url{https://www.tng-project.org/}. The EAGLE simulations are publicly available at \url{https://icc.dur.ac.uk/Eagle/}.

\bibliography{references}

\appendix
\section{Choice of matching algorithm}
\label{sec:apndx-matching}
\begin{figure}
    \includegraphics[width=\linewidth]{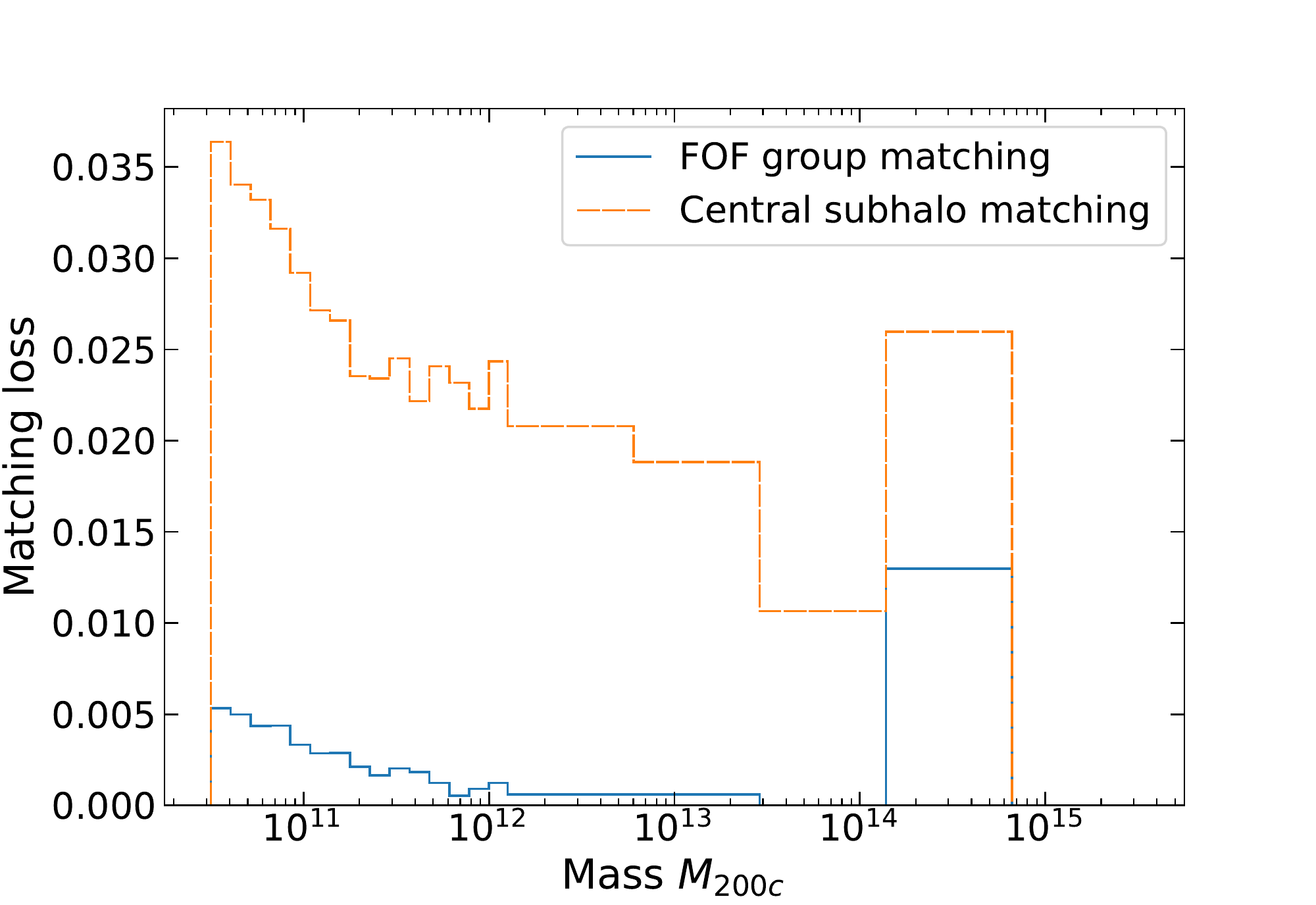}
    \caption{Fraction of haloes in the TNG300-1 hydrodynamical simulation that have not found a match is shown as a function of mass $M_{200c}$.}
    \label{fig:efficiency-mass}
\end{figure}

In this work, we studied the response of dark matter halo to galaxy by comparing the haloes in hydrodynamical simulations to their counterparts in gravity-only runs. We described the matching procedure in \secref{sec:methods-match}, here we discuss some of the specific choices in that procedure.
For this purpose, let us consider the FOF group haloes in TNG300 simulation with $\log M(\Mh)>10.5$.  There are 543588 FOF groups satisfying this criterion in the hydrodynamical run of TNG300, with each of them having more than 500 particles within their $R_{\rm vir}$ in the highest resolution run.
Following the matching procedure described in \secref{sec:method} (requiring only that the matching fraction of the hydrodynamical halo with respect to the gravity-only halo is greater than 0.5), we get a matching halo in the gravity-only run for 541594 of them, leaving out only 1994 haloes unmatched, that is a negligible 0.4\% spread across the mass range (\figref{fig:efficiency-mass}).
However, if we follow the same procedure using matching fraction between the central subhaloes instead of FOF group themselves, then we get a order of magnitude more unmatched haloes as can be seen in the \figref{fig:efficiency-mass}.  

\begin{figure}
\centering
\includegraphics[width=\linewidth]{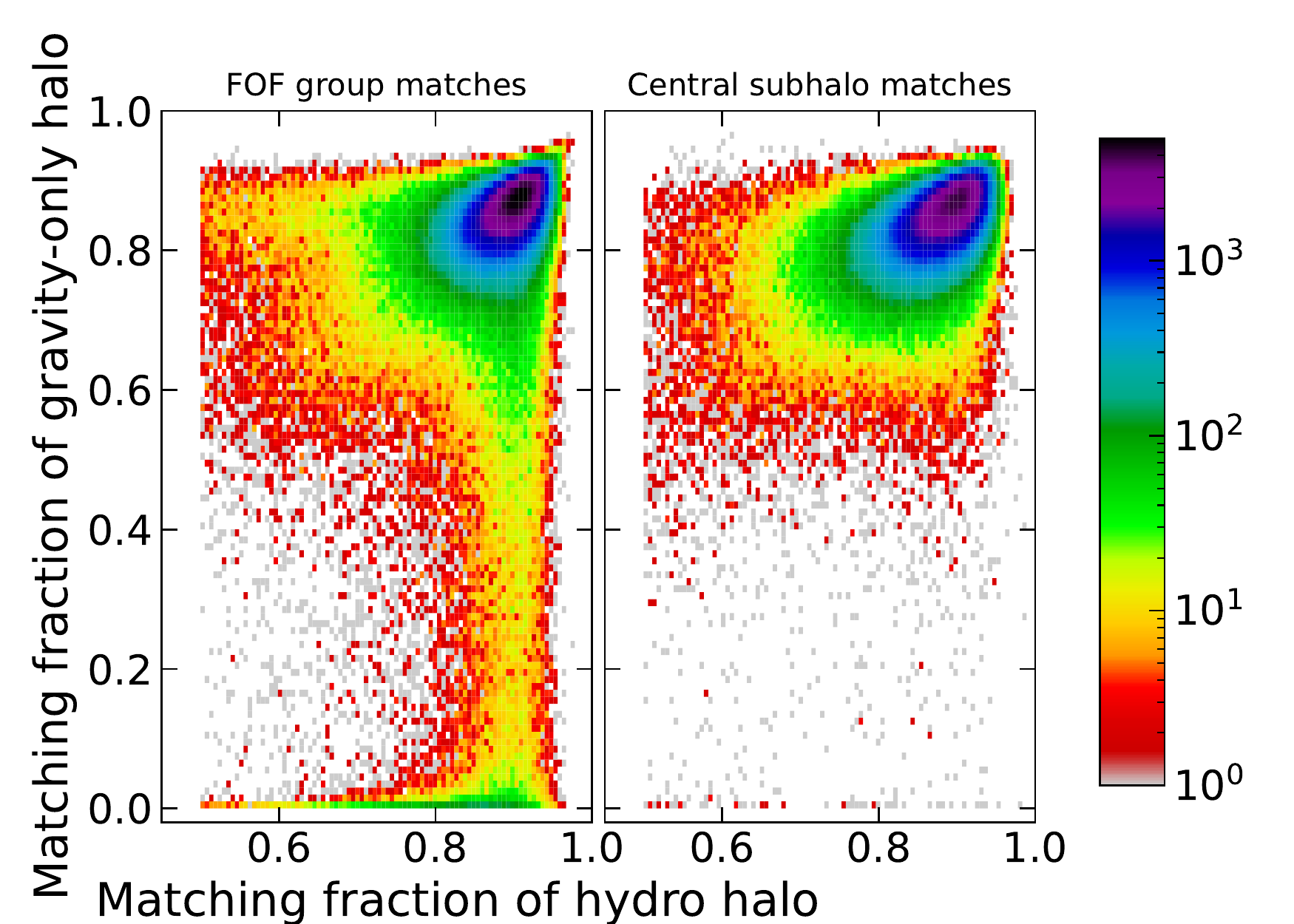}
\caption{Histogram of matching accuracy of haloes in the FOF group matched catalogue (left) and central subhalo matched catalogue (right). The x-axis is the fraction of dark matter particles in the hydrodynamical halo that is also in the gravity-only halo.}
\label{fig:accuracy-hist2d}
\end{figure}

\begin{figure}
    \includegraphics[width=\linewidth]{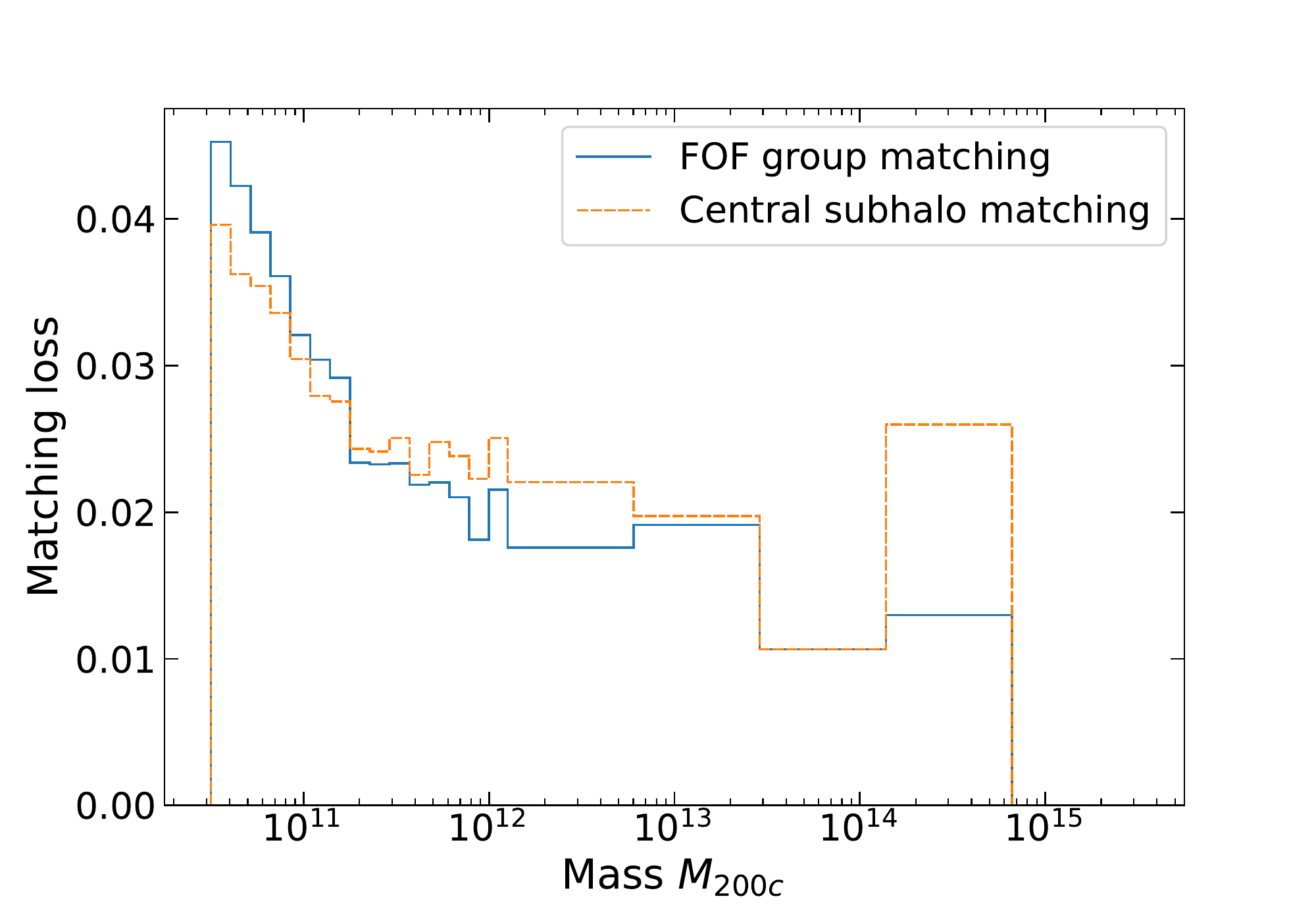}
    \caption{Fraction of haloes in the TNG300-1 hydrodynamical simulation that have not found a match is shown in left panel as a function of mass $M_{200c}$ after removing asymmetrical matches (see text in appendix \ref{sec:apndx-matching}.}
    \label{fig:efficiency-mass-rev}
\end{figure}

While obtaining matches for as many haloes as possible, we also have to ensure the quality of match. To check how well the haloes in each of the pairs in our catalogue are matching we look at the particle matching fraction of each halo in the pair with respect to the other as shown in \figref{fig:accuracy-hist2d}. By definition the gravity-only halo in every pair has atleast 50\% of the dark matter particles of the hydrodynamical halo in that pair. But note that in FOF group matching based catalogue, for a significant number of pairs the hydrodynamical halo in the pair has less than 10\% of the gravity-only halo's particles. By visually inspecting some of those halo pairs we find that, they represent haloes with ongoing merger events. This explains the significant loss in matching efficiency when we used central subhalo matching. 

In the \figref{fig:efficiency-mass-rev}, matching loss as a function of halo mass is shown after removing all those pairs in which the particle matching fraction of the gravity-only halo with respect to hydrodynamical halo is less than 50\%. Since this symmetrical matching condition produces consistent matched catalogue of haloes, we apply this additional matching condition  but stick to matching whole FOF groups. With this procedure, our final matched catalogue for TNG300 contains 524841 pairs, leaving 18747 out of 543588 hydrodynamical haloes unmatched. The additional unmatched haloes primarily reside in dense environments, and we can't find match for those haloes because of the inherent issues with 3d FOF algorithm in dealing with mergers.

\section{Mock particles in a Galaxy-Halo system}
\label{appen:Mock}
Here we validate our method used in extracting the relaxation relation (see \secref{sec:methods-relx-reln}) using mock galaxy-halo systems.
In this, we use Hernquist profile for the initial unrelaxed dark matter halo and similar analytical mass profiles for the baryon components \citep[see appendix of][]{2021MNRAS.507..632P}. We then get the relaxed dark matter profile using quasi-adiabatic relaxation model with different values of $q$. Mock particle data is then generated with these mass profiles for different components of hydrodynamical halo and the corresponding gravity-only halo.

\begin{figure*}
    \centering
    (a) Mass profiles\\
    \includegraphics[clip,trim={0.2cm 0.5cm 0.4cm 0cm}, width=0.49\linewidth]{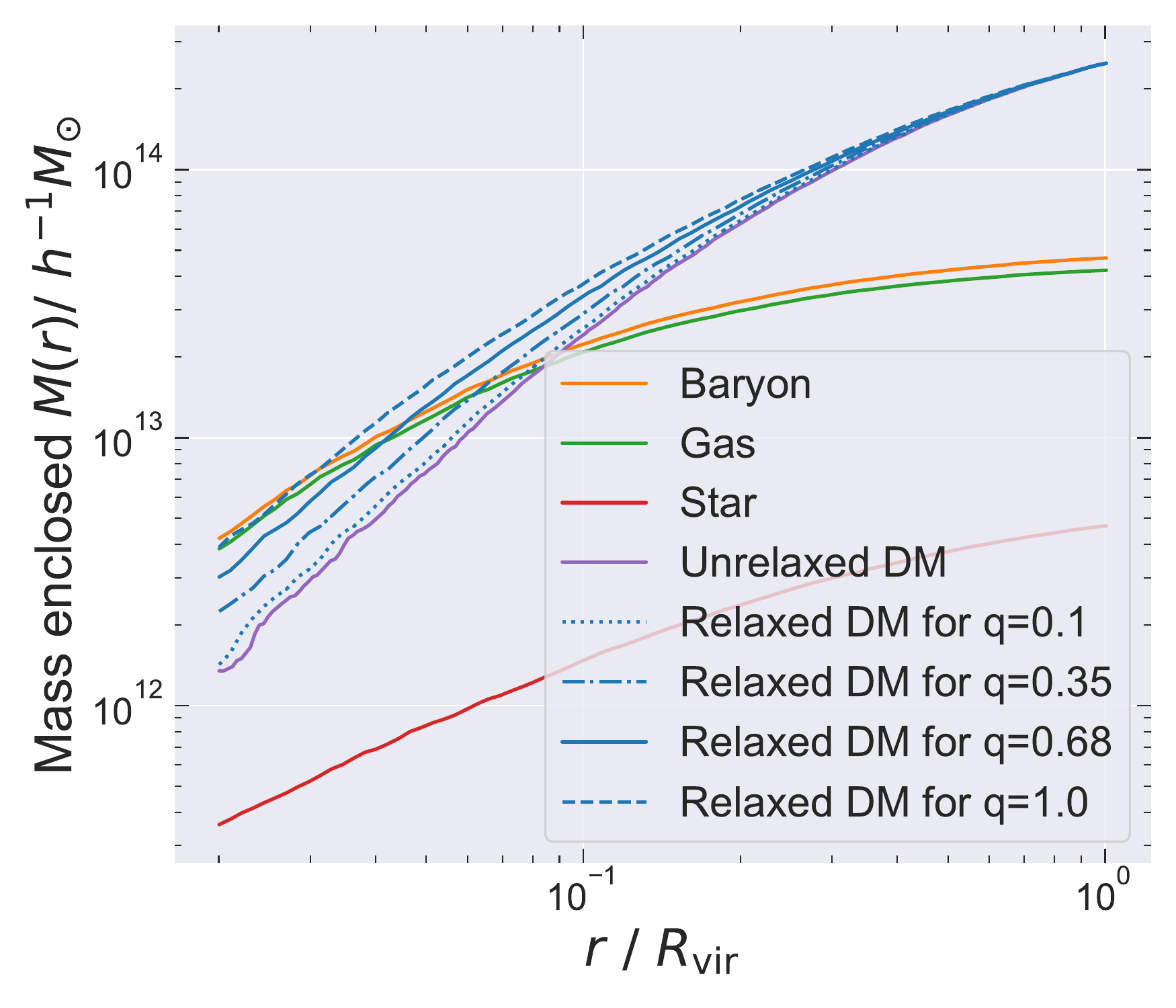}
    \includegraphics[clip,trim={0.2cm 0.5cm 0.4cm 0cm}, width=0.49\linewidth]{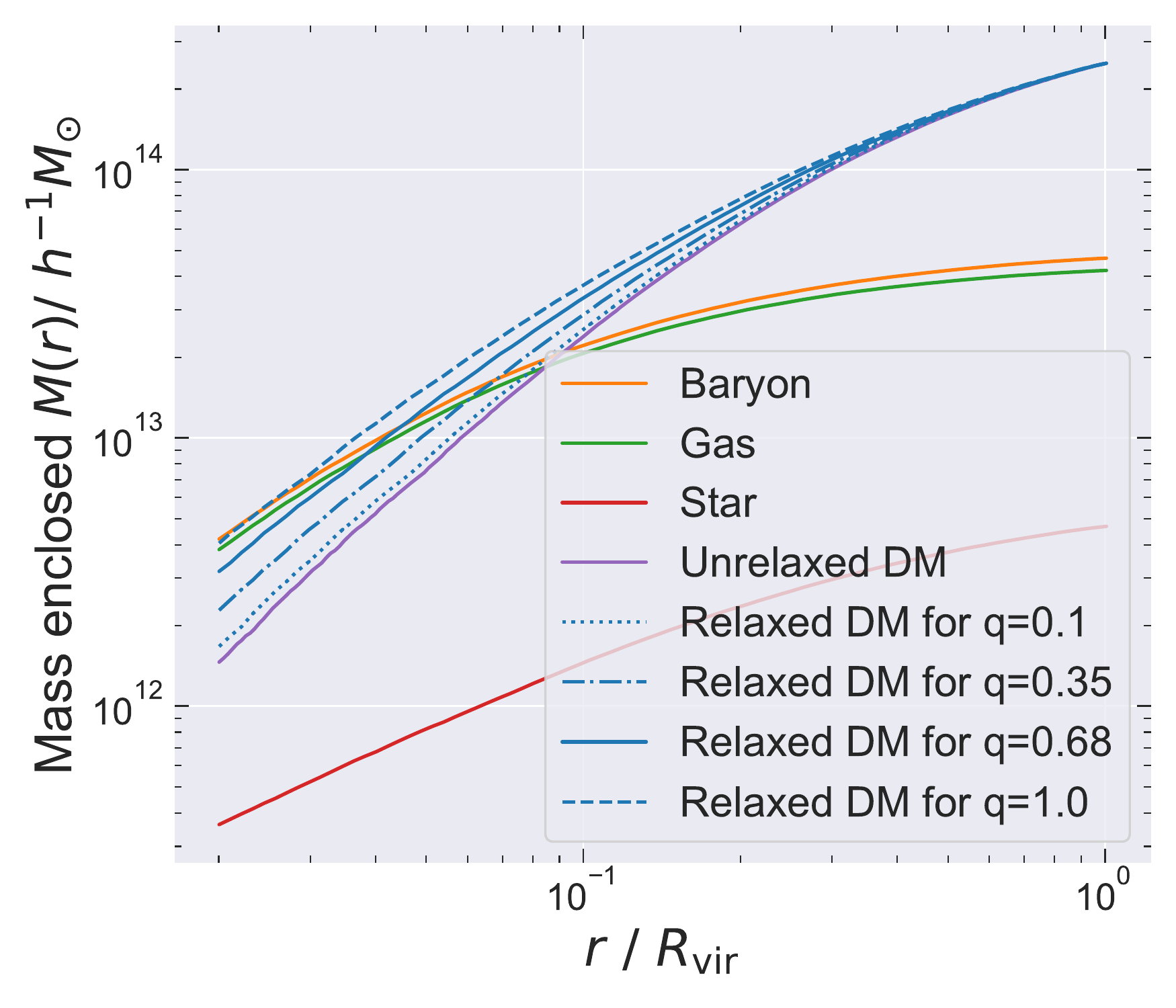}
    (b) Relaxation ratio\\
    \includegraphics[clip,trim={0.2cm 0.5cm 0.4cm 0cm}, width=0.49\linewidth]{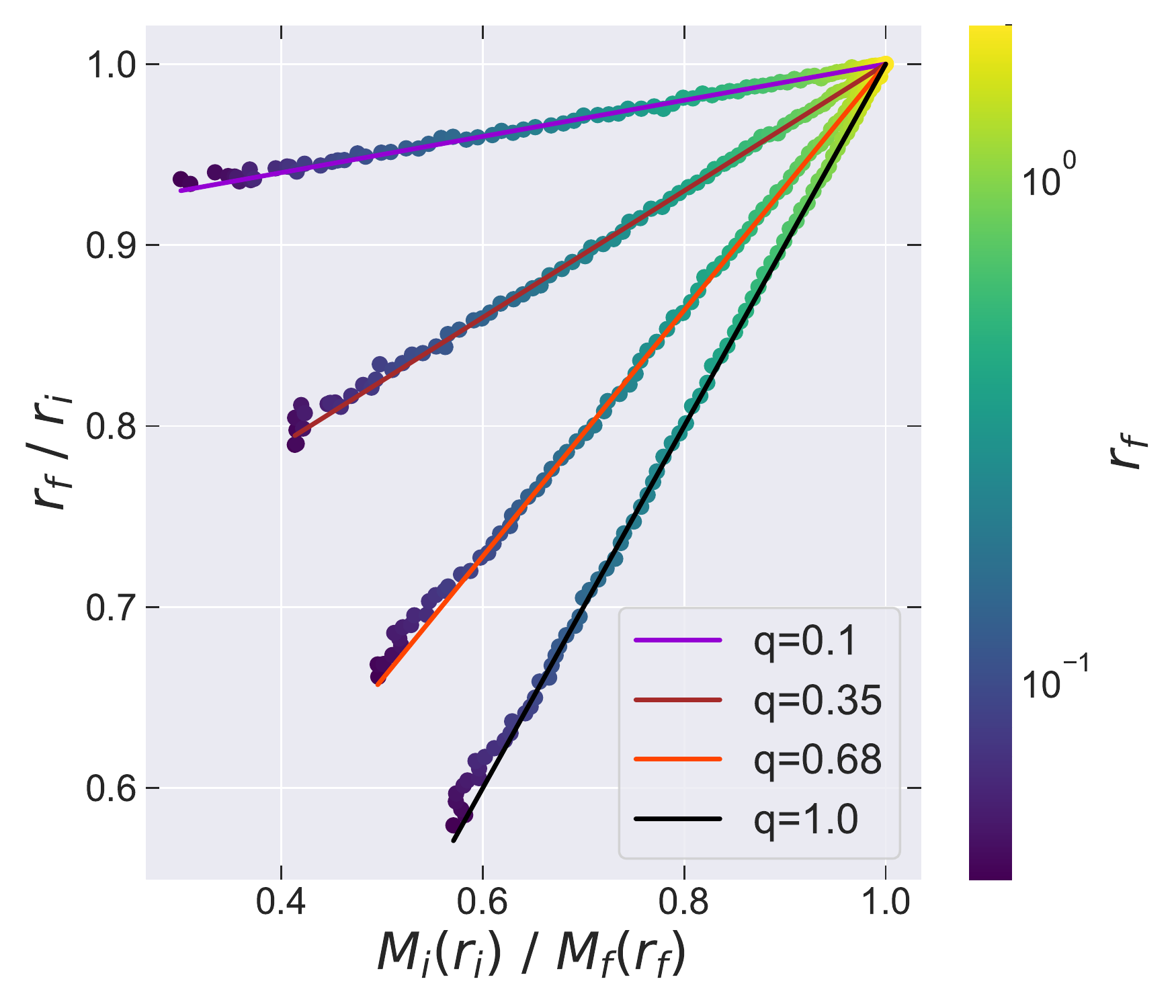}
    \includegraphics[clip,trim={0.2cm 0.5cm 0.4cm 0cm}, width=0.49\linewidth]{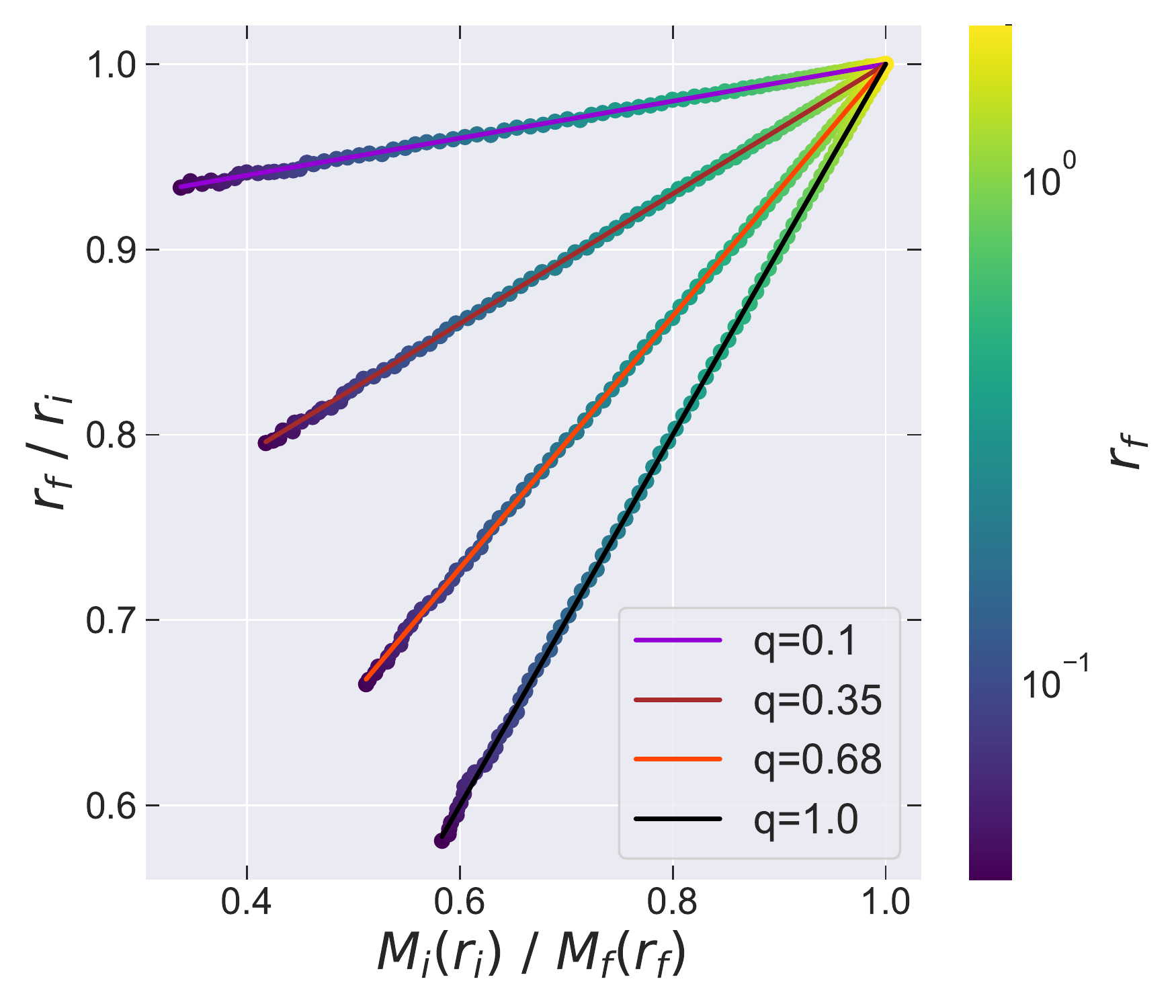}
    \caption{Mock halo with Hernquist profile for unrelaxed dark matter and with the analytical baryon mass profile from \citet{2021MNRAS.507..632P} for gas and star components. The relaxed dark matter profile is modelled by quasi-adiabatic relaxation with different choices of q parameter. Haloes are sampled by \textbf{$10^4$} particles in left panel and \textbf{$10^5$} particles in the right panel.
        (a) Mass enclosed within spherical shell as a function of the shell radius for different components. 
        (b) Relaxation ratio as a function of the ratio of total mass enclosed. The linestyle indicate the different values for the model parameter $q$ used in generating the mock data, while the colourbar shows the final radius. While the solid lines represent the expected relaxation relation for each case, the colored markers show the computed relaxation relation from the mock particle data parametrized by the relaxed radius in units of the virial radius.}
    \label{fig:mock}
\end{figure*}

For those mock halo pairs, we compute the mass profiles and obtain the relaxation ratio as a function of the dark matter shell radius as discussed in \secref{sec:method}. We then repeat this for mock halo pairs generated with different particle resolutions. We find that the computed relaxation relation shown by colored markers matches with expected relaxation relation shown in solid lines; however, the choice of radial bins is limited by the particle resolution (see \figref{fig:mock}).

\section{Using the three parameter model}
\label{sec:apndx-demo}
\begin{figure*}
    \centering
    \includegraphics[width=0.85\linewidth]{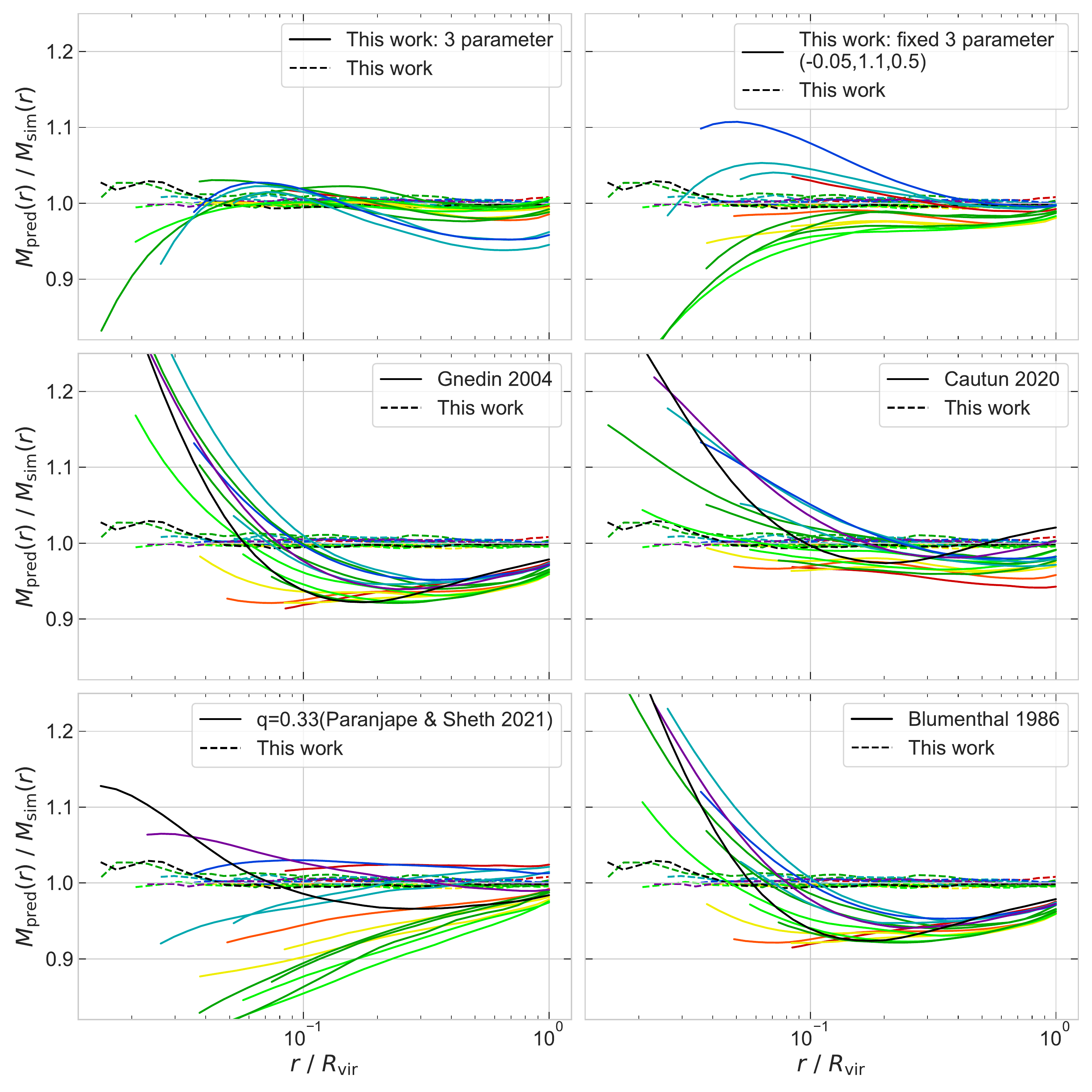}
    \caption{Ratio of the relaxed dark matter mass profile predicted by various models to the dark matter profile found in the hydrodynamical simulation IllustrisTNG is shown by solid curves. Here the mass profiles are stacked across multiple haloes selected by their mass and the color coding follows \figref{fig:mass_bin_label}. Results from our three parameter model \eqn{eq:q3-model} is shown in the upper panel, while the left panel accounts for mass dependence in the parameters, the right panel assumes fixed values namely $q_0\simeq-0.05$, $q_{10}\simeq1.1$ and $q_{11}\simeq0.5$. In the rest of the panels, the corresponding ratio is shown for the mass profiles predicted by few existing models as mentioned in the appendix \ref{sec:apndx-demo}. The result from this work, as shown in bottom panel of \figref{fig:demo-fit} is shown in dashed curves for reference.}
    \label{fig:relxn_models_compare}
\end{figure*}

Here we show the relaxed mass profile predicted by our 3-parameter model \eqn{eq:q3-model} for the haloes with mass, $M\leq10^{13}\Mh$. We follow a similar procedure as described in \secref{sec:mass-prof-demo}; once again we consider the response as a function of only the halo mass and ignore the dependence on other halo properties. For a given halo we obtain the values of the three parameters namely $q_0$, $q_{10}$ and $q_{11}$ by simply interpolating the fitted parameters shown in \figref{fig:3-param-mass-only} as a function of mass. We find that accounting for radial dependence through a simple \eqn{eq:q3-model}, gives a mass profile that is within $10\%$ of the simulation even upto $2 \%$ of virial radii (see upper left panel of \figref{fig:relxn_models_compare}) and within $3 \%$ for the low mass haloes. The upper right panel of \figref{fig:relxn_models_compare}, shows the corresponding profiles assuming a mass independent fixed values for the parameters $q_0\simeq-0.05$, $q_{10}\simeq1.1$ and $q_{11}\simeq0.5$. In addition to these, we also compare with mass profiles predicted by the standard adiabatic relaxation \citealp{1986ApJ...301...27B} and few other relaxation models from \citet{2004ApJ...616...16G}, \citet{2021MNRAS.507..632P} and \citet{2020MNRAS.494.4291C} for the IllustrisTNG haloes.

\section{Dependence on galaxy properties in the cluster scale}
\begin{figure*}
    \centering
    \includegraphics[width=0.32\linewidth]{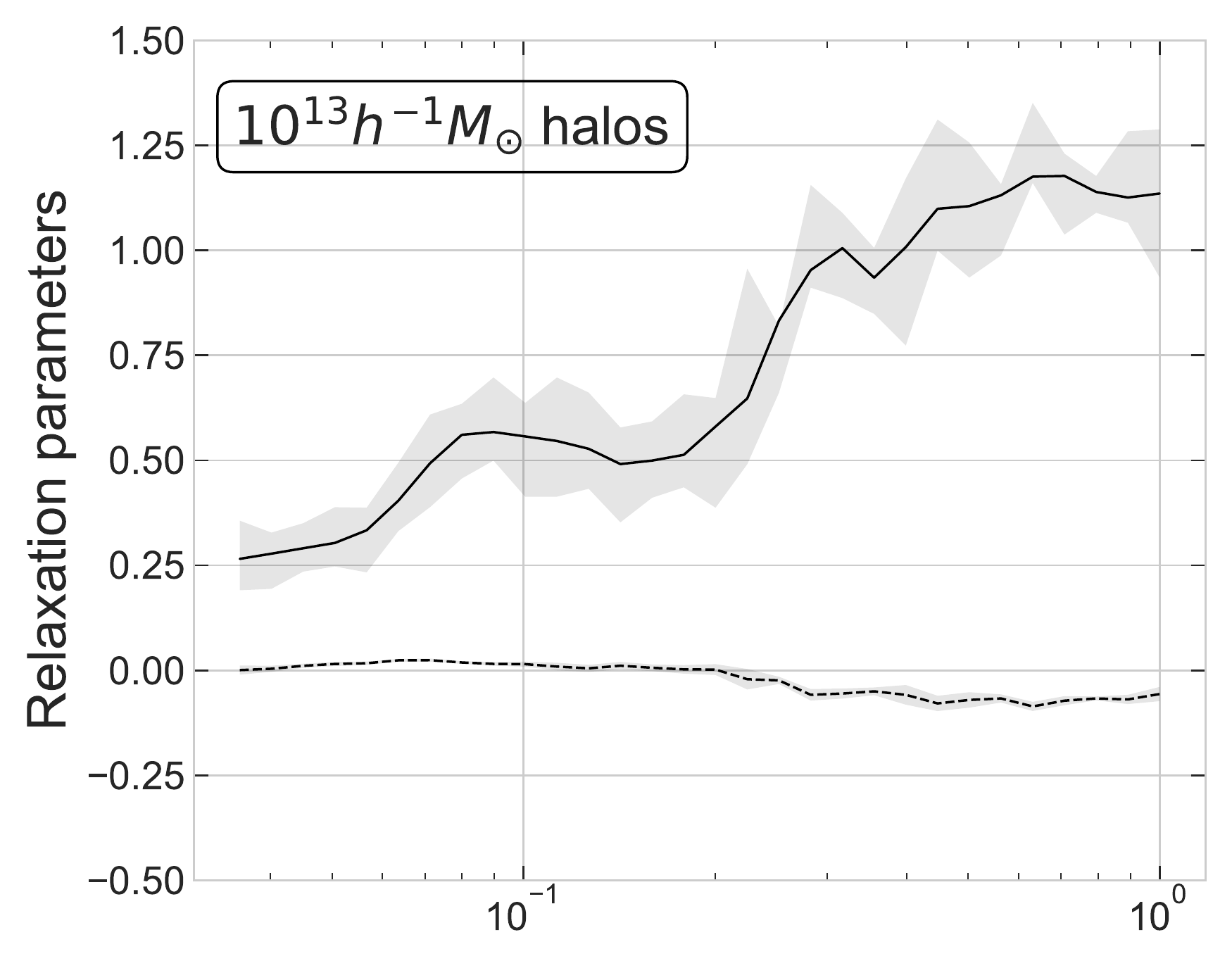}
    \includegraphics[width=0.32\linewidth]{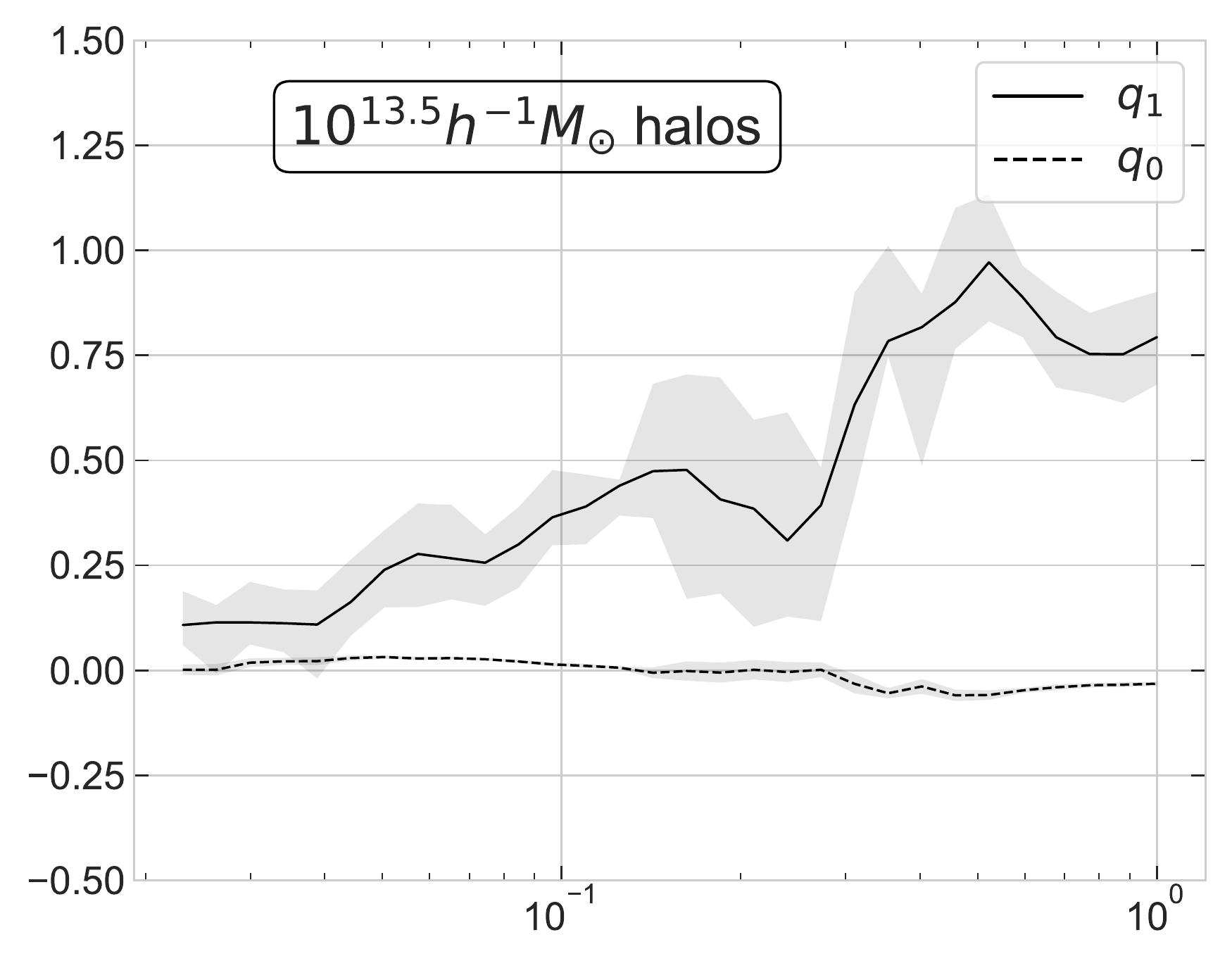}
    \includegraphics[width=0.32\linewidth]{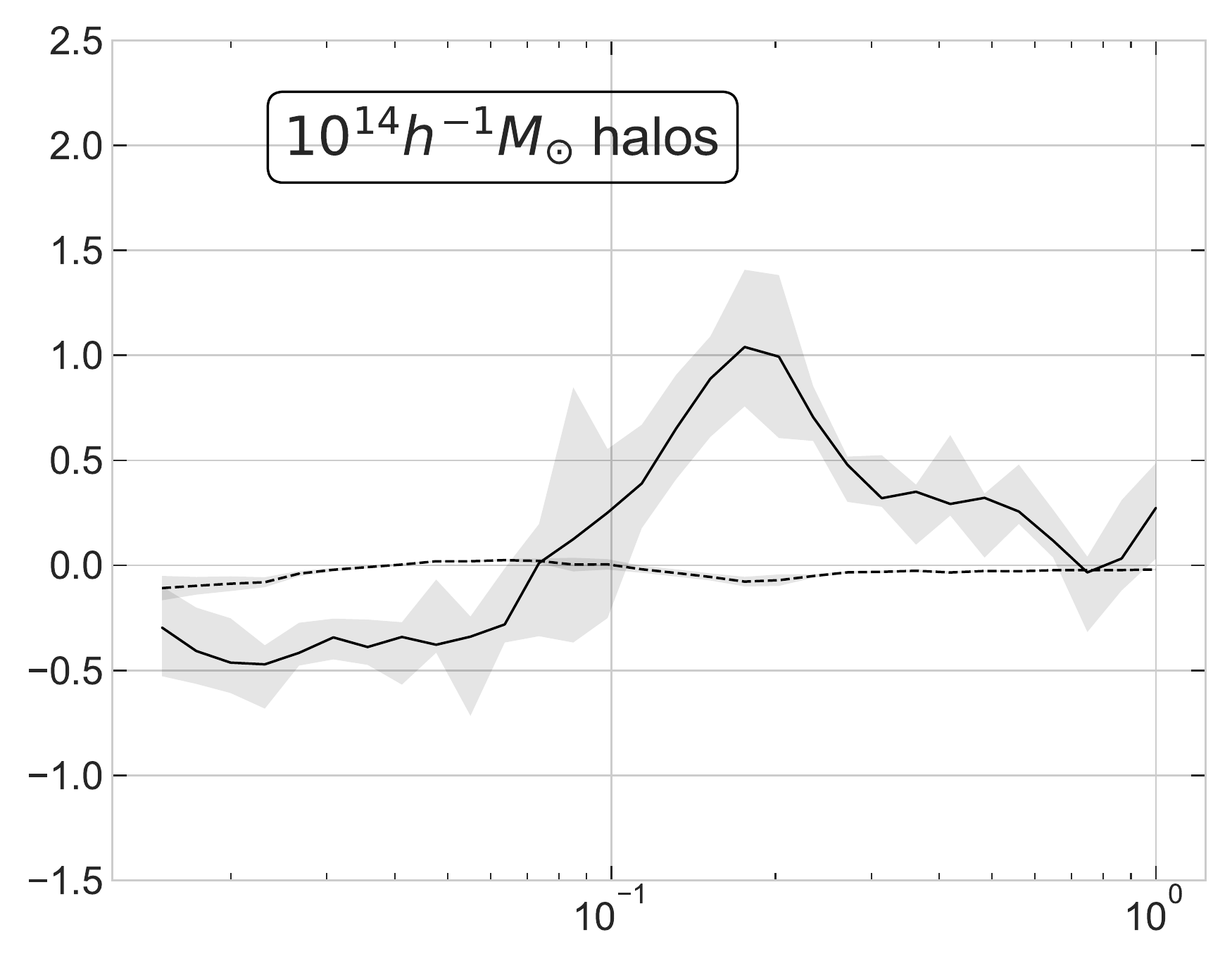}
    
    \includegraphics[width=0.32\linewidth]{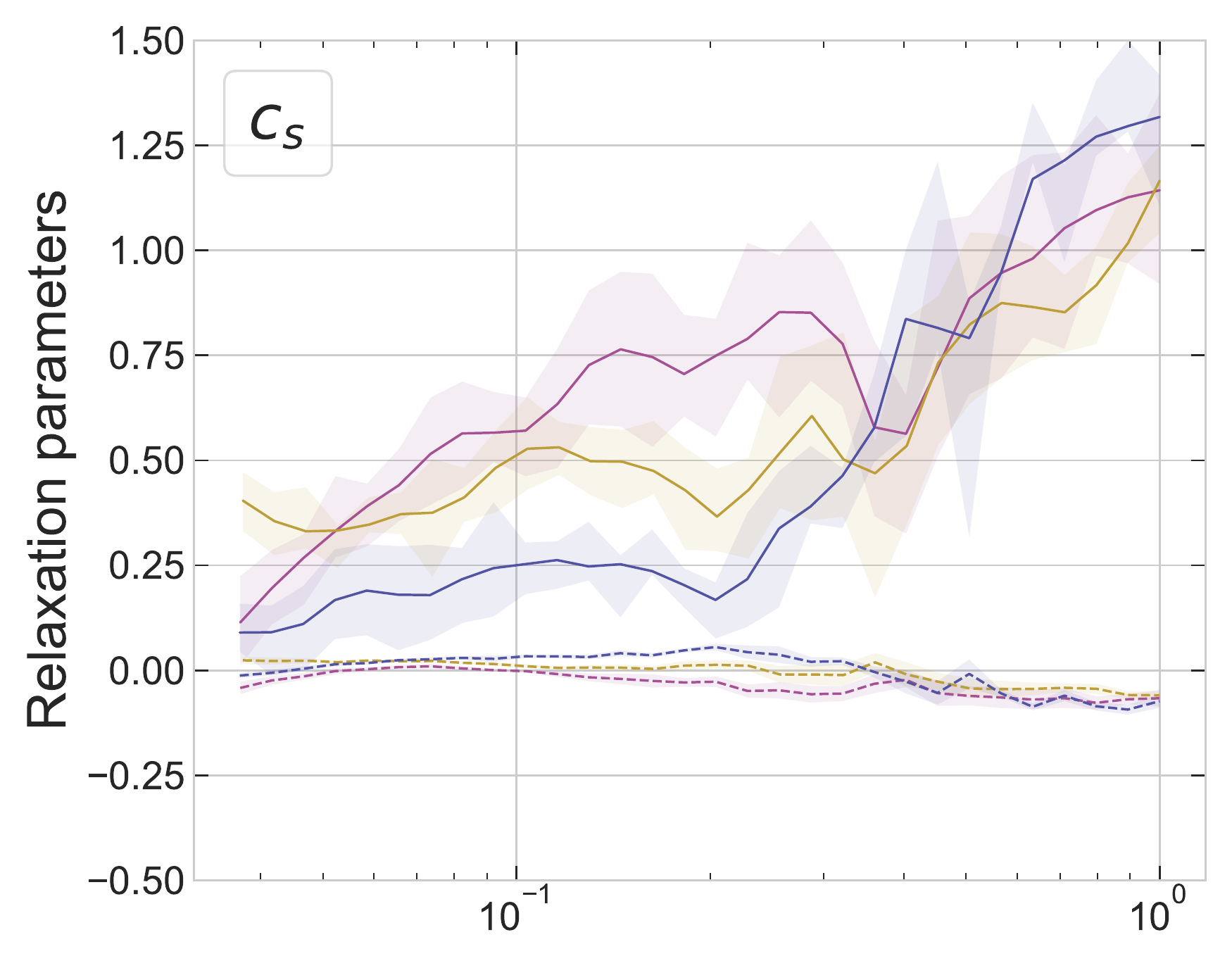}
    \includegraphics[width=0.32\linewidth]{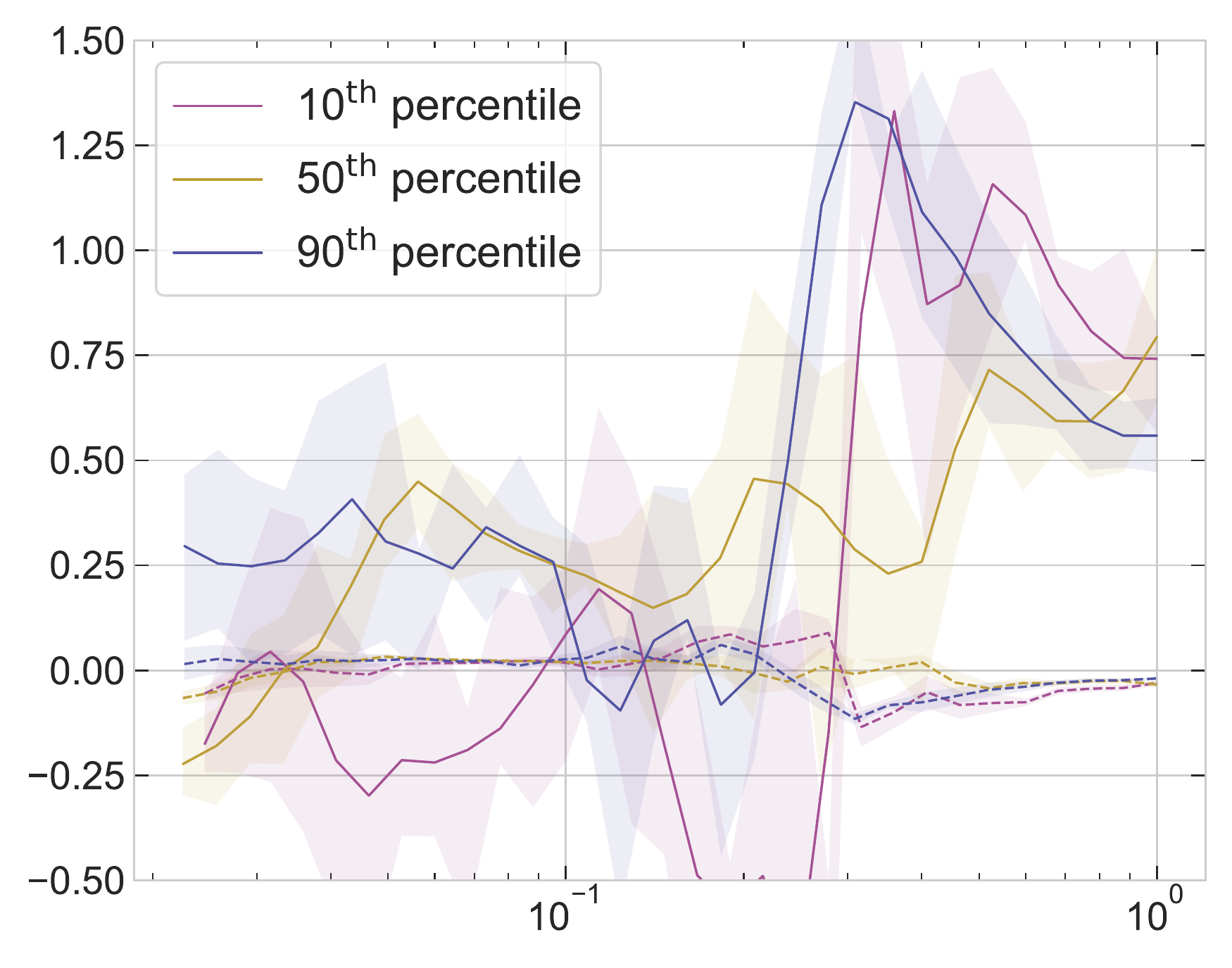}
    \includegraphics[width=0.32\linewidth]{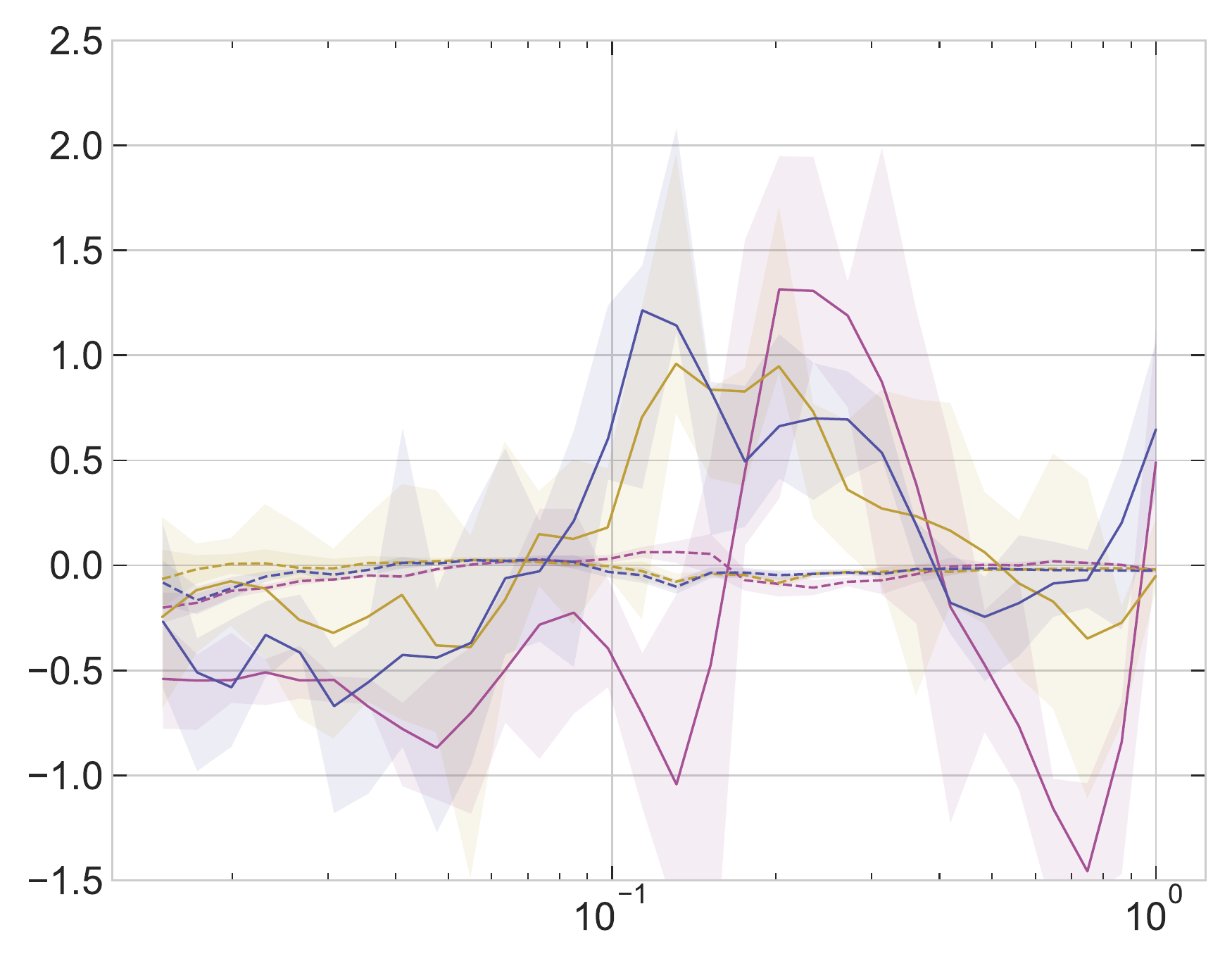}
    
    \includegraphics[width=0.32\linewidth]{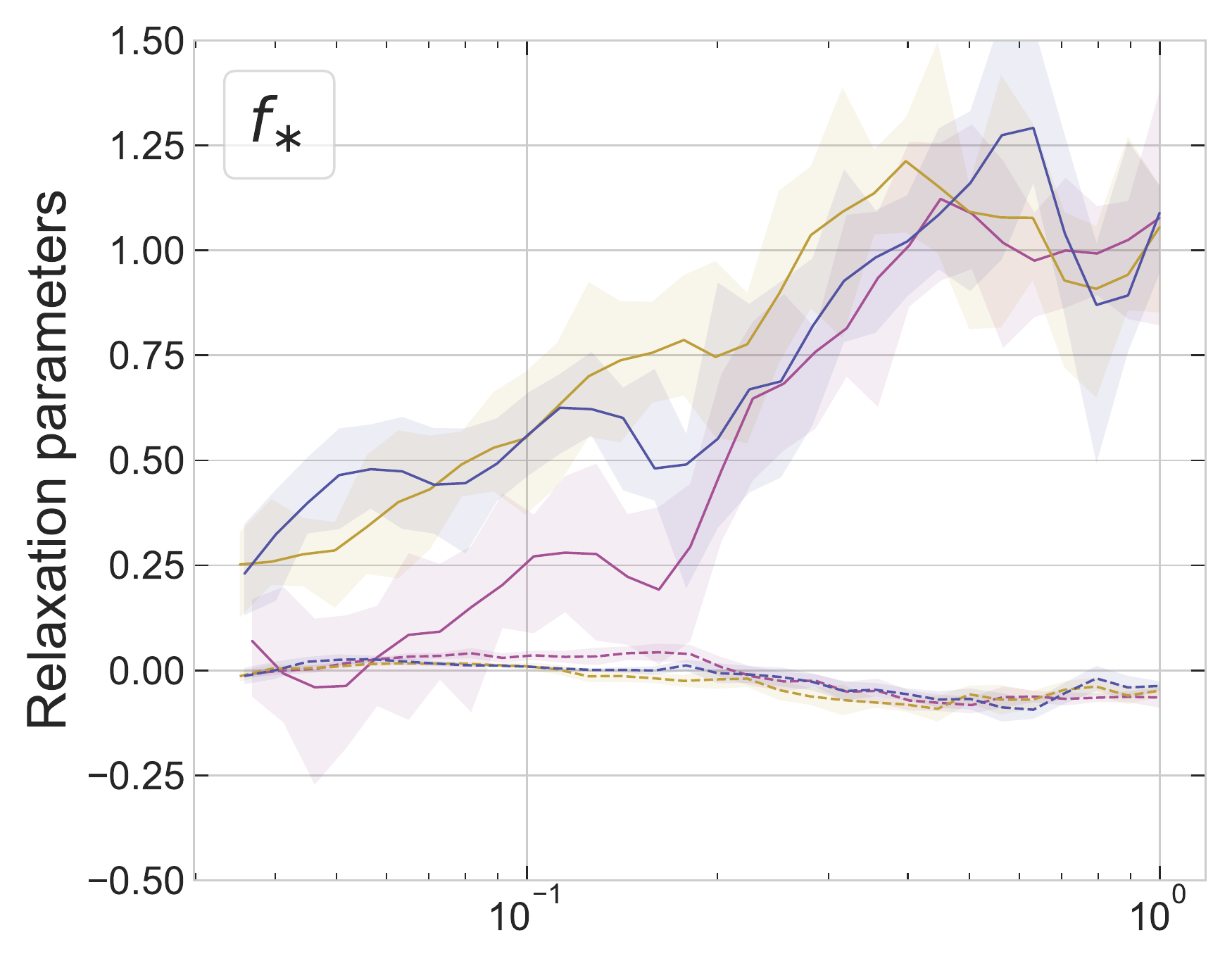}
    \includegraphics[width=0.32\linewidth]{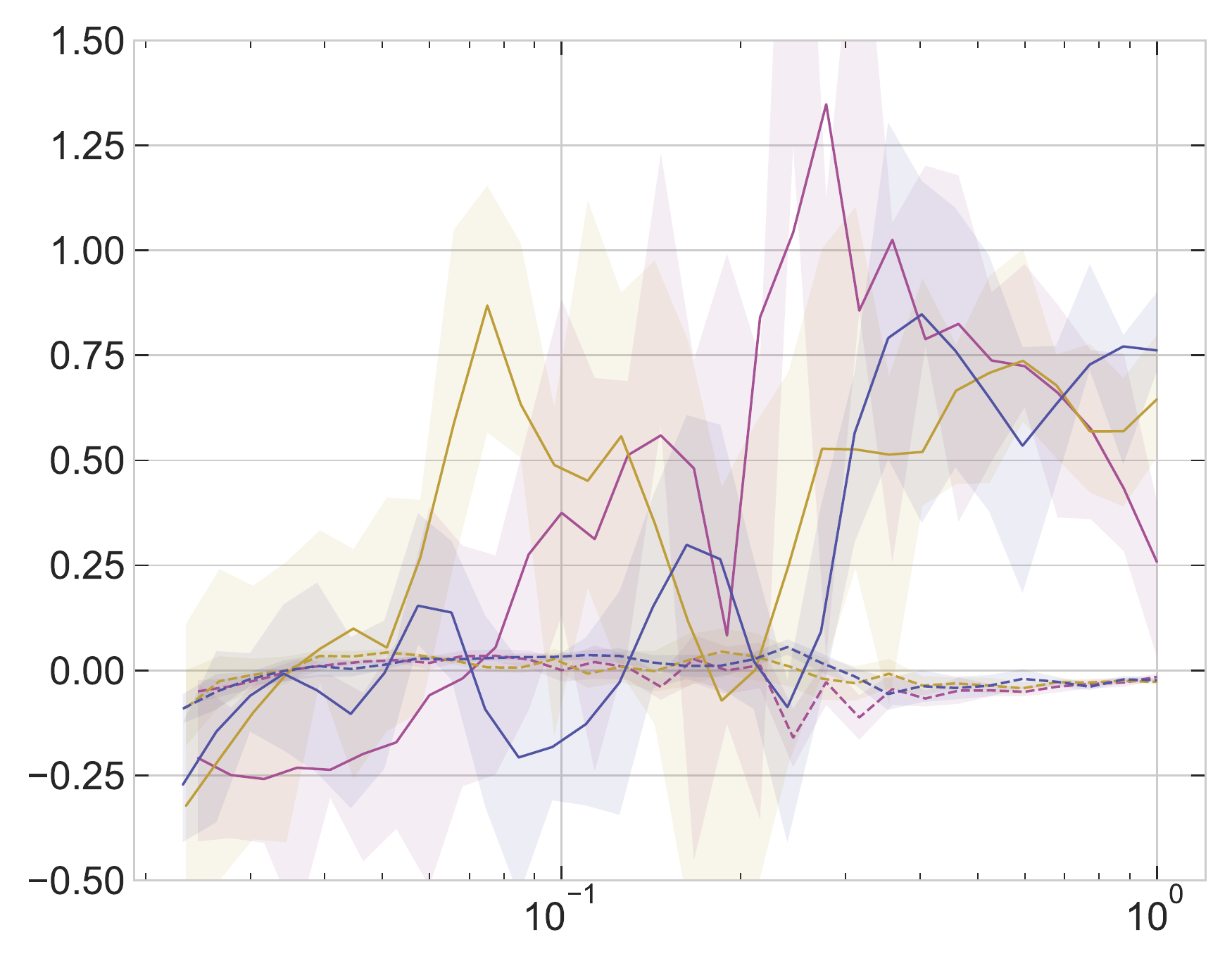}
    \includegraphics[width=0.32\linewidth]{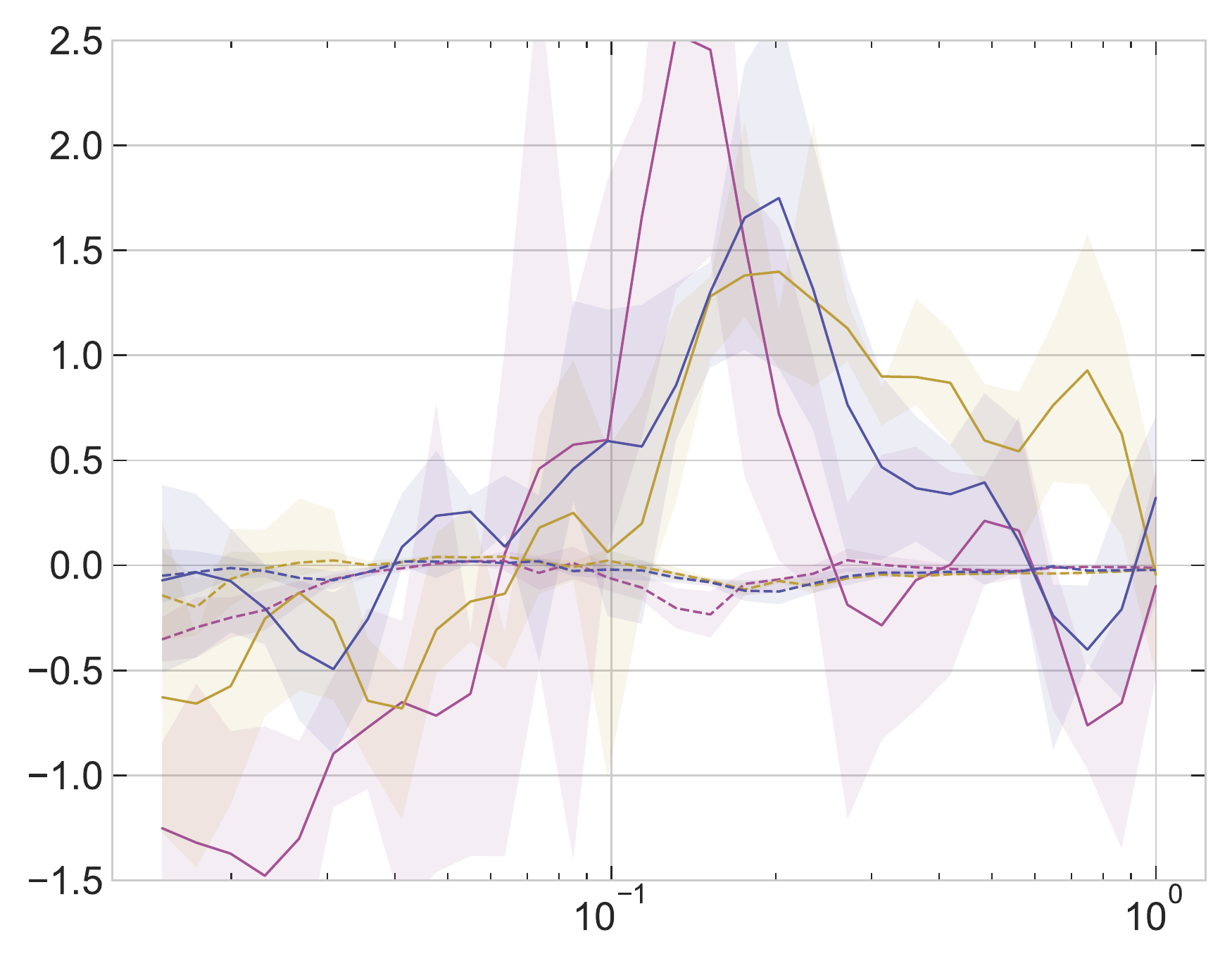}
    
    \includegraphics[width=0.32\linewidth]{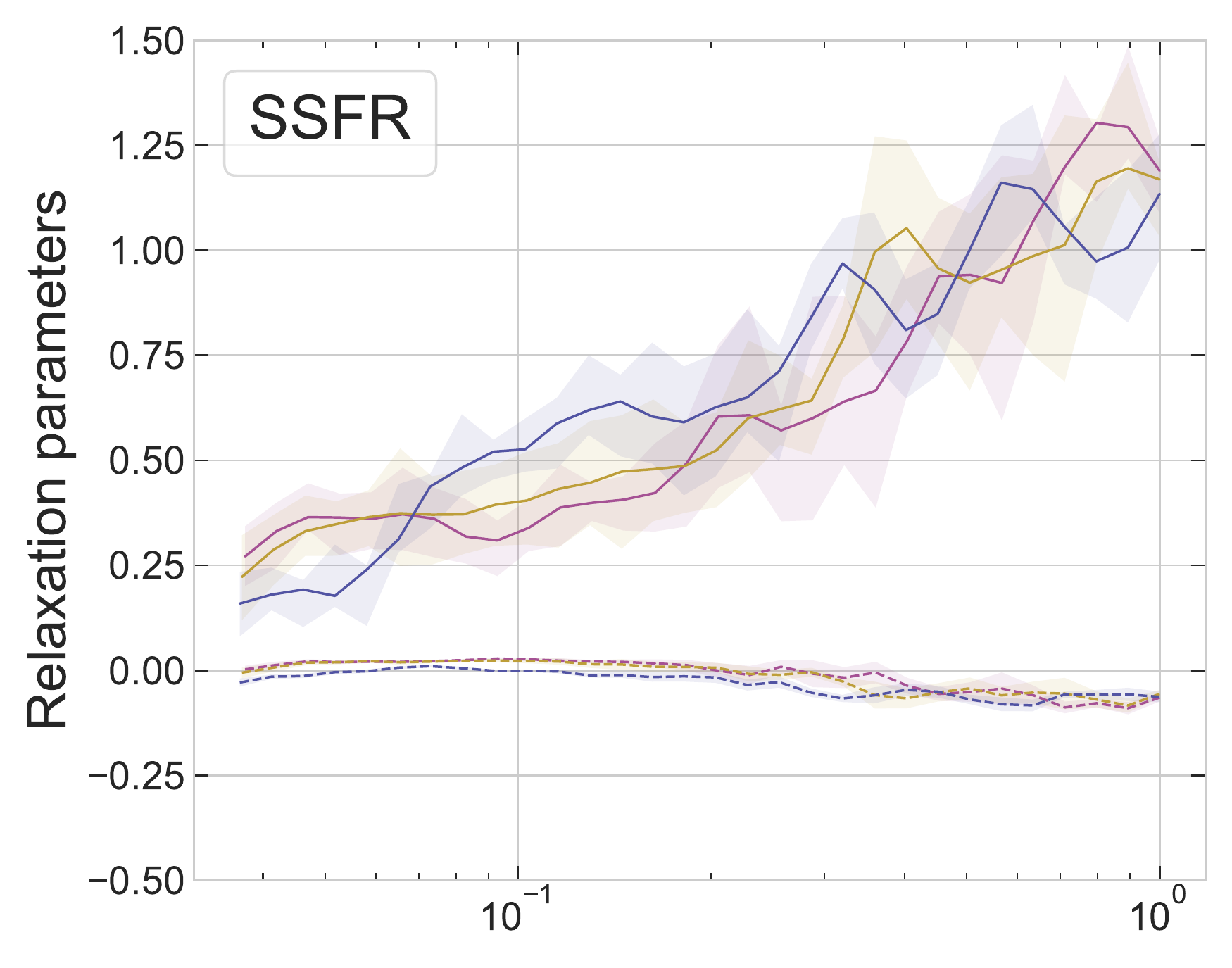}
    \includegraphics[width=0.32\linewidth]{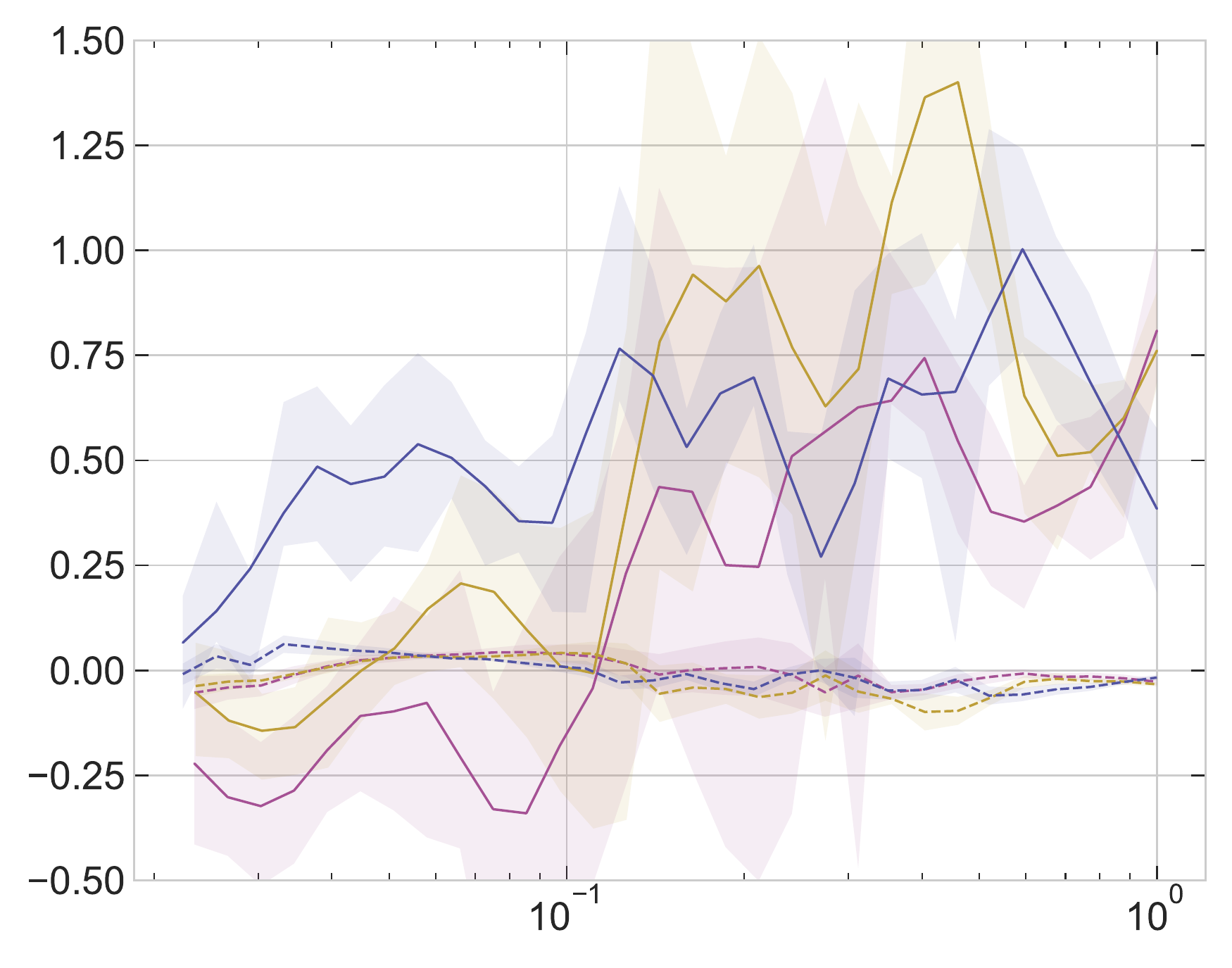}
    \includegraphics[width=0.32\linewidth]{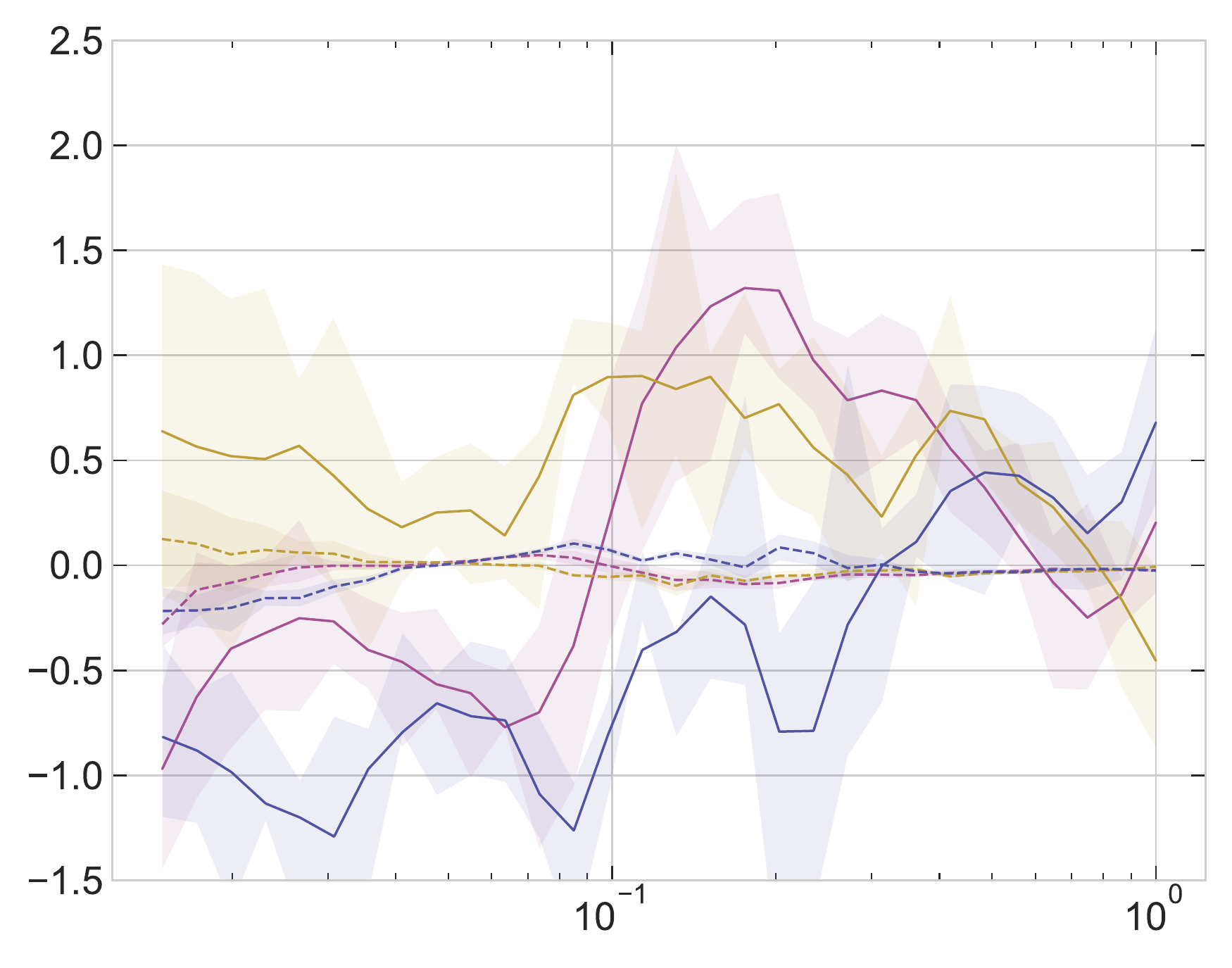}
    
    \includegraphics[width=0.32\linewidth]{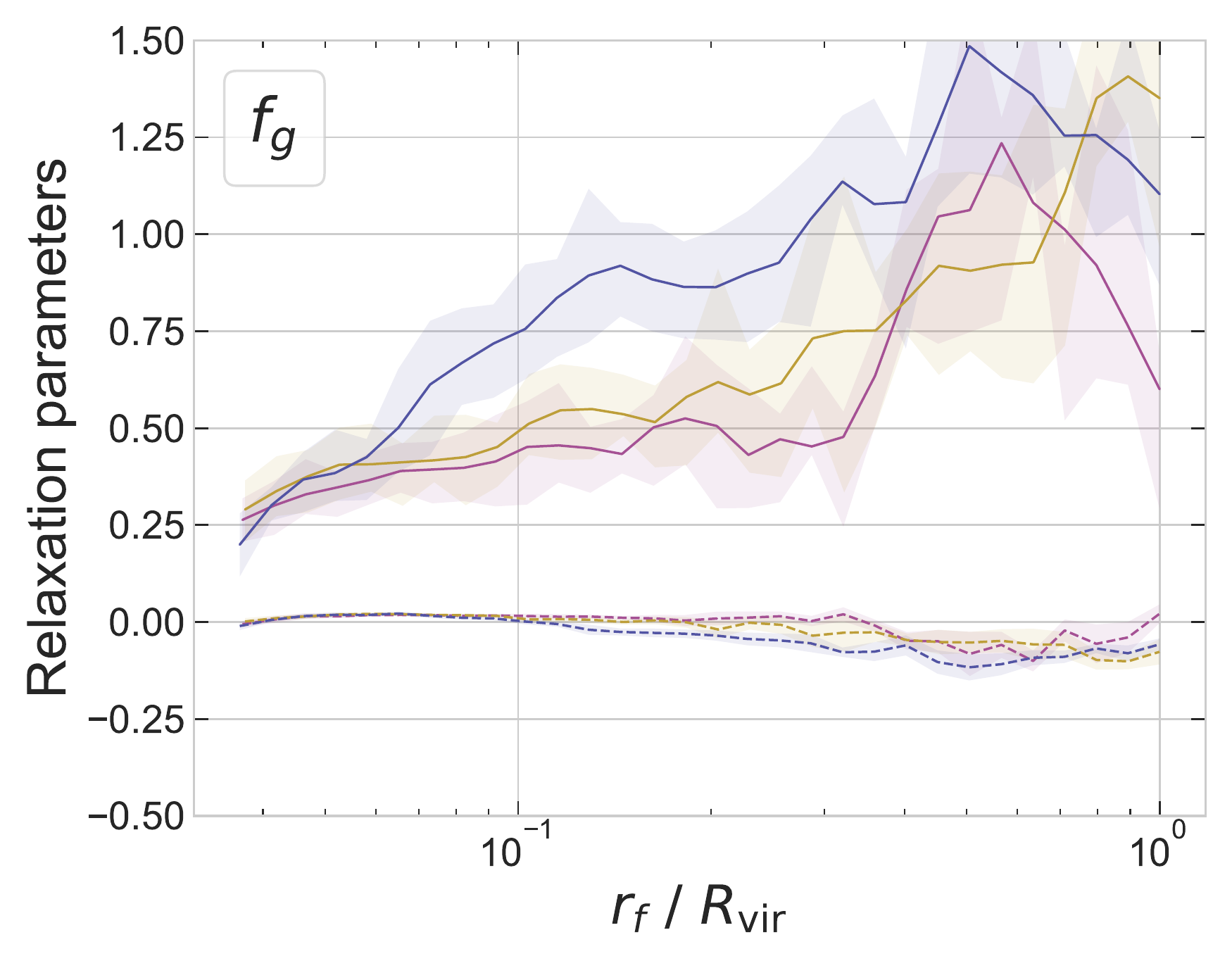}
    \includegraphics[width=0.32\linewidth]{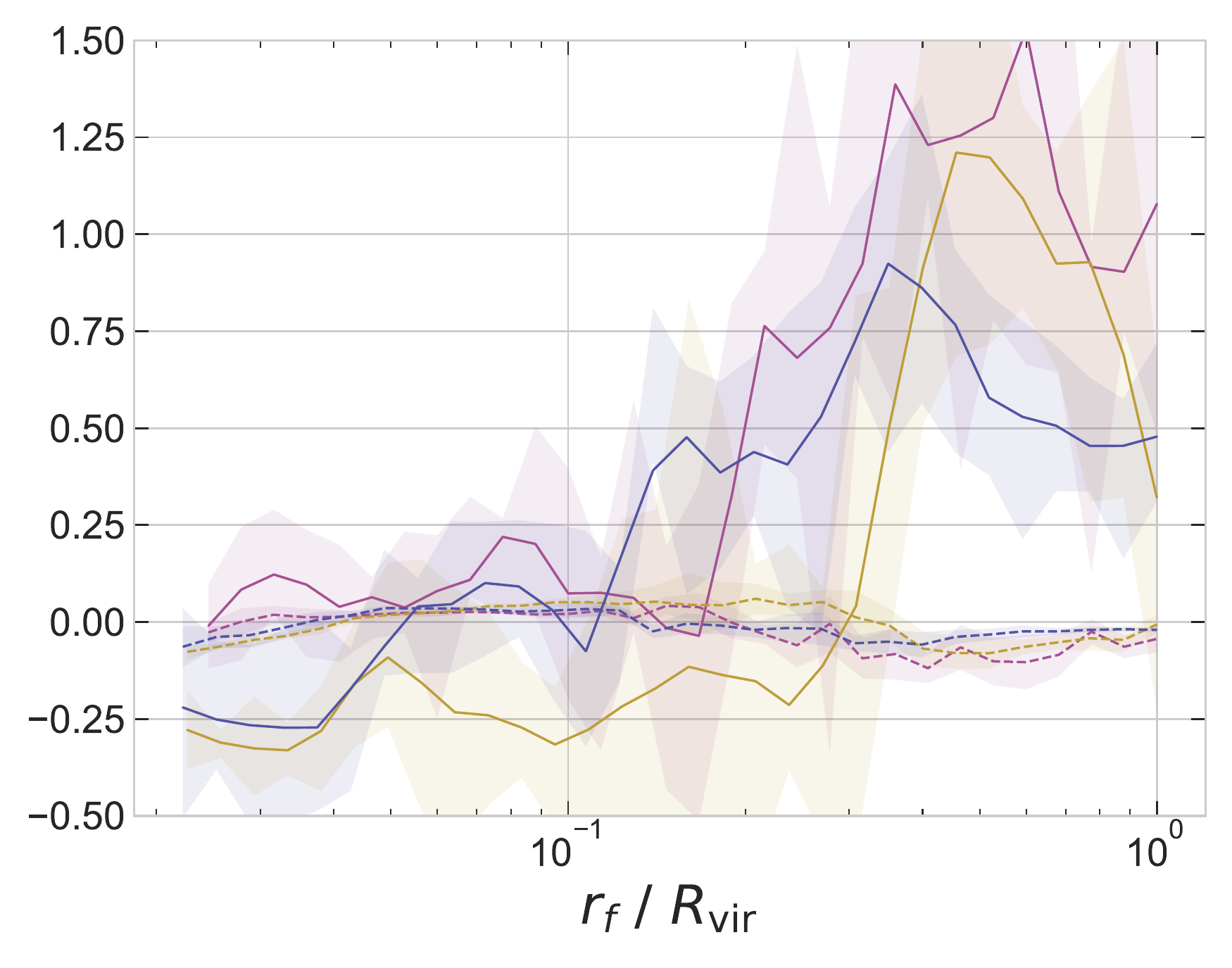}
    \includegraphics[width=0.32\linewidth]{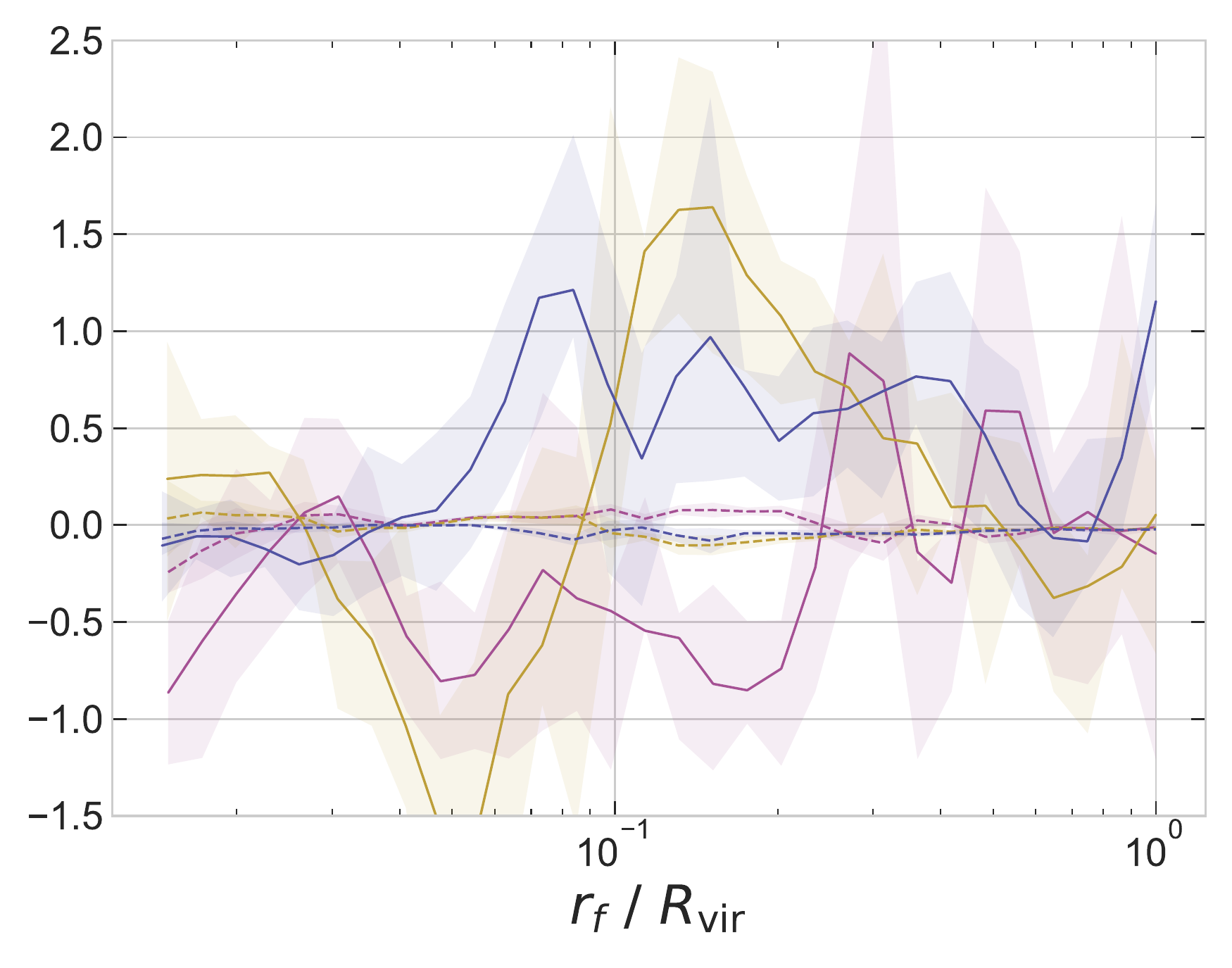}
    
    \caption{Similar to upper panel of \figref{fig:rf-fit-params} with cluster-scale haloes further split by other properties. In top row, only halo mass dependence is shown, whereas in the next three rows, we further show the dependence on halo concentration, stellar mass fraction, specific star formation rate and gas fraction in terms of percentiles respectively.} 
    \label{fig:fit-func-rf-13514}
\end{figure*}

The effect of different halo and galaxy properties on the response of the dark matter is presented here; the relaxation parameters $q_0(r)$ and $q_1(r)$ is shown in \figref{fig:fit-func-rf-13514} as our three parameter description fails for these cluster scale haloes (see section \ref{sec:dep-on-hal-gal-props} for details).

\label{lastpage}

\end{document}